\documentclass{aa}  
\usepackage{lineno}

\linenumbers

\usepackage{placeins}
\usepackage[hidelinks]{hyperref}
\usepackage[dvipsnames]{xcolor}
\usepackage{tabularx}
\hypersetup{colorlinks, linkcolor={blue!70!black}, citecolor={blue!70!black}, urlcolor={blue!70!black}}
\usepackage{graphicx}

\usepackage{txfonts}
\usepackage{array}

\begin{document}
\setlength{\extrarowheight}{5pt}

   \title{Chronology of our Galaxy from \textit{Gaia} colour-magnitude diagram fitting (ChronoGal)}

   \subtitle{II. Unveiling the formation and evolution of the kinematically selected\\ Thick and Thin Discs .}

   \author{Emma Fern\'andez-Alvar\inst{1, 2}, Tom\'as Ruiz-Lara\inst{3,4}, Carme Gallart\inst{1, 2}, Santi Cassisi\inst{5,6}, Francisco Surot\inst{2}, Yllari K. Gonz\'alez-Koda\inst{3}, Thomas M. Callingham\inst{7}, Anna B. Queiroz\inst{1,2}, Giuseppina Battaglia\inst{1,2}, Guillaume Thomas\inst{1,2}, Cristina Chiappini\inst{8}, Vanessa Hill\inst{9}, Emma Dodd\inst{7}, Amina Helmi\inst{7}, Guillem Aznar-Menargues\inst{1}, Alejandro de la Cueva\inst{3}, David Mirabal\inst{1}, M\'onica Quintana-Ansaldo\inst{1}, Alicia Rivero\inst{1,2}.}

   \titlerunning{asdf}
   \authorrunning{asdf}
   
   \institute{
    Departamento de Astrof\'isica, Universidad de La Laguna, E-38205 La Laguna, Tenerife, Spain
   \and 
   Instituto de Astrof\'isica de Canarias, E-38200 La Laguna, Tenerife, Spain
   \and 
   Dpto. de F\'{\i}sica Te\'orica y del Cosmos, University of Granada, Facultad de Ciencias (Edificio Mecenas), E-18071, Granada, Spain
    \and 
    Instituto Carlos I de F\'\i sica Te\'orica y Computacional, Universidad de Granada, E18071, Granada, Spain
   \and
   INAF -- Astronomical Observatory of Abruzzo, via M. Maggini, sn, 64100 Teramo, Italy
   \and
   INFN, Sezione di Pisa, Largo Pontecorvo 3, 56127 Pisa, Italy 
   \and
   Kapteyn Astronomical Institute, University of Groningen, Landleven 12, 9747 AD Groningen, The Netherlands
   \and
   Leibniz-Institut f\:ur Astrophysik Potsdam (AIP), An der Sternwarte 16, 14482 Potsdam, Germany
   \and
   Universit\'e C\^ote d'Azur, Observatoire de la C\^ote d'Azur, CNRS, Laboratoire Lagrange, Nice, France}

   \date{Received --, --; accepted --, --}

 
  \abstract
   {The investigation of the formation, origin and evolution of the dichotomy of the Milky Way's thin and thick disc components has been a focal point of research since it is key to understand the formation of our Galaxy. One of the difficulties is that the populations defined based on their morphology or kinematics show a mix of chemically distinct populations. The age is then a key parameter to understand the disc evolution.}
   {We aim to derive age and metallicity distributions of the kinematic thick and thin discs in order to reveal details of the duration, intensity and relation between the star formation episodes that led to the current kinematic thick/thin disc configuration.}
   {We apply the CMDft.Gaia pipeline based on CMD-fitting technique to derive the dynamically-evolved star formation history (deSFH) of the kinematically selected thin and thick disks. The analysis is based on Gaia DR3 data within a cylindrical volume centered on the Sun, with a radius of 250 pc and a height of 1 kpc.}
   {Our analysis shows that the kinematically selected thick disc is predominantly older than 10 Gyrs and underwent a rapid metallicity enrichment through three main episodes. The first occurred over 12 Gyrs ago, peaking at [Z/H] $\sim$ -0.5 dex; the second around 11 Gyrs ago saw a rapid increase in metallicity up to [Z/H] = 0.0 and a broad spread in [$\alpha$/Fe] from $\sim$0.3 to solar values; and the third, just over 10 Gyrs ago, reached supersolar metallicities. In contrast, the kinematic thin disc stars began forming about 10 Gyrs ago, coinciding with the thick disk's star formation end, characterized by supersolar metallicity and low-[$\alpha$/Fe]. The transition between the kinematic thick and thin discs aligns with the Milky Way's last major merger: the accretion of Gaia-Sausage Enceladus (GSE). We also identify a small population of kinematically selected thin disc stars with high/intermediate-[$\alpha$/Fe] abundances, slightly older than 10 billion years, indicating a kinematic transition from thick to thin disc during the Milky Way's high/intermediate-[$\alpha$/Fe] phase. The kinematic thin disk's age-metallicity relation reveals overlapping star formation episodes with distinct metallicities, suggesting radial mixing in the solar neighborhood, with the greatest spread around 6 Gyrs ago. Additionally, we detect an isolated thick disc star formation event 6 Gyrs ago at solar metallicity, coinciding with the estimated first pericenter of the Sagittarius satellite galaxy.
    }
   {These findings provide high precision age and metallicity distributions from complete samples of stars, in addition to the star formation rates, which are crucial pieces of evidence for chemical evolution models and cosmological simulations, laying the groundwork for further analyses in subsequent studies. }


   \maketitle
%

\section{Introduction}
\label{sec:intro}

Understanding galaxies is crucial for grasping the fundamental principles that govern the Universe. As the primary building blocks of the cosmos, their formation and evolution are intimately connected to the factors that drive cosmic development, like the mechanisms that initiate and sustain star formation. Thus, observing and characterizing galaxies provide valuable data that is critical for testing and refining cosmological models.

The processes that either trigger or quench star formation in galaxies leave distinctive imprints on the chemical and dynamical properties of their stars \citep{freeman02}. One of the morphological characteristics observed in disc galaxies including our own, the Milky Way, is the presence of both a thick and a thin stellar disc (\citealt{T79}; \citealt{GR83}; \citealt{morrison94}; \citealt{dalcanton2002}; \citealt{YD08}; \citealt{comeron11}). For some time analysis of their properties pointed to that thick discs should be among the oldest stellar components of galaxies (\citealt{seth05}; \citealt{elmegreen06}; \citealt{yoachim06}; \citealt{comeron14}; \citealt{haywood13}; \citealt{bovy16}). However, more recent studies combining observations with cosmological simulations consider that star formation in some thick discs could be sustained during longer times \citep{pinna24a}.  The study of thick and thin discs and their relation provides key insights into the conditions and processes prevailing during the early stages of galaxy formation.

Our Galaxy offers a unique opportunity to study the formation and evolution of the disc in far greater detail than is possible in any other galaxy. In external edge-on disc galaxies, a thin and a thick disc component are usually required to fit the stellar light \citep{YoachimDalcaton2006} and the same is true for the Milky Way, where the number density distribution of stars in the solar neighborhood is well fit by two exponential discs (thin and thick) with scale heights of 300 and 900 pc~\citep[][although see more recent measurements in \citealt{khannaa24}]{Juric2008}.  
This constitutes a {\it geometric} definition of the thin and thick disks, which is in fact the only currently possible in the case of external galaxies. In the Milky Way, it was found that thick disc stars tend to move in more eccentric and hotter orbits than thin disc stars, lagging behind the local standard of rest by about 50 km s${^-1}$ in the solar neighborhood \citep{Soubiran2003, bensby03}, and thus, a {\it kinematic} definition of the two components is sometimes used. Finally, the chemical composition of thick disc stars is overall more metal-poor and with a higher content of $\alpha$-elements with respect to iron than that of thin disc stars \citep{Fuhrmann1998, bensby03, Adibekyan2012, Hayden2015_chemicalStructure, Queiroz2020, Fuhrmann2021}, leading to a {\it chemical} definition of the thin and thick disc populations that is possibly the most commonly used today. It is important to bear in mind that geometrically, kinematically and chemically defined thin and thick discs do not result in identical populations \citep{KawataChiappini2016}. In fact, studies exploring stellar populations defined based on their chemical composition show the overlapping of these populations in kinematic and geometric space \citep{anders18}. Therefore, it is necessary to specify how a disc population has been defined in any detailed discussion of their properties and possible origin. 

Even so, the overall hotter kinematics and larger $\alpha$-elements content of Milky Way thick disc stars in any of these definitions are usually interpreted as indicative of they having formed early in the history of the galaxy, during a time of high dynamical activity (based on predictions of cosmological simulations, e.g., \citealt{brook04}) and in an environment enriched by Type II supernovae over a period of rapid star formation before Type Ia supernovae became significant contributors (based on predictions of chemical evolution models \citealt{matteucci86, pagel95, chiappini97, kobayashi11, snaith14, grisoni17, Grand2018_chemdisks, spitoni21}). Stellar age determinations have broadly confirmed this scenario \citep{bensby03, haywood13, Feuillet2018, gallart19, miglio21,  sahlholdt22, xiang22, queiroz23, 2024A&A...687A.168G, Pinsonneault2024arXiv}, showing that high-$\alpha$ stars are typically older than 9-10 Gyr while low-$\alpha$ stars are overall younger covering a broader range of ages. However, there is no total consensus on the idea that the thick disc is a separate, discrete component of the Milky Way and some authors advocate for a continuity in the properties of thick and thin disks, and thus in their formation process \citep{bovy12, Recio_Blanco2014, sharma21, prantzos23}.

Since the advent of all sky surveys like SDSS/SEGUE \citep{yanny09}, SDSS/APOGEE \citep{majewski17}, RAVE \citep{rave20}, GALAH \citep{galah21}, Gaia-ESO \citep{gaiaeso22} or \textit{Gaia} \citep{gaia16b} the characterization of Milky Way stellar populations has significantly improved, but the evidences provided are still insufficient to fully answer fundamental open questions. For instance, there is a great uncertainty about the exact timing and duration of star formation in each disk, how the observed properties have varied over time, and what was the role of mergers, gas accretion and secular evolution to shape both disks.  

To understand the true sequence of events that led to the present-day configuration of the Milky Way, precise age determinations for large and unbiased samples of stars are necessary. Until recently, precise stellar ages were achieved using techniques like isochrone fitting (\citealt{sahlholdt22}; \citealt{xiang22}, \citealt{queiroz23}) or asteroseismology (\citealt{chaplin13}; \citealt{lebreton14}; \citealt{silva15}; \citealt{aldo17}; \citealt{mackereth19}; \citealt{miglio21}), but they were available for relatively small samples, usually tailored to spectroscopic surveys affected by restrictive selection functions, thus limiting the information that their analysis could provide. A promising pathway to overcome these limitations has been opened with the ChronoGal project \citep{2024A&A...687A.168G}, which uses the color-magnitude diagram (CMD) fitting technique \citep{Tolstoy1996_method, gallart96b, Hernandez1999, Dolphin2002MNRAS_method, aparicio04, cignonitosi2010} on \textit{Gaia} data to obtain precise age and metallicity distributions of vast samples of Milky Way stellar populations. Previously successful in studies of Local Group dwarf galaxies \citep[e.g.][]{gallart99, noel07, bernard07, hidalgo09, monelli10, Weisz2012, hidalgo13, Cignoni2013, weisz14, Cole2014, gallart15, Skilman2017, McQuinn2024}, this technique was not widely used for the Milky Way\footnote{Some early attempts from Hipparcos and Gaia DR1 data are those of e.g. \citet{Hernandez2000_Hipparcos, bertelli2001AJ....121.1013B, cignoni2006, Bernard2018IAUS}} due to the lack of precise distance measurements for significant samples of stars until the advent of the \textit{Gaia} mission, which is now providing unprecedented precision in parallax and photometry for stars within a large Milky Way volume. 

After the pioneer works by \cite{gallart19} and \cite{ruizlara20}, a more refined and optimised suite of codes, CMDft.Gaia \citep{2024A&A...687A.168G}, has been developed to deliver the required age precision (better than 10$\%$) and accuracy ($\sim$6$\%$) for complete samples of stellar populations in the Galaxy, providing a crucial missing piece for reconstructing our Galaxy's history. The Milky Way stellar samples analysed in this way contain a mix of populations, both formed locally and migrated from their birth site via dynamical processes. For this reason, we refer to the SFHs derived by ChronoGal as "dynamically evolved star formation histories" (deSFH), providing the amount of mass, per unit time and metallicity, that has been transformed into stars somewhere in the galaxy to account for the stars that are populating today the studied volume. From the deSFH, the distribution of ages and metallicities of the stars in the analyzed sample can be derived.

A detailed explanation of CMDft.Gaia is presented in the first paper of this series (\citealt{2024A&A...687A.168G} - Paper I) which presents and discusses the most detailed deSFH ever inferred for the solar neighborhood, within a spherical volume of 100 pc from the Sun, using the exquisite CMD of the \textit{Gaia} Catalogue of Nearby Stars \citep{GCNS21}. In the present paper, the second of the series, we aim to extend this analysis by deriving the deSFHs for the thin and thick disks, selected kinematically over a broader volume, to shed new light on their origins and evolutionary pathways.

This paper is organized as follows: Section \ref{sec:data} details the data analyzed, the kinematic selection criteria applied, and the methodology used to derive the deSFHs. Section \ref{sec:results} presents the resulting deSFHs, with a detailed evaluation of the age-metallicity distributions of the thin and thick discs and their correlation with chemical abundance trends derived from spectroscopic databases. Section \ref{sec:discussion} discusses the implications of these findings for our understanding of the formation of the thick and thin disks. Finally, Section \ref{sec:conclusions} summarizes the main conclusions of the study.

\section{Data and methodology}
\label{sec:data}

\subsection{Volume definition and quality cuts.}

We select stars within a cylindrical volume with a radius of 250 pc and a total height of 1 kpc, centered on the Sun. This very local volume is chosen to ensure at the same time: i) high precision in astrometric parameters and photometry, enabling us to derive accurate CMDs; and ii) the required number of stars (increased by two orders of magnitude compared to the sample analysed in Paper I) to separate our final sample into the three main stellar populations it contains (thin disk, thick disk, and halo) and accurately infer the deSFH for each disc component separately.

An additional reason for this selection is that a nearby sample minimizes the effects of extinction across the Galactic plane. However, extinction is not entirely negligible even within 250 pc of the Sun. To address this, we correct the \textit{Gaia} DR3 photometry using state-of-the-art extinction maps. Specifically, we apply both the map by \cite{lallement22} and \cite{vergely22} (L22 from now on), and the map by \cite{green19} (G19). For the L22 map, to enhance efficiency in determining the extinction for each star, we pre-compute mean extinction values within healpixels across Galactic latitude and longitude at various distances from the cubes provided at three different resolutions by the authors. For the G19 map, we use the \textit{dustmaps-bayestar}\footnote{https://dustmaps.readthedocs.io/en/latest/index.html} Python package, and adjust some values where extinction was reported as zero, as these cases were likely underestimated (private communication). The results obtained with each of the extinction maps are similar and consistent. Thus, from now on we will focus on the results obtained when applying L22 extinction corrections as this map provides extinction measurements for all the sky while G19 only covers the North Hemisphere. 

To convert the extinction in E(B-V) to $A_{G}$, we follow the prescription suggested by the \textit{Gaia} Collaboration, as detailed in \cite{Fitz19}. Consistent with \cite{gallart19} and \cite{ruizlara20}, we restrict our selection to stars with relatively low extinction, specifically $A_{G}$ < 0.5. 

We also clean our sample from stars that have unreliable photometry, by selecting those verifying 
\begin{equation*}
    0.001 + 0.039 \times \tt{bp\_rp} < \log(\tt{phot\_bp\_rp\_excess\_factor})
\end{equation*}
 and 
 \begin{equation*}
     \log(\tt{phot\_bp\_rp\_excess\_factor}) < 0.12 + 0.039 \times \tt{bp\_rp}
 \end{equation*}

 Finally, we compute distances as the inverse of the parallax, selecting only stars with a relative parallax uncertainty of less than 20\%. This criterion ensures that the inverse parallax provides a reliable approximation of the distance. We account for individual zero-point offsets as estimated by \cite{lindegren21} and include a systematic uncertainty of 0.015 mas in the zero-point \citep{lindegren21}.

\subsection{Kinematic selection}
\label{sec:method_selection}

\begin{figure*}
\centering
\includegraphics[width=1.1\textwidth, trim=0 50 0 30]{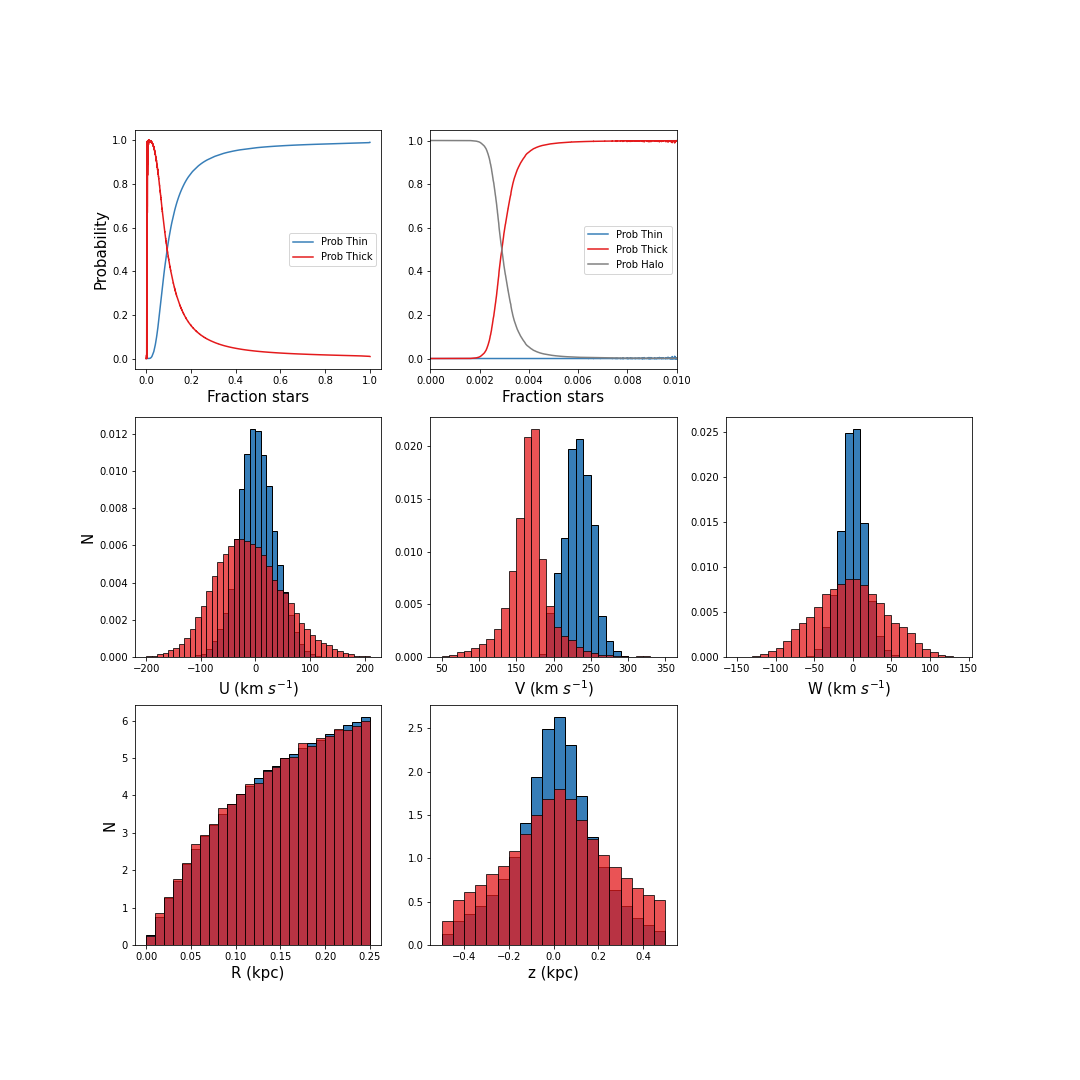}
\caption{Top panels: Probability of stars of belonging to the \textit{ks\_thick} (red), \textit{ks\_thin} (blue) disc or \textit{ks\_halo} as a function of fraction of stars (left), and a zoom to small fraction of stars to help visualizing the halo probabilities (right). Middle panels: Velocity components U (left), V (middle) and W (right) distributions of selected \textit{ks\_thick} (red) and \textit{ks\_thin} (blue). Bottom panels: Distribution of the Galactocentric radius projection over the Galactic plane, R, (left), and distance from the plane, z, (right) of the selected \textit{ks\_thick} (red) and \textit{ks\_thin} (blue).}
\label{prob_veloc_Rz}
\end{figure*}

\begin{figure}
\includegraphics[scale=0.55, trim=0 0 0 0]{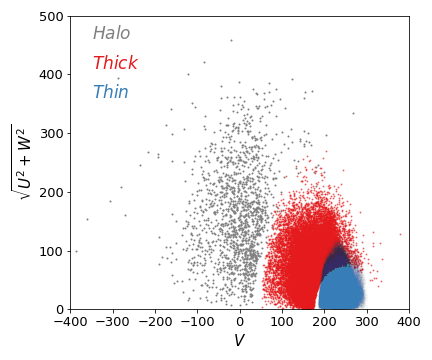}
\caption{Toomre diagram of the selected \textit{ks\_thin} disc (blue), \textit{ks\_thick} disc (red) and halo (grey) stellar samples, considering quality selection criteria. }
\label{kine}
\end{figure}

We use {\it Gaia} DR3 parallaxes, proper motions and line-of-sight velocities to calculate the cartesian velocity components with respect to the Galactic Reference Frame, U,V, W (considering the sun position with respect to the Galactic center at 8.178 kpc (\citealt{gravity19}) and motion with respect to the local standard of Rest ($U_\sun$, $V_\sun$, $W_\sun$) = (11.1,12.24,7.25) km\ $\rm s^{-1}$ \citep{Schonrich2010}, and positive values in the direction towards the Galactic Center, the sense of rotation of the Sun around the Galactic Center and the direction towards the North Galactic Pole). Based on these velocities, it is possible to explore the kinematic stellar populations present in our sample. To this end, we have adopted the procedure by \citet{bensby03} to calculate  probability distribution functions for each star to be a member of the thin, the thick disc or the halo by taking into account their velocity and their position within the Galaxy. Notice that this is not identical to the chemical thick and thin disks, selected based on their [$\alpha$/Fe]. For this reason we will refer to our samples as \textit{ks\_thin} and \textit{ks\_thick}. Note also that within the \textit{ks\_thick} disc kinematically selected population there might be also stellar populations born in the chemical thin disc that have been heated to thick disc kinematics (e.g., \cite{anders18}, \cite{FT21}, \cite{FT22}). We assumed that the velocity components follow Gaussian distributions with means and standard deviations equal to $(U\pm \sigma_U,V\pm \sigma_V,W\pm \sigma_W)$ = ($0\pm39$,$236\pm20$,$0\pm20$) km\ $\rm s^{-1}$ for the \textit{ks\_thin} disc \citep[the rotational velocity is from][]{reid19}, $(U\pm \sigma_U,V\pm \sigma_V,W\pm \sigma_W)$ = ($0\pm63$,$206\pm39$,$0\pm39$) km\ $\rm s^{-1}$ for the \textit{ks\_thick} disc, following \citet{Soubiran2003} for the velocity dispersions, and $(U\pm \sigma_U,V\pm \sigma_V,W\pm \sigma_W)$ = ($0\pm141$,$0\pm106$,$0\pm94$) km\ $\rm s^{-1}$ based on \cite{chiba2000} for the halo. We accounted for the exponential decrease of the stellar density of the \textit{ks\_thin} and \textit{ks\_thick} discs using the expression:

\begin{equation}
\rho(z,R) = \rho_0*e^{-|z|/h_Z}*e^{-R/h_l}
\end{equation}

taking into account stellar fractions, $\rho_0$, at solar position of 0.90 and 0.08 (as in \citealt{ramirez13}), scale heights ($h_Z$) of 0.3 and 0.9 kpc, and scale lengths ($h_l$) of 2.6 and 2 kpc, for the \textit{ks\_thin} and the \textit{ks\_thick} discs, respectively (\citealt{BG16}). For the halo, we considered a power law of the form:

\begin{equation}
\rho(r) = \rho_0*r^\alpha
\end{equation}

considering $\rho_0$ equal to 0.02 and $\alpha$ equal to -2.5.
The probability to belong to one of the kinematic components is:
\begin{equation}
p = \rho*\frac{1}{\sqrt{2\pi}\sigma_U \sigma_V \sigma_W}* e ^{-0.5((\frac{U}{\sigma_U})^2+(\frac{(V-V_{LSR})}{\sigma_V})^2+ (\frac{W}{\sigma_W})^2)}
\end{equation}

Top panels in Figure \ref{prob_veloc_Rz} show the resulting probabilities of belonging to the halo (grey), the thick (red) and the thin (blue) discs as a function of the fraction of stars of the total sample. A star would be assign to a particular Galactic component if the probability of belonging to that component is higher than 50\%. The Figure shows that the great majority of the sample ($\sim$ 90$\%$) is classified as \textit{ks\_thin} disc, as expected within this volume. Only $\sim$10$\%$ of the total sample have more probability of belonging to the \textit{ks\_thick} disc, and a very small fraction to the halo. We select stars with a probability higher than 75\% (to avoid stars within the kinematic boundaries of the two disc components) which include 94\% of \textit{ks\_thin} disc stars and 67\% of \textit{ks\_thick} disc stars. Middle panels of Figure \ref{prob_veloc_Rz} exhibit the velocity component distribution of the selected \textit{ks\_thick} and \textit{ks\_thin} discs (i.e., with probabilities higher than 75$\%$), and the bottom panels the distribution of both samples in Galactocentric radius projected on the Galactic plane, R, and the distance from the plane, z.

After applying the quality selection cuts explained above, our final samples comprise 582,956, 37,936  and 1,479 stars for the \textit{ks\_thin} disc, \textit{ks\_thick} disc and halo, respectively (down to $M_{G} < 5.5$). Figure \ref{kine} shows the location of our final \textit{ks\_thin}, \textit{ks\_thick} discs and halo selected stars on the Toomre diagram. The reader can notice a small overlap in the velocity space between the \textit{ks\_thick} and the \textit{ks\_thin} discs selected samples. This is due to the fact that the probability also takes into account the position and not only the velocity of each star. The gaps between the components arise from the fact that we are not considering all the stars in the sample, but only those with probabilities higher than 75\%. In Appendix \ref{sec:other_probs} we discuss the deSFHs of stellar samples with lower and higher probabilities and how they compare with our final selection.

\subsection{Derivation of the star formation histories.}
\label{sec:method_sfh}

We derive the deSFHs of the \textit{ks\_thick} and \textit{ks\_thin} discs separately using \textit{CMDft.Gaia}, which is a suite of procedures specifically designed to perform CMD fitting using \textit{Gaia} data. This software is presented in great detail in Paper I. Briefly, it comprises three modules performing the main steps required in the total process: i) ChronoSynth, which computes the synthetic \textit{mother} CMDs (see Paper I) from a given stellar evolution library; ii) DisPar-\textit{Gaia}, that simulates in the \textit{mother} CMD the observational errors and completeness of the observed one; a complete description of the procedure is presented in the Appendix \ref{sec:Appendix} of this paper; and iii) \textit{Dir}SFH, which performs the statistical comparison of the observed CMDs and a large number of model CMDs obtained as linear combination of single stellar populations (SSP; sample of stars covering a small range of age and metallicity) extracted from the \textit{mother} CMD, searching for the solution CMD that best reproduces the observed one. 

We perform the analysis using a \textit{mother} CMD computed with ChronoSynth, based on the BaSTI-IAC stellar evolution library \citep{hidalgo18}. It contains 120 million stars with $M_{G} \leq$ 5. A binary fraction $\beta$ = 0.3, a minimum mass ratio $q_{min}$ between secondary and primary stars equal to 0.1, and a Kroupa IMF \citep{kroupa93} have been adopted. The distribution of stellar ages and metallicities in this mother CMD is flat within the ranges 0.02 < age < 13.5 Gyrs and 0.0001 < Z < 0.039. S age bins (see Paper I for a definition of the age bins applied) and 0.1 dex metallicity bins have been adopted by $Dir$SFH. We have also taken into account the small offsets of -0.035 mag and 0.04 mag in ($BP-RP$) color and $M_{G}$ magnitude, respectively, between the \textit{Gaia} photometric system and the BaSTI-IAC theoretical framework in the same photometric passbands (see Paper I for more details). We use a bundle with a faint limit $M_{G}$ = 4.1 and weight of each CMD pixel calculated as the inverse of the variance of the stellar ages in that pixel. We encourage the reader to consult Paper I for a deeper insight on the details of the method.

\begin{figure*}
    \centering
    \includegraphics[width=1\textwidth]{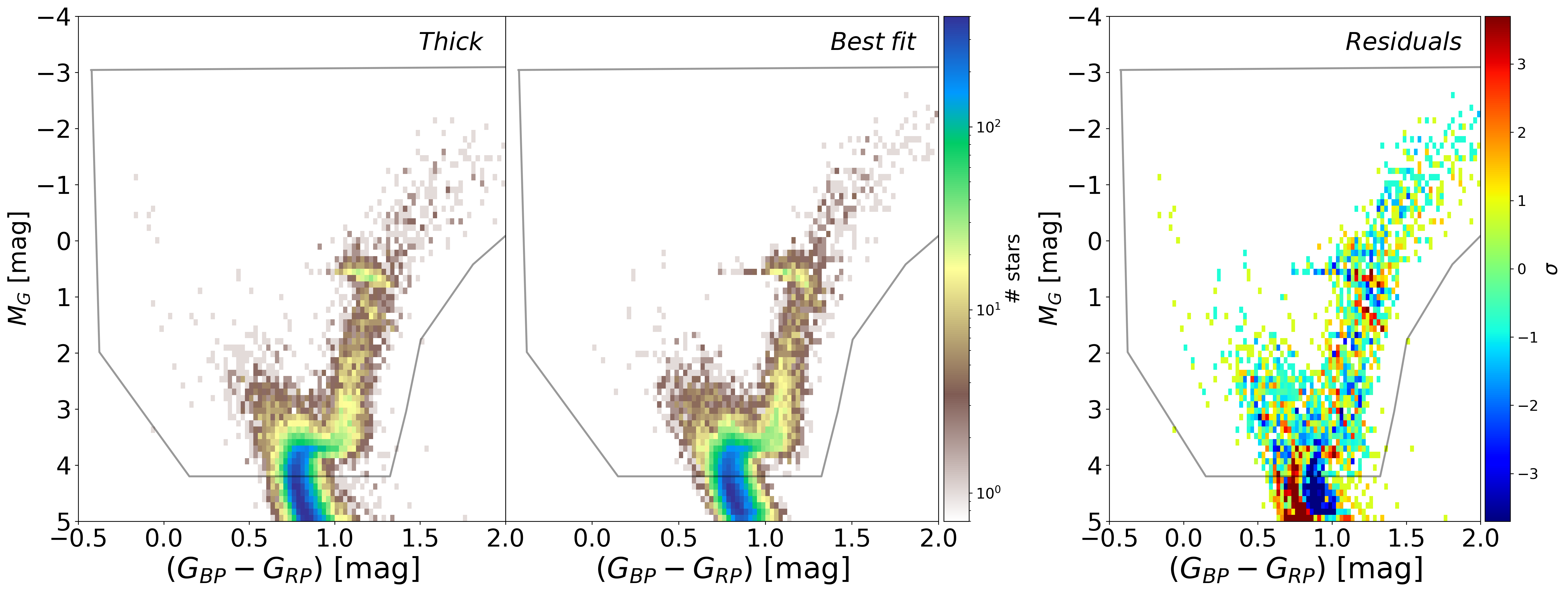}
    \includegraphics[width=1\textwidth]{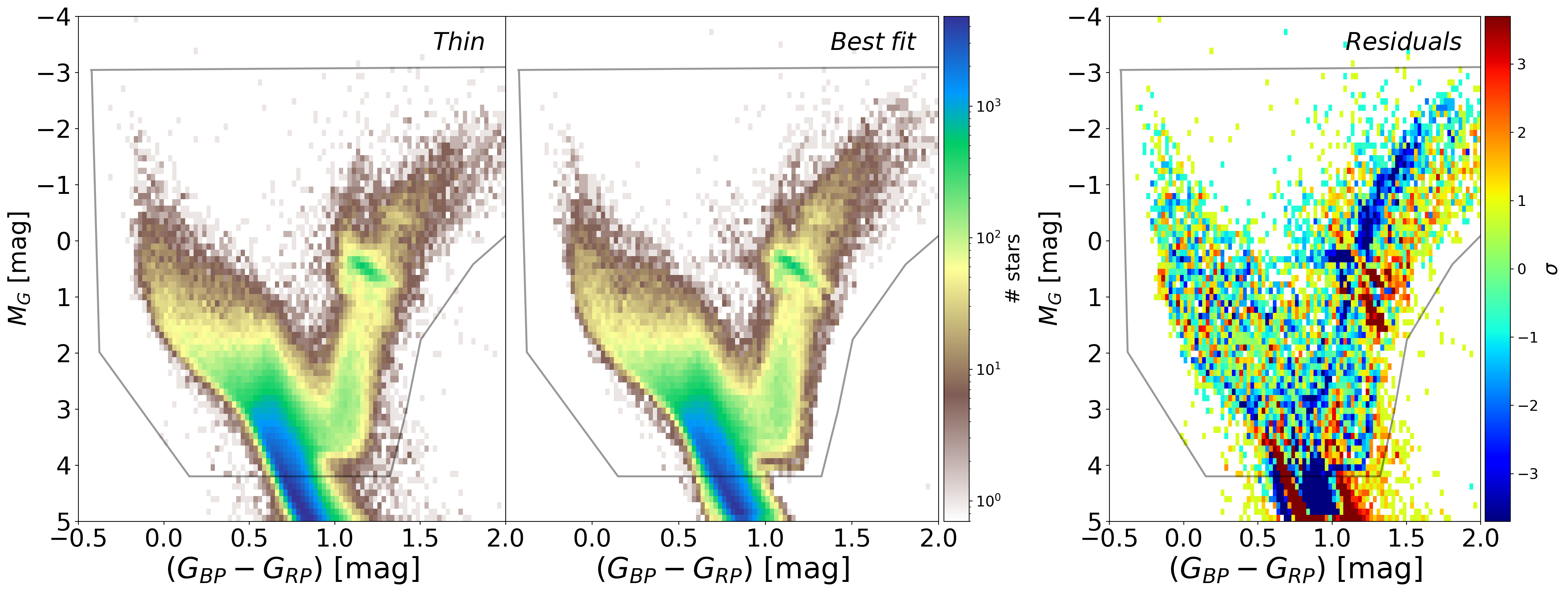}
    \caption{Observed (left) and best fit solution (middle) CMDs and the residuals (right) for the \textit{ks\_thin} disc (top) and \textit{ks\_thick} disc (bottom). The region used in the fit is plotted enclosed by a black polygon.}
    \label{sol}
\end{figure*}

Figure \ref{sol} shows the observed CMDs for the \textit{ks\_thick} disc and the \textit{ks\_thin} disc (left panel on both top and bottom panels), with the bundle region used to perform the analysis enclosed by a black polygon. The best CMD solutions and the corresponding residuals are displayed on the middle and right graphics of the Figure. Note the overall good match of the solution CMDs, which faithfully reproduce the shapes of the different stellar sequences in the observed CMDs, and also the number counts within $\pm 3 \sigma$ with no relevant systematic residual features within the area used for the fit. The most noticeable difference between the observed and the solution CMD occurs for the red-giant branch and red-clump region of the \textit{ks\_thin} disk. These two stellar evolutionary phases that can be affected, from a theoretical point of view, by non negligible uncertainties mainly in the effective temperature, as due for instance to the actual efficiency of superadiabatic convection \citep[see, e.g. the discussion in][ and references therein]{Creevey2024}, and to the bolometric corrections used for transferring model predictions from the H-R diagram to the Gaia photometric system. However, we wish to note that these regions of the CMD have a relatively low weight in the overall fit, both due to the lower number of stars compared to the main sequence and to the weighting scheme applied. Outside the bundle, in the main sequence below the old main sequence turnoff, the residuals of the fit exceed 3$\sigma$ and show some systematics. Note that in this region there is a very large density of stars and any small mismatch in the error simulation, in the slope of the main sequence or in the reddening corrections can lead to residuals over 3 $\sigma$ which however have had no influence in the fit.

\section{Results} \label{sec:results}
In this section we will describe the resulting deSFH for the local, kinematically selected, Milky Way \textit{ks\_thick} and \textit{ks\_thin} disks, and we will explore their chemical properties based on chemical abundance data from the literature. Owing to radial migration, this local sample is composed both by stars that have formed within its boundaries, and by stars born both at inner and outer radii and currently inside our volume. Also, a fraction of stars born within the volume boundaries are currently outside them. This has to be considered in the interpretation of our findings, and for this reason we use the term of deSFH or deSFR\footnote{deSFR: dynamically evolved star formation rate, that is, the marginalization over metallicity of the deSFH} when describing the results.

\subsection{The age-metallicity distributions.}
\begin{figure}
\includegraphics[scale=0.4]
{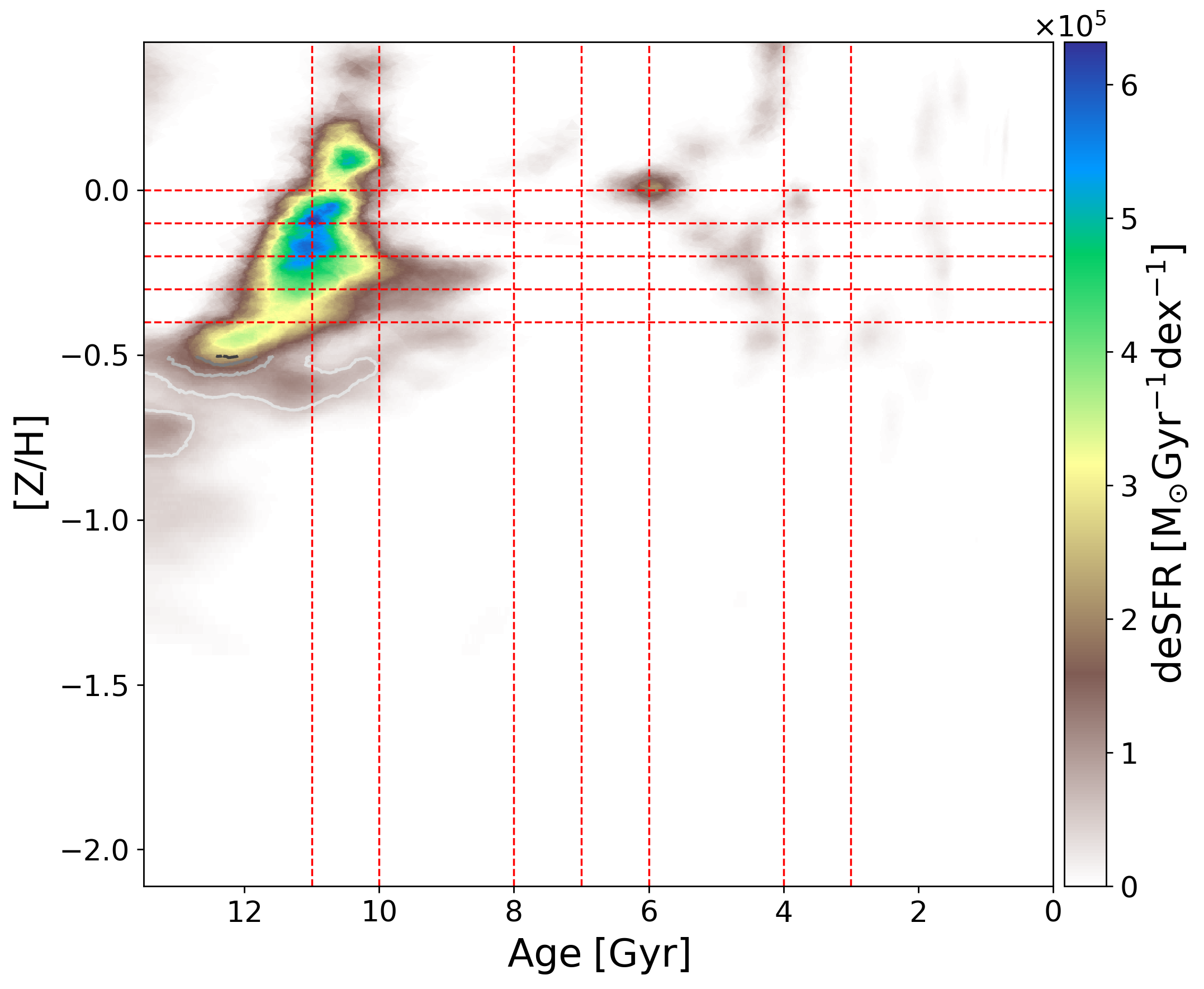}
\includegraphics[scale=0.4]{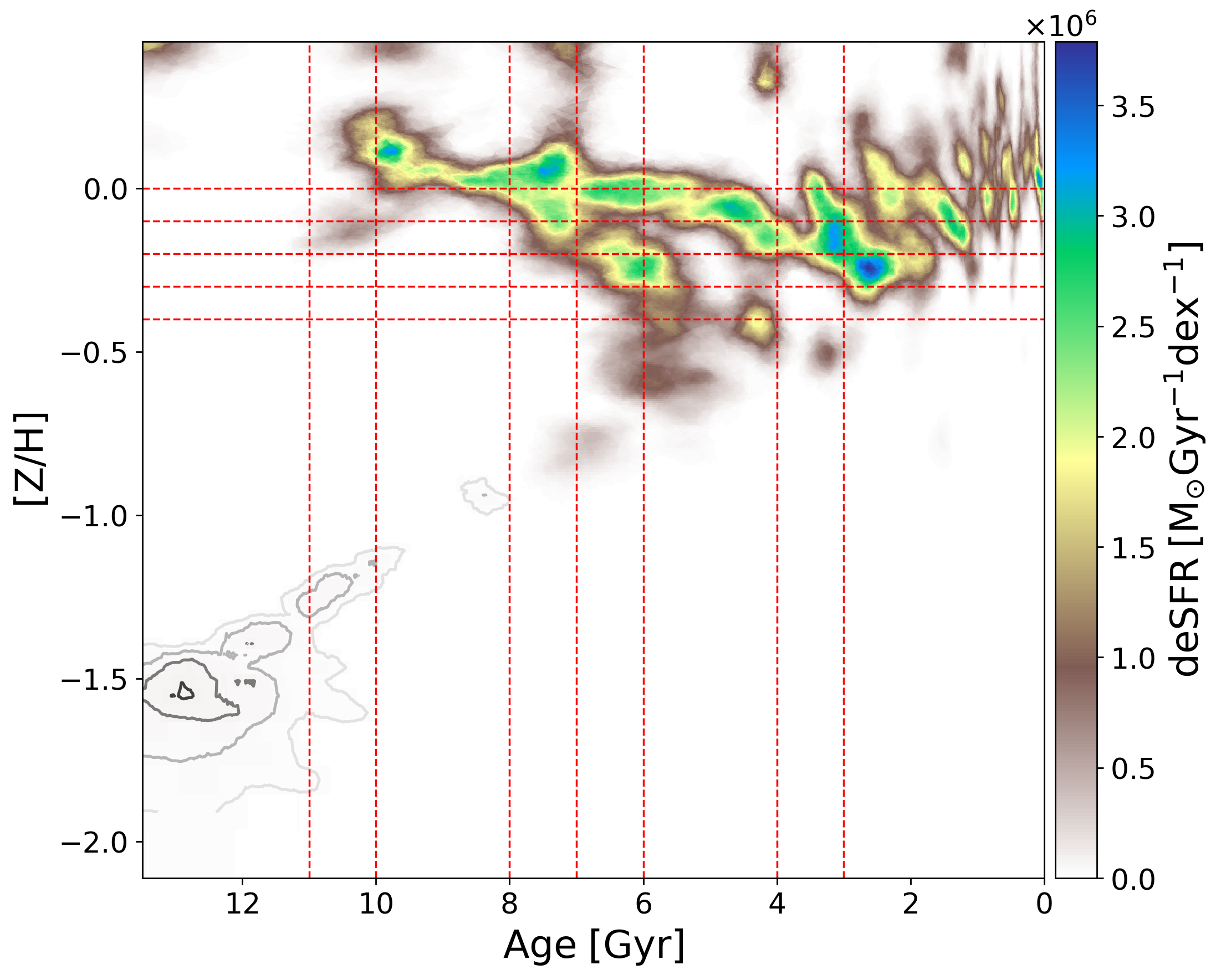}
\caption{Dynamically evolved star formation rates (deSFR, as indicated in the color scale) as a function of age and metallicity for the \textit{ks\_thick} (top) and \textit{ks\_thin} (bottom) discs derived with CMDft.Gaia. Red dotted lines mark key age and metallicity values: 11, 10,8,7,6, 4 and 3 Gyrs; metallicity values from 0 to -0.4 in steps of 0.1 dex.}
\label{both_thinthick}
\end{figure}

\begin{figure}
    \centering
    \includegraphics[width=1\linewidth]{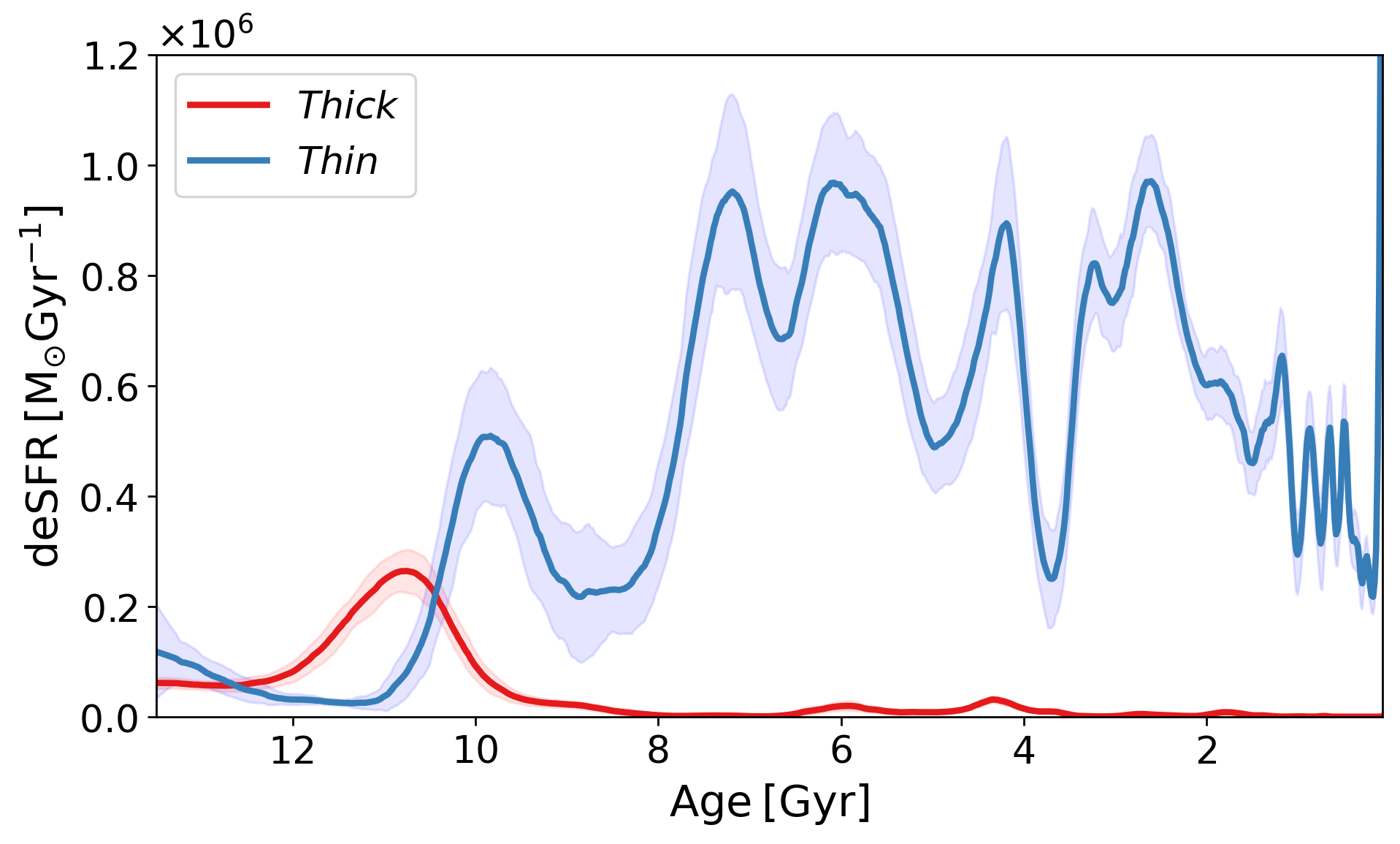}
    \caption{deSFR as a function of stellar age derived with CMDft.Gaia for the \textit{ks\_thick} (red) and \textit{ks\_thin} (blue) discs.}
    \label{fig:SFR}
\end{figure}

Figure \ref{both_thinthick} shows the deSFH derived for our kinematic \textit{ks\_thick} disc (top panel) and \textit{ks\_thin} disc (bottom panel). The graphics represent the age-metallicity distribution of the stellar mass formed (SFR) as a function of lookback time (age) and total metallicity ([Z/H]). 

Our results reveal that the \textit{ks\_thick} disc is mainly older than 10 Gyrs and underwent a fast chemical enrichment in a short period of time. Its deSFH is characterized by three main episodes of star formation with discontinuities and peaks distributed as follows: one older than 12 Gyrs with its maximum of star formation at [Z/H] $\sim$ -0.5 dex, but also with a tail towards lower metallicities at older ages; a second one, that produced most stars, reaches its peak around 11 Gyr ago and lasts $\sim$1 Gyr while increasing fast in metallicity from [Z/H] = -0.5 to [Z/H] = 0.0; and a third one which is very concentrated around $\sim$10.5 Gyrs ago and it has already supersolar metallicities. These three main episodes show a reduction in the SFR around [M/H]=-0.4 and at solar metallicities, approximately. Apart from this major age-metallicity trend, another noticeable peak of star formation stands out in isolation at 6 Gyrs and solar metallicity. 

Surrounding these main episodes of star formation our analysis returns some other minor populations: for instance, a parallel sequence at slightly lower metallicities and ages younger than the main episodes, following also an increasing metallicity trend as the age decreases down to 8 Gyr; at super-solar metallicities, there is some signal at 13, 10 and 4 Gyrs approximately (which will be discussed in an upcoming paper of the same series, Ruiz-Lara et al. in prep); finally, minor stellar populations younger than 6 Gyrs with a wide range of metallicities between -0.5 and 0.5 dex are also visible. Their low significance does not allow us to make strong conclusions about them since they could be residuals of the analysis or contamination from halo or \textit{ks\_thin} disc stars.
 
The \textit{ks\_thin} disc age-metallicity distribution basically starts where those of the \textit{ks\_thick} disc ends, that is, around 10 Gyr ago and with super-solar metalicity. The age-metallicity distribution is narrow down to $\sim$8 Gyrs while at younger ages it splits in several populations, increasing the range of metallicity at a particular age as time pass by, with its maximum $\sim$6 Gyrs ago. Another interesting characteristic is that the metallicity decreases with age until $\sim$3 Gyrs ago when it starts increasing again. This is contrary to what would be expected from a simple chemical evolution (in a close box), reflecting the complex evolutionary history of the Milky Way disc that will be discussed in detail in Section \ref{sec:discussion}.

In addition to this general trend, some minor stellar groups are also visible. One of them is an old metal-poor population, with [Z/H] < -1, highlighted with contour lines. This population was also observed in the deSFH of the stars within 100 pc of the Sun (Paper I) and has particular interest because there have been several works showing evidence of the existence of low metallicity stars in disc orbits. Their origin is currently under debate since it is puzzling to see stars so metal-poor moving in cool orbits (e.g., \citealt{sestito20}; \citealt{FA24}; \citealt{BK22}; \citealt{dillamore24}; \citealt{zhang24}, \citealt{nepal24disk}, among others). We will discuss this population in further detail in Section \ref{sec:discussion}. On the metal-rich side, we also see discrete populations of supersolar metallicities at 13, 10, 7, 4 and 1 Gyr (Ruiz-Lara et al. in prep), some of them coinciding with those present in our \textit{ks\_thick} disc sample. Between 11 and 10 Gyr we see a minor population connected with the super solar star formation episode but at lower metallicities, which resembles the trend observed in the \textit{ks\_thick} disc at such metallicities and slightly older ages. Finally, at intermediate ages, $\sim$7-6 Gyrs, where the maximum spread in metallicity is observed, there is a weak tail of stars with metallicities down to [Z/H]$\sim$-0.8, which is in line with the classical low metallicity end of the bulk of thin disc stars \citep{bensby04}. 

Figure \ref{fig:SFR} shows the deSFR for the \textit{ks\_thick} (red) and \textit{ks\_thin} (blue) discs as a function of time, essentially collapsing the [Z/H] dimension from Figure \ref{both_thinthick}. This representation highlights the continuity between the star formation in the \textit{ks\_thick} and \textit{ks\_thin} disks, and the fact that star formation is non-zero at early times for both components. It also helps appreciating the relative values of the deSFR of both components as defined in this paper. However, displaying the deSFR only as a function of time conceals some of the structure and gaps in the deSFH, which are only apparent when considering the deSFR's dependence on both age and [Z/H].

\subsection{Metallicity distribution functions.}
\label{sec:MDF}

Our methodology enables us to determine precise metallicity distribution functions (MDFs) using photometric data from the \textit{Gaia} mission, which provides more comprehensive coverage than spectroscopic surveys, overcoming selection biases. The accuracy of the metallicity values can be verified comparing our MDFs with those derived from high-precision spectroscopic measurements like those obtained from high-resolution spectroscopic surveys such as APOGEE DR17 \citep{abdurro}, GALAH DR3 \citep{galah21}, and \textit{Gaia} RVS \citep{RB23_RVS}. For this purpose, we cross-matched these spectroscopic databases with our \textit{ks\_thin} and \textit{ks\_thick} discs samples. We cleaned them from not reliable abundance determinations following the same criteria applied by \cite{FA24}. In the case of \textit{Gaia} RVS we choose the sample provided by \citep{RB_golden} (see their Appendix B) that has extremely precise sample of stellar chemo-physical parameters and iron abundance (\texttt{mh\_gspspec}) measurements with uncertainties lower than 0.05 dex.

It is important to note that these surveys provide measurements of iron abundance, not the total stellar metallicity (i.e., the sum of all elements heavier than helium), even when they use the [M/H] notation to describe them. Therefore, we must calculate the total metallicity from the spectroscopic chemical abundances to make a valid comparison with our metallicity estimates. We estimate the global metallicity [Z/H] from the values of [Fe/H] and [$\alpha$/Fe] by adopting the rescaling law provided by \citet{salaris93}, whose coefficients have been updated in \citet{pietri21} in order to take into account the updated reference solar heavy element distribution adopted in the BaSTI-IAC stellar library:

\begin{equation}
    \rm [Z/H] = [Fe/H] + \log(0.694 x 10^{[\alpha/Fe]}+0.301)
\end{equation}

To perform the comparison appropriately, we replicated the selection function of these spectroscopic surveys on the solution CMD. In each bin of color and magnitude across the solution CMD, we randomly selected a number of stars matching the fraction observed by each spectroscopic survey relative to the \textit{Gaia} photometric observations. We call the resulting CMD the \textit{masked} solution CMD. We calculated a \textit{masked} solution for each of the spectroscopic surveys explored. A more detailed explanation of the procedure and an example of a \textit{masked} solution can be found in Appendix \ref{sec:Appendix_mask}.

Figure \ref{mdf_both} shows the spectroscopic measurements from each of the three databases evaluated (in red or blue, for \textit{ks\_thick} and \textit{ks\_thin} disks, respectively) with the MDFs inferred from our methodology after applying the \textit{mask} (in black), both normalized to the total area under the curve. The shape of the MDF changes from one survey to another due to the differences of the selection functions. The comparison shows a very good agreement, with the peak and dispersion of our resulting MDF following closely the spectroscopic ones. This manifests i) that our metallicity estimations have similar precision to the independent and reliable measurements from high-resolution spectroscopic surveys; ii) that the metallicity scales are also very similar; and iii) that the solution CMD has been properly treated to take into account the incompleteness of each spectroscopic survey. Our \textit{ks\_thin} disc MDF seems to be affected by a very small [Z/H] offset, with a deficiency of supersolar metal-rich stars and an excess in the sub-solar metallicity side. Apart of this small difference, the comparison confidently assures that our metallicity values are reliable and in the same scale.

Figure \ref{mdf_both} also shows that the \textit{ks\_thick} disc MDF is broader than the \textit{ks\_thin} disc MDF, extended down to [Z/H] $\leq$ -1.5 with a peak at [Z/H] $\sim$-0.25, approx., which corresponds to a [Fe/H] $\sim$ -0.5 since this population is $\alpha$-enriched. The \textit{ks\_thin} disc has its maximum at [Z/H]$\sim$0 and extends down to [Z/H] $\sim$-1, although there are also a few stars down to [Z/H]$\sim$-2. On their metal-rich side, both the \textit{ks\_thick} and the \textit{ks\_thin} disc MDFs reach supersolar metallicities.

\begin{figure*}
\includegraphics[scale=0.45]{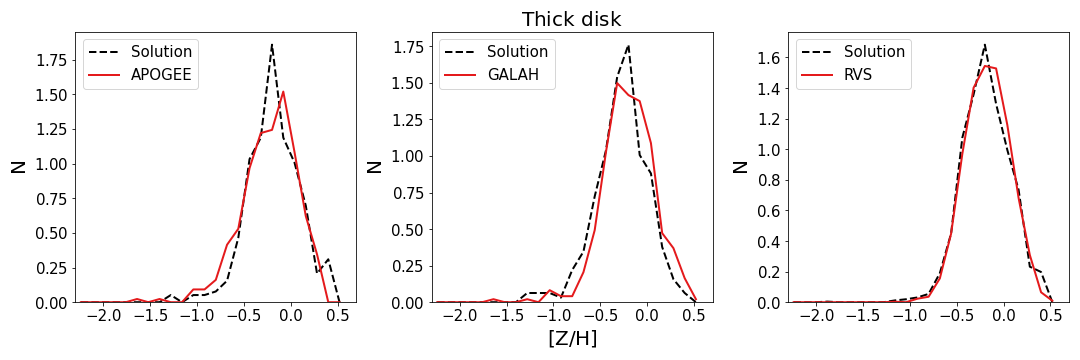}
\includegraphics[scale=0.45]{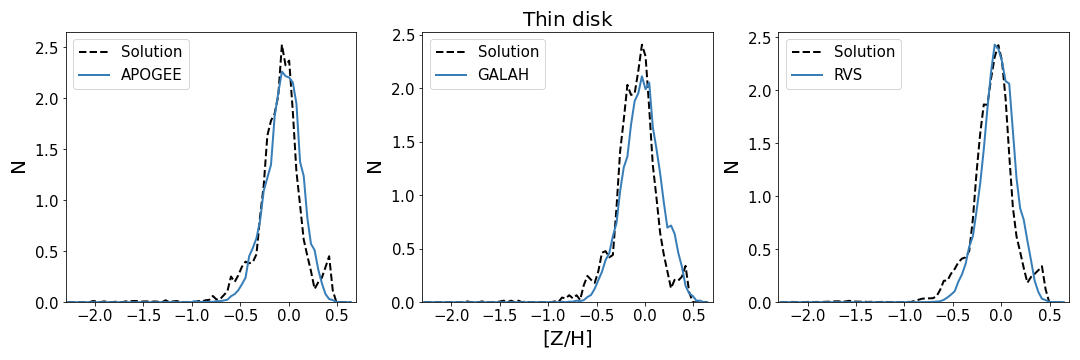}
 \caption{Metallicity distribution functions for the \textit{ks\_thick} disc (top panels) and the \textit{ks\_thin} disc (bottom panels) derived from our solution CMDs compared with those obtained from the spectroscopic surveys APOGEE (left), GALAH (middle) and the Golden Sample of \textit{Gaia}-RVS (right). The red or blue curves correspond to the spectroscopic MDFs and the black lines display the MDF distribution from each corresponding \textit{masked} solution CMD (see text for details).}   
   \label{mdf_both} 
\end{figure*}

\subsection{The [$\alpha$/Fe] distribution of the Milky Way discs across time.}
\label{Section_afe}

\begin{figure*}
    \includegraphics[scale=0.5]{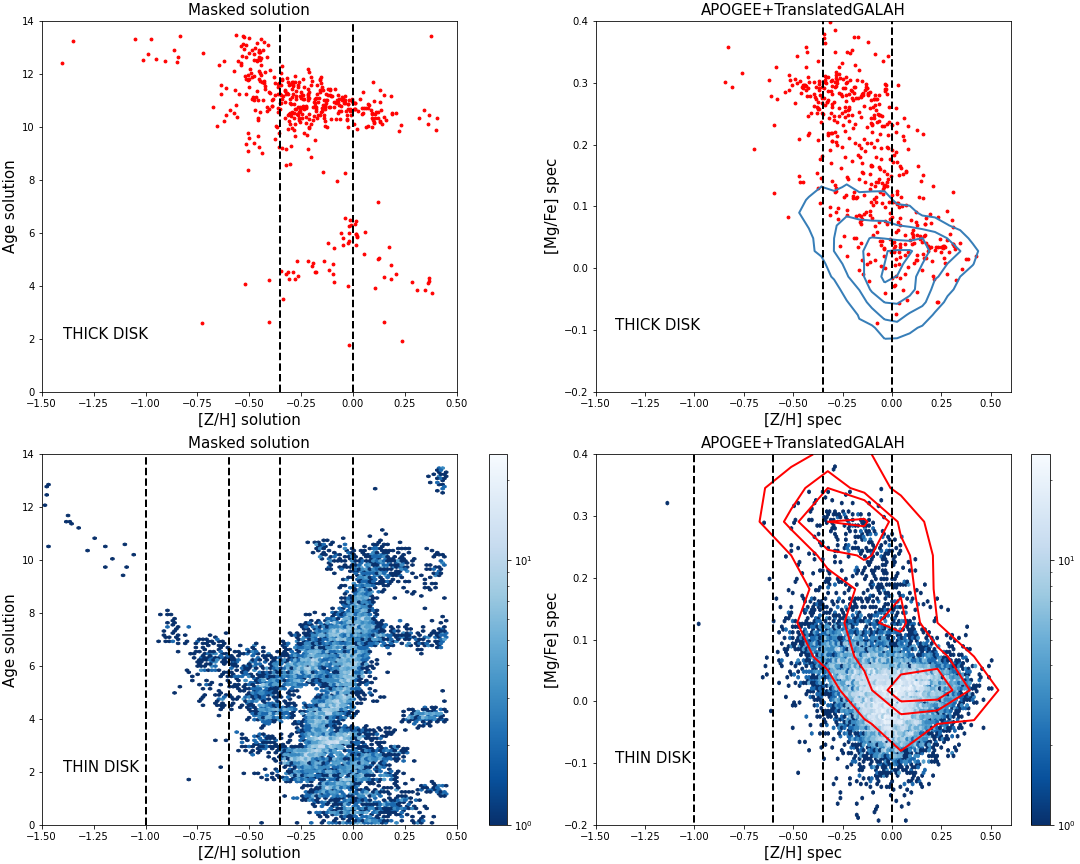}     \caption{Left panels: Age-metallicity distribution of the masked solutions for \textit{ks\_thick} disc stars (top, represented by individual red points) and \textit{ks\_thin} disc stars (bottom, shown as a blue-scale hexbin density diagram). This is the same solution as Figure \ref{both_thinthick}, but this time, instead of showing the deSFR as a funtion of age and metallicity the plots correspond to the distribution of synthetic stars from the best fit. Right panels: [Mg/Fe] versus [Z/H] (global metallicity, see Section \ref{sec:MDF} in the main text) based on spectroscopic measurements from APOGEE DR17 and GALAH DR3, homogenized using The SpectroTranslator. \textit{ks\_thick} disc stars are displayed at the top (with individual red points), while \textit{ks\_thin} disc stars appear at the bottom (as a blue-scale hexbin density diagram). The isocontours in red (blue) display the 30, 60 and 90$\%$ percent of the stellar distribution in the chemical space corresponding to the \textit{ks\_thin} disc (\textit{ks\_thick} disk) on top of the \textit{ks\_thick} disc (\textit{ks\_thin} disk) stars.  The vertical black dashed lines indicate the metallicity ranges of interest discussed in the main text.}
    \label{afeplot}
\end{figure*}

After confirming that our metallicity scale aligns well with the spectroscopic scales, we will evaluate the [$\alpha$/Fe] distribution observed for stars in the spectroscopic surveys with the age distributions obtained in this work at the same metallicity bins where we see particular episodes of star formation. We evaluate the \textit{masked} solution to compare the same fraction of stars in the age-metallicity space with the ones observed spectroscopically in the [$\alpha$/Fe] vs. [Z/H] space.

We perform this comparison using abundances from APOGEE and GALAH, with the latter transformed to the APOGEE scale through the SpectroTranslator\footnote{https://research.iac.es/proyecto/spectrotranslator/} methodology, recently introduced by \cite{thomas24}. We focus on the [Mg/Fe] vs. [Z/H] distribution, as magnesium (Mg) is the element that best distinguishes the two [$\alpha$/Fe] trends observed in the Milky Way disk. After applying the recommended flags, we cross-match our spectrosopic sample with the one that we use to derive the deSFH, resulting in 519 and 13,582 stars in common with our \textit{ks\_thick} and \textit{ks\_thin} disc samples, and with reliable chemical abundances from SpectroTranslator. The \textit{Gaia} RVS extremely precise sample of stellar chemo-physical parameters sample lacks sufficient Mg abundance determinations to draw robust conclusions, so we limit our analysis to the SpectroTranslator abundances.

Figure \ref{afeplot} shows, on the left, our age-metallicity distributions, now with the metallicity on the x-axis and age on the y-axis, obtained from the \textit{masked} solution CMD corresponding to the APOGEE+GALAH SpectroTranslator sample. In these plots we are representing individual stars (top, \textit{ks\_thick}) or stellar density (bottom, \textit{ks\_thin}) of the \textit{masked} solution CMD. Plots on the right show the [Mg/Fe] vs. [Z/H] distribution, represented with red dots in the case of the \textit{ks\_thick} disc and with a density colormap in blue for the \textit{ks\_thin} disc (because of the larger number of stars).  The isocontours in blue (red) display the 30, 60 and 90\% percent of the stellar distribution in the chemical space corresponding to the \textit{ks\_thin} disk (\textit{ks\_thick} disc) on top of the \textit{ks\_thick} disk (\textit{ks\_thin} disk) stars. These plots highlight the fact that the kinematically selected \textit{ks\_thick} and \textit{ks\_thin} discs do not correspond exactly to the chemically selected thick and thin disks, i.e., they do not result in the typical clean chemical separation between high-$\alpha$ and low-$\alpha$ populations.

The \textit{ks\_thick} disc main episodes of star formation cover three distinct well defined metallicity ranges: [Z/H] < -0.35, -0.35 < [Z/H] < 0.1, and [Z/H] > 0.1. We split the sample in these ranges in order to evaluate the [Mg/Fe] abundances separately for each of them. The [$\alpha$/Fe] abundances of the oldest (> 12 Gyr) \textit{ks\_thick} disc stars are overall high-$\alpha$. The metallicity range covered by the most prominent epoch of star formation between 12 and 10 Gyr ago (-0.35 < [Z/H] < 0.1), shows a very large [Mg/Fe] dispersion, although the contour lines corresponding to the \textit{ks\_thick} disc (over-imposed on the \textit{ks\_thin} disc plot) reveal that a majority of the stars in this range have high [Mg/Fe] and also a global decreasing trend of [Mg/Fe] as a function of metallicity from a high-$\alpha$ plateau down to [Mg/Fe] $\sim$ 0 at [Z/H] $\sim$ 0. Finally, the third population at supersolar metallicites is characterized by solar [Mg/Fe] abundances. 

A high-$\alpha$ plateau followed by a decreasing [Mg/Fe] while metallicity increases is a chemical trend well explained by the chemical enrichment of the gas by SNe with a massive star progenitor contributing early at low metallicities, and then by low-mass stars exploding as SNIa later, increasing the metallicity while the [Mg/Fe] ratio decreases \citep{matteucci86}. Our results clearly show that the Milky Way started forming stars very early and enriched the ISM very fast: approximately 12 Gyr ago, the Milky Way had reached already a metallicity of [Z/H] $\sim$-0.5 and, based on their high-$\alpha$ abundance ratios, most of the stars did not have received yet the contribution of the nucleosynthetic products of low-mass stars. 

Even though most stars born in the second epoch (12-10 Gyr ago) are $\alpha$-rich, the decrease in [Mg/Fe] is noticeable already at [Z/H]$\sim$-0.35, i.e., at ages younger than 12 Gyrs. This indicates that after $\simeq$1.5 Gyr of star formation there would have been enough time for SNIa from low-mass stars to explode and enrich the ISM with Fe. However, the \textit{knee} of this population is not well-defined. This may be due in part to uncertainties in spectroscopic metallicities. But note also that a number of intermediate-age stars with age $\simeq$ 6 Gyrs old and solar metallicities, which would have a low-$\alpha$ signature, enter in this metallicity range and increase the dispersion of [Mg/Fe]. Additionally, we propose that the observed dispersion in both metallicity and [Mg/Fe] could have a physical origin related to the Milky Way's accretion history during this period. If the Milky Way accreted a satellite galaxy at this time, the mixing of gas and stars could have enhanced the variability in metallicity and $\alpha$-element abundances of the stars formed afterward. We will explore this scenario in more detail in Section \ref{sec:discussion}. Lastly, the chemical analysis of the most metal rich episode of star formation in the \textit{ks\_thick} disc shows that 10 Gyrs ago the Milky Way already reached super-solar metallicities and low-$\alpha$ ratios.

The deSFH of the \textit{ks\_thin} disc is more complex, with star formation episodes at several metallicities overlapping in age. We again compare the age-metallicity distribution with the [Mg/Fe] vs. [Z/H] trend in ranges of metallicity. As mentioned earlier, the majority of the \textit{ks\_thin} disc began forming 10 Gyrs ago, already exhibiting super-solar metallicities. The comparison with the [Mg/Fe] vs. [Z/H] trends shows that these stars correspond to the metal-rich tip of the low-$\alpha$ sequence or chemical thin disk. The bulk of these stars show an overall [$\alpha$/Fe] $\sim$0, with some dispersion ranging between -0.1 and 0.1. This metallicity range is also populated by the youngest star formation and the very metal-rich discrete populations observed at different ages.

As mentioned above, the \textit{ks\_thin} disc age-metallicity trend decreases in metallicity as time goes by, up to around 4-3 Gyrs ago when the metallicity increases again. At the same time, there is an overlap of stellar populations with different metallicities at the same age. Or equivalently, at a particular metallicity range there are stellar populations covering a broad range of ages. This is specially the case for the metallicity range between $-0.35$ < [Z/H] < $0$, in which there are stars ranging from 11 Gyrs to the youngest ones. At the same time there is a very large dispersion in [Mg/Fe], from sub-solar up to thick-disk-like high-$\alpha$ ratios. This large spread might be explained by the superposition of these several populations at different ages that evolved differently with time and, consequently, reached distinct [Mg/Fe] ratios at the same [Z/H]. In particular, it seems that the stars with the highest [Mg/Fe] ratios correspond to stars between 11 and 10 Gyrs old, if we consider the comparison with the thick-disk chemical trend.

Lower metallicities correspond to the low metallicity tip of the low-$\alpha$ sequence or chemical thin disk. Traditionally, the metal-poor boundary of the thin disc has been set around [Fe/H] $\sim$ $-0.7$, equivalent to [Z/H] $\sim$ $-0.6$ when considering [Mg/Fe] $\sim$ 0.1. Stars ranging between $-0.6$ < [Z/H] < $-0.35$ show a moderate $\alpha$ enhancement, mostly concentrated at [Mg/Fe] $\sim$ 0.1, following a decreasing [Mg/Fe] trend with increasing [Z/H]. In our age-metallicity space, these stars clearly cover an age range between 7 and 3-2 Gyr. 

Recently, several studies have provided increasing evidence for stars in thin-disk-like orbits with metallicities below [Z/H] $\simeq -0.6$ (e.g., \citealt{sestito19}, \citealt{sestito20}, \citealt{sestito21}; \citealt{FA21}, \citealt{FA24}; \citealt{yuan23}; \citealt{bellazzini24}; \citealt{nepal24disk}; \citealt{GZ24}; although see also \citealt{zhang24}). Our findings support these results, as our analysis also identifies stellar populations with [Z/H] < -0.6 and even [Z/H] < -1. In our age-metallicity distribution, these metal-poor stars correspond to old and intermediate-age stars, spanning an age range from approximately 13.5 Gyrs down to around 8 Gyrs ([Z/H] < -1) and around 5-4 Gyr ([Z/H] < -0.6), drawing a trend of increasing metallicity with age. Interestingly, despite the similar age-metallicity trend, there appears to be a gap or disconnection between stars older than 8 Gyrs and those younger, which could suggest a different origin for stars with metallicities above and below -1. However, the limited number of stars in our sample prevents us from drawing a definitive conclusion. The [Mg/Fe] vs. [Z/H] plot only shows a handful of these stars, some with high-$\alpha$ ratios and others with lower-$\alpha$ content. The only star with [Z/H] < -1 is most likely older than 10 Gyrs as has a clear high-$\alpha$ ratio. The apparent discrepancy in the number of low metallicity stars ([Z/H]< -0.6) inferred from our deSFH and that in the spectroscopic sample may have a number of causes and its origin is beyond the scope of this work. The fraction of stars in this metallicity range on our solution is very small (2.5\% of the total \textit{ks\_thin} disk), but their presence seems to be a robust finding \citep[this tail of low metallicity stars is also found in the deSFH for the 100 pc sample studied by][]{2024A&A...687A.168G}, which will be discussed in detail by Queiroz et al. (in prep).  While it is usually assumed that these stars are old based on their low metallicity, in this work we are actually confirming this assumption by dating this stellar population.

\section{Discussion}
\label{sec:discussion}

\subsection{The \textit{ks\_thick} disc star formation history.}
\label{subsec_thickSFH}

This work clearly demonstrates that the \textit{ks\_thick} disc formed very early, around 13 Gyrs ago, undergoing rapid metallicity enrichment and reaching supersolar values approximately 10 Gyr ago, at which point subsequent star formation in the Milky Way disc appears to have occurred in colder, thin-disk-like orbits. This is not the first time that evidence in this sense has been presented. Based on ages inferred from exquisite asteroseismic constraints on RGBs observed by APOGEE \citet{miglio21} showed that the chemical thick disk, i.e., the $\alpha$-rich population, has ages around 11 Gyr with an spread of $\sim$1.5 Gyr. \citet{xiang22} published an age-metallicity relation for Milky Way stars, derived from LAMOST subgiant stars, for which precise individual stellar ages could be determined. Their results also indicated that the thick disk\footnote{More specifically, they discuss a high-$\alpha$, J$_\phi$<1500 kpc km s$^{-1}$ stellar sample formed in an 'early phase' of the Milky Way, that can be related to the \textit{ks\_thick} disk.} began forming about 13 Gyrs ago, with the majority of its stars being approximately 11 Gyrs old. Similarly, \cite{sahlholdt22} analyzed GALAH DR3 main sequence turn-off and subgiant stars, finding that the kinematically hot stars formed along a single age-metallicity sequence, which ended around 10 Gyrs ago, when the low-$\alpha$ sequence began forming and stars with lower vertical dispersions became dominant. Finally, \cite{queiroz23} show that the chemical thick disc for LAMOST, APOGEE and GALAH peak at approximately 11.3 Gyrs with a dispersion of 1.5 Gyrs.

However, our age-metallicity distribution exhibits significantly less dispersion at a given age compared to those inferred based on individual star age determinations (see also figures 13, C.3, C.4 and C.7 in Paper I). This, together with the large statistics in our samples \footnote{Note that our samples are basically complete within the considered magnitude limit, and contain stars in all evolutionary phases, while studies involving individual stellar ages typically concentrate in particular evolutionary phases, like the subgiant branch when ages are derived through isochrone fitting, of red-giant branch stars in asteroseismology studies. Additionally, the amount of data needed for these dating methods is larger than for CMD-fitting, as additional to photometry and distances, metallicities (and asteroseismic data) are necessary. Since these types of datasets are in general not complete, the number of available stars is substantially reduced. }, allows revealing details hidden within the uncertainties in previous studies, like the fact that the \textit{ks\_thick} disc experienced three distinct episodes of star formation, centered approximately 12, 11, and 10.5 Gyrs ago, with the most significant peak occurring around 11 Gyrs ago. Additionally, and unlike in these previous studies, our age-metallicity distribution can be considered virtually complete, as the very mild incompleteness that affects our stellar sample has been carefully simulated in the mother CMD as discussed in Appendix~\ref{sec:Appendix}. Finally, the deSFH shown in Figure~\ref{both_thinthick} accounts for all stellar mass formed throughout time, including stars that are no longer present due to having completed their life cycles. 

Figure~\ref{both_thinthick} shows that the \textit{ks\_thick} disc appears to have experienced its last major star formation episode around 10.5 Gyrs ago, coinciding with the timing of the last major merger in the Galaxy: the accretion of the \textit{Gaia}-Sausage-Enceladus (GSE) galaxy (\citealt{belokurov18}; \citealt{helmi18}; \citealt{vincenzo19}; \citealt{DM19}; \citealt{gallart19}\footnote{In fact \citet{2024A&A...687A.168G} infer a conspicuous episode of star formation in their geometric \textit{ks\_thick} disc sample (defined as stars with |Z|>1.1 Kpc located within 2 Kpc of the Sun) coinciding with the accretion time of GES, which they consider to have occurred $\simeq$ 10 Gyr ago, when a sharp cutoff is observed in the stellar age distribution of a kinematically selected halo population within the same volume. The agreement with the current study is very good, considering that the data and the methodology have been thoroughly updated: Gaia DR2 vs Gaia EDR3, different sample definition (geometric vs. kinematic \textit{ks\_thick} disk), BaSTI \citep{pietri21} vs. BaSTI-IAC stellar evolution models, updated CMD fitting methodology (TheStorm, \citealt{Bernard2018bulge_TheStorm}, vs. CMDft.Gaia) and more careful error and completeness simulation}; \citealt{bonaca20}; \citealt{montalban21}). Cosmological simulations predict that the accretion of galaxies can trigger star formation in both the host galaxy and the accreted satellite (\citealt{dicintio21}; \citealt{orkney22}). The pericenter passages of the accreted satellite compress the gas in the host galaxy through gravitational interactions. This compression increases the gas density leading to conditions favorable for star formation, with peaks aligning with the pericentric passages of the satellite.

Given this, it is plausible that the GSE accretion also stimulated star formation in the proto-Milky Way. Therefore, we speculate that the star formation episodes observed in the \textit{ks\_thick} disc around 11 and 10.5 Gyrs ago may be linked to the GSE merger. \cite{queiroz23} concluded that the high-[$\alpha$/Fe] population, which they call the genuine thick disc, formed before the GSE accretion. In our study the high-[$\alpha$/Fe] population would be the first star formation episode ($\le$ 12 Gyr), and probably part of the second one ($\sim$ 11 Gyr). The metallicity range cover by the second and third star formation events show a significant dispersion in [$\alpha$/Fe], yet an overall decrease in [$\alpha$/Fe] with increasing [Z/H] within a relatively narrow [Z/H] range. Two and a half Gyr of chemical evolution could be in principle sufficient to lower the [$\alpha$/Fe] abundances, and produce a well defined 'knee' in the [$\alpha$/Fe] vs. [Z/H] chemical space which, however, we don't observe. Yet, if these starbursts were triggered by a pericenter passage of GSE, leading to the accretion of more metal-poor gas from this satellite into the Milky Way, the newly accreted gas would dilute the pre-existing metal-enriched gas. This dilution would result in a slower overall metallicity increase, counteracting the rise in Fe from Type Ia supernova explosions. Additionally, the starburst would lead to the formation of new massive stars, which would quickly explode as supernovae, keeping the [$\alpha$/Fe] relatively high and likely contributing to the observed abundance dispersion.

\cite{BK22} analysed the azimuthal velocity spin-up of stars within the APOGEE DR17 and \textit{Gaia} DR3 databases to conclude that the transition from halo to thick disc orbits occurred between -1.3<[Fe/H]<-0.9\footnote{As our age-metallicity distribution is expressed in overall metallicity [Z/H], in the following the transformation between [Z/H] and [Fe/H] for an $\alpha$-enriched population (eq. 4) will be taken into account. For example, [Fe/H] = -0.9 corresponds to [Z/H]=-0.7 for [$\alpha$/Fe]=0.3.}, after what the Galaxy settles into a coherently rotating thin disk. We see that the first peak of star formation occurred $\sim$12 Gyrs ago with a metallicity of [M/H]$\sim$-0.5, but we also detect a tail towards lower metallicities down to at least [M/H] < -1 at ages < 13 Gyrs. We are thus identifying in our deSFH a well defined event of star formation that can be associated with that early disc component, and thus we are able to date this spin-up as having occurred indeed very early on in the history of the Galaxy, around 12 Gyr ago, or even earlier.

Another intriguing feature in the upper panel of Figure \ref{both_thinthick} is the sharp peak of star formation around 6 Gyrs ago. This coincides with the estimated timing of the first pericenter passage of the Sagittarius dwarf galaxy (\citealt{law10}; \citealt{laporte18}). As previously suggested, such a passage could have triggered star formation in the Milky Way, as other studies have also proposed \citep{ruizlara20}. By this time, the \textit{ks\_thin} disc was already established, as indicated by its SFH. The observation of this stellar population in orbits characteristic of the \textit{ks\_thick} disc suggests that gas and/or stars may have been heated during the merger with the Sagittarius dwarf galaxy, contributing to the observed stars with kinematic characteristics of the \textit{ks\_thick} disk. This is similar to the explanation proposed by \cite{gallart19} and \cite{belokurov20} to account for star with chemically thick-disk-like properties found in halo orbits.

The presence of young stars (ages $<$8 Gyr) in the \textit{ks\_thick} population might also be attributed to the phenomenon of young alpha-rich stars \citep{Chiappini2015, Martig2015, Grisoni2024}. Such stars, characterized by high abundances of alpha-elements and young apparent ages, have been identified in multiple datasets and exhibit kinematics consistent with those of thick disk stars \citep{Silva-Aguirre2018, Lagarde2021, queiroz23}. This phenomenon is likely a consequence of mass accretion in binary systems \citep{Jofre2016, Jofre2023}, which can cause stars to appear younger than their true ages.  

\subsection{The transition from the \textit{ks\_thick} disc to the \textit{ks\_thin} disk.}
\label{subsec_transition}

The majority of the \textit{ks\_thin} disk's stellar content spans from 10 Gyr ago to the present. Notably, \textit{ks\_thin} disc stars began forming at the same time and metallicity at which the \textit{ks\_thick} disc halted its primary star formation—around 10 Gyr ago, at slightly supersolar metallicities. Even more intriguingly, this transition from \textit{ks\_thick} to \textit{ks\_thin} disc formation coincided with the time range inferred by most studies for the accretion of GSE (\citealt{belokurov18}; \citealt{helmi18}; \citealt{vincenzo19}; \citealt{DM19}; \citealt{gallart19}; \citealt{bonaca20}; \citealt{montalban21}). Our findings suggest that sustained star formation in the settled \textit{ks\_thin} disc could only begin after the last major merger had concluded, ushering in a more stable period. This provides independent evidence for the timing of the Milky Way's last major merger. We also observe that the transition from \textit{ks\_thick} to \textit{ks\_thin} disc orbits occurred at supersolar metallicities, corresponding to the metal-rich end of the low-$\alpha$ stellar population. This was already pointed out by \cite{haywood13} were they showed that the oldest stars of the chemical thin disc have a metallicity as high as the youngest stars of the chemical thick disk, and a subsequent increase in metallicity dispersion is found for younger thin disc stars.

However, there are also a small fraction of stars with high-$\alpha$ abundances that exhibit thin-disc kinematics. The fact that these stars are slightly older than 10 Gyr suggests they could be among the earliest formed in thin-disk-like orbits, during a period when the Galaxy was accumulating angular momentum. \cite{anders18} and \cite{ciuca21} discuss a chemically distinct population often referred to as 'transition' or 'bridge' stars, which exhibit chemical abundances that fall between those of the chemical thin and thick disc populations. These stars are more prevalent in the inner regions of the galaxy. The earlier formation phase that we detect in the \textit{ks\_thin} disc in the metallicity range where we observe a large range of [Mg/Fe] could potentially be linked to the chemically identified transition population, reflecting a significant period of disc evolution when these 'bridge' stars emerged. 

Although the majority of the \textit{ks\_thin} disc stars have ages younger than 10 Gyr, our analysis also reveals the presence of very old stars (age > 12 Gyr) with both very low ([Z/H] < -1) and very high metallicities ([Z/H] > 0). This finding is consistent with previous reports of very old, metal-rich stars (\citealt{nepal24disk}, \citealt{RC24}) and the already discussed presence of very metal-poor stars. However, these populations appear disconnected from the bulk of the \textit{ks\_thin} disc stars, suggesting they may have a different origin than the majority of the \textit{ks\_thin} disk present in the analyzed volume, whose older populations are linked to the transition from thick to thin orbits.

\subsection{The \textit{ks\_thin} disc star formation history.}
\label{subsec_thinSFH}

The deSFHs inferred represent that of the stars currently present in the solar neighborhood. Stars on cool orbits, such as those in the \textit{ks\_thin} disk, are more influenced by radial migration than stars on \textit{ks\_thick} disc orbits. As a result, many of these stars were likely born in different regions of the Galaxy (\citealt{haywood08}; \citealt{minchev13}; \citealt{minchev14}). Radial migration is expected to have brought stars from distant locations, preferentially the inner regions. Also stars do not move in perfectly circular orbits but also stars from both the inner and outer disc might be crossing the solar radius at the present time as they reach their apocenters and pericenters. Therefore, the SFH inferred from \textit{ks\_thin} disc stars reflects star formation not only in the solar neighborhood but also in the regions where these migrated stars originally formed.

One of the characteristic signatures of radial mixing is the presence of substantial metallicity dispersion at a given radius (\citealt{haywood08}; \citealt{minchev13}). Previous analysis of the stellar and interstellar medium abundances across the thin disc have demonstrated a metallicity gradient with respect to Galactocentric distance, with the inner regions being more metal-rich than the outer regions \citep{tissera16, esteban18, lemasle18, MendezDelgado2022, lian23, CarbajoHijarrubia_2024_OCCASO}. In contrast, the \textit{ks\_thin} disc deSFH inferred from our analysis shows distinct star formation episodes at several metallicities overlapping in age, which increase the total metallicity range and hint to the superposition of several stellar populations. This is specially noticeable between 8 and 3 Gyr ago and suggests the presence in the studied volume of stars that have migrated from different regions of the Galaxy, each having undergone distinct SFHs. For ages younger than 3 Gyrs, the metallicity dispersion decreases, consistent with the notion that these younger stars have had less time to migrate, and that less mergers inducing star formation and migration have occurred at later times. 

Based on the observed interstellar medium gradient  \citep{MendezDelgado2022} and assuming that a qualitatively similar gradient has been present over the lifetime of the Milky Way, we would expect that the most metal-rich stars in our sample originated from the inner disk, while the metal-poor stars came from the outer disk. Consequently, the inferred decrease in metallicity with decreasing age, which seems to hold for the two main metallicity branches, may result from the combined effect of the inside-out growth of the disc and the radial mixing between stars from inner and outer radii. Similar conclusions have been drawn by \citet{xiang22}, \citet{sahlholdt22} and \citet{2024A&A...687A.168G}, who highlighted the role of migrating stars in shaping their derived age-metallicity relations. In this scenario, the lowest metallicity populations, extending between 8-7 and 3 Gyr ago,  can be used to constrain the epoch of formation of the outer disk, indicating it cannot be older than 8 Gyrs, as an earlier formation would have allowed time for these stars to migrate to the solar neighborhood.

An interesting scenario for this possible formation of the outer disc around 8 Gyrs ago has been presented by \cite{renaud21a}, \cite{renaud21b}, based on their analysis of the Vintergatan simulation \citep{agertz21}. This simulation, which models a Milky Way-like galaxy, presents an [$\alpha$/Fe] dichotomy and multiple branches in the age-metallicity plane, similar to the ones observed in our Galaxy. According to the simulation, star formation in the outer disc was triggered when a satellite galaxy began merging with a Milky Way-like galaxy (Vintergatan) along a filament. The tidal forces associated with this gravitational interaction compressed gas in the outer disk, thereby boosting star formation, which occurred at a lower [Fe/H] compared to the inner disc that had already been previously enriched in an epoch characterized by frequent merger events. \cite{renaud21a}, \cite{renaud21b} demonstrated how this last major merger initiated the formation of the outer disk, the low-$\alpha$ sequence, and the \textit{ks\_thin} disk. They predicted that this process would result in a dichotomy of [Fe/H] at around 8 Gyrs ago, corresponding to the low and high metallicity ends of the low-$\alpha$ sequence.

Our results indicate a split in [Z/H] starting around 8 to 7 Gyr ago. However, we also infer the presence of stars with supersolar metallicities and \textit{ks\_thin} disc kinematics 10 Gyrs ago, the approximate time when the chemistry of the stars transitions to solar-like abundances. This suggests that star formation in thin disc orbits began in the inner regions earlier than in the outer disk. Unlike in the Vintergatan simulation, here the onset of the $\alpha$ poor population and the appearance of the low metallicity branch seem to be disconnected, maybe pointing to more than one merger event as responsible. We have already discussed the likely relation between the GES merger and the \textit{ks\_thick}/\textit{ks\_thin} disc transition occurring $\simeq$ 10 Gyr ago. The emergence of the lowest metallicity branch around 8-7 Gyr ago could be linked to a different merger event, possibly that of the Helmi Streams, dated to have occurred around that time \citep{Kepley2007HS, Koppelman2019HS, Ruiz-Lara2022HS}, while later signatures (like the \textit{ks\_thick} disc stellar overdensity and clearly defined double branch 6 Gyr ago, or the features at three different metallicities 4 Gyr ago) could point to the Sgr dwarf accretion as the culprit \citep{ruizlara20, 2024A&A...687A.168G}. It is important to note, however, that our sample consists of stars currently populating a volume of radius 250 parsecs around the Sun, which may not adequately represent the outer disk. This limitation affects our ability to provide a robust characterization of that region.\\


\section{Summary and conclusions}
\label{sec:conclusions}

We have derived the dynamically evolved star formation histories (deSFH) of the \textit{ks\_thick} and \textit{ks\_thin} disks, selected according to position and kinematics, within a cylindrical volume with a radius of 250 pc and a height of 1 kpc, centered on the Sun. These results provide unprecedented insights into the star formation and evolution of the \textit{ks\_thick} and \textit{ks\_thin} disks. While confirming previous findings, they crucially resolve lingering questions about specific star formation episodes, their duration, associated chemical enrichment, and the timing of the transition from thick to thin disc kinematics. Our main conclusions can be summarized as follows:

i) Our results show that the \textit{ks\_thick} disc is primarily older than 10 Gyrs and experienced rapid metallicity enrichment in a short period. Its SFH reveals three key episodes: i) the first, older than 12 Gyrs ago, peaks at [Z/H] $\sim$ -0.5 dex and allows to quantitatively date for the first time the very early disc spin-up as having occurred over 12 Gyr ago; ii) a second, more intense episode occurred around 11 Gyrs ago, with metallicity quickly rising from [Z/H] = -0.5 to 0.0; finally, a third short period took place just over 10.5 Gyrs ago, at supersolar metallicities. 

ii) The bulk of \textit{ks\_thin} disc stars began forming around 10 Gyr ago with supersolar metallicity and low-[$\alpha$/Fe] content. Our analysis indicates that the low-$\alpha$ sequence started at super solar metallicities around 10 Gyr ago and extended to its metal-poor end ([Z/H] $\sim$ -0.8) by 8-7 Gyr ago. Stars at intermediate metallicities span ages from 10 Gyr ago to present time.

iii) There is a minor population of\textit{ks\_thin} disc stars with high-$\alpha$ abundances that are slightly older than 10 Gyr. This result evidences that when the Galaxy was transitioning form high-$\alpha$ to low-$\alpha$ and from sub-solar to supersolar metallicities, between 11 and 10 Gyrs ago, a kinematical transition was happening too. This transition coincides with the estimated epoch at which the Milky Way underwent its last major merger with the accretion of GSE.

iv) The age-metallicity relation of the \textit{ks\_thin} disc remains narrow until about 8 Gyr ago, after which it diverges into overlapping star formation episodes at the same age with distinct metallicities, likely indicative of radial mixing. This leads to an increased metallicity range that peaks around 6 Gyr ago. Given that stars in the low-metallicity branch are likely originated in the outer disk, our results suggest the outer disc formed around 8 Gyr ago, possibly coinciding with subsequent merger events.  

v) The average \textit{ks\_thin} disc metallicity starts to increase approximately 3 Gyr ago, likely reflecting the true local chemical enrichment, since stars may not be able to migrate substantially in this short period of time, in which additionally, no substantial merger events have occurred.

vi) Our analysis also uncovers minor populations within the \textit{ks\_thick} and \textit{ks\_thin} disc that appear disconnected from the main age-metallicity trends, echoing similar findings in the literature. These include vi.i) a very old (age > 10 Gyrs), metal-poor ([Z/H] < -1) stellar population with \textit{ks\_thin} disc kinematics; vi.ii) a notable 6 Gyr old population with solar metallicity and \textit{ks\_thick} disc kinematics, possibly associated with the first pericenter of the Sagittarius satellite galaxy; vi.iii) several distinct populations of very metal-rich ([Z/H] $\sim$ 0.5) stars with various ages, likely migrated from the inner galaxy, consistent with recent findings \citep{Nepal2024youngbar, RC24} .

The exquisite details in the deSFH of the \textit{ks\_thick} and \textit{ks\_thin} disk, discussed in detail above, clarify unresolved questions regarding the specific star formation episodes occurred in both disks, their duration and consequent chemical enrichment, and allow us to date the kinematic settlement of the Milky Way disc. These results constitute important observational constraints that need to be contrasted with chemical evolution models and cosmological simulations to understand the physical processes that led to such SFHs characteristics. Further papers in this series will keep exploiting the ability of CMDft.Gaia to dissect the Milky Way as never before.

\begin{acknowledgements}
EFA acknowledges support from HORIZON TMA MSCA Postdoctoral Fellowships Project TEMPOS, number 101066193, call HORIZON-MSCA-2021-PF-01, by the European Research Executive Agency. EFA, CG, ABQ, GB and GT also acknowledges support from the Agencia Estatal de Investigaci\'on del Ministerio de Ciencia e Innovaci\'on (AEI-MCINN) under grants “At the forefront of Galactic Archaeology: evolution of the luminous and dark matter components of the Milky Way and Local Group dwarf galaxies in the {\it Gaia} era” with references PID2020-118778GB-I00/10.13039/501100011033 and PID2023-150319NB-C21/10.13039/501100011033. TRL acknowledges financial support by the research projects AYA2017-84897-P, PID2020-113689GB-I00, and PID2020-114414GB-I00, financed by MCIN/AEI/10.13039/501100011033, the project A-FQM-510-UGR20 financed from FEDER/Junta de Andaluc\'ia-Consejer\'ia de Transformaci\'on Econ\'omica, Industria, Conocimiento y Universidades/Proyecto and by the grants P20-00334 and FQM108, financed by the Junta de Andaluc\'ia (Spain), as well as Juan de la Cierva fellowship (IJC2020-043742-I). SC acknowledges financial support from PRIN-MIUR-22: CHRONOS: adjusting the clock(s) to unveil the CHRONO-chemo-dynamical Structure of the Galaxy” (PI: S. Cassisi) funded by European Union – Next Generation EU, and Theory grant INAF 2023 (PI: S. Cassisi). AH and ED acknowledge financial support from a Spinoza grant.

\end{acknowledgements}

\bibliographystyle{aa} 
\bibliography{references.bib}

\begin{thebibliography}{161}
\expandafter\ifx\csname natexlab\endcsname\relax\def\natexlab#1{#1}\fi

\bibitem[{{Abdurro'uf} {et~al.}(2022){Abdurro'uf}, {Accetta}, {Aerts}, {Silva
  Aguirre}, {Ahumada}, {Ajgaonkar}, {Filiz Ak}, {Alam}, {Allende Prieto},
  {Almeida}, {Anders}, {Anderson}, {Andrews}, {Anguiano}, {Aquino-Ort{\'\i}z},
  {Arag{\'o}n-Salamanca}, {Argudo-Fern{\'a}ndez}, {Ata}, {Aubert},
  {Avila-Reese}, {Badenes}, {Barb{\'a}}, {Barger}, {Barrera-Ballesteros},
  {Beaton}, {Beers}, {Belfiore}, {Bender}, {Bernardi}, {Bershady}, {Beutler},
  {Bidin}, {Bird}, {Bizyaev}, {Blanc}, {Blanton}, {Boardman}, {Bolton},
  {Boquien}, {Borissova}, {Bovy}, {Brandt}, {Brown}, {Brownstein}, {Brusa},
  {Buchner}, {Bundy}, {Burchett}, {Bureau}, {Burgasser}, {Cabang}, {Campbell},
  {Cappellari}, {Carlberg}, {Wanderley}, {Carrera}, {Cash}, {Chen}, {Chen},
  {Cherinka}, {Chiappini}, {Choi}, {Chojnowski}, {Chung}, {Clerc}, {Cohen},
  {Comerford}, {Comparat}, {da Costa}, {Covey}, {Crane}, {Cruz-Gonzalez},
  {Culhane}, {Cunha}, {Dai}, {Damke}, {Darling}, {Davidson}, {Davies},
  {Dawson}, {De Lee}, {Diamond-Stanic}, {Cano-D{\'\i}az}, {S{\'a}nchez},
  {Donor}, {Duckworth}, {Dwelly}, {Eisenstein}, {Elsworth}, {Emsellem},
  {Eracleous}, {Escoffier}, {Fan}, {Farr}, {Feng}, {Fern{\'a}ndez-Trincado},
  {Feuillet}, {Filipp}, {Fillingham}, {Frinchaboy}, {Fromenteau}, {Galbany},
  {Garc{\'\i}a}, {Garc{\'\i}a-Hern{\'a}ndez}, {Ge}, {Geisler}, {Gelfand},
  {G{\'e}ron}, {Gibson}, {Goddy}, {Godoy-Rivera}, {Grabowski}, {Green},
  {Greener}, {Grier}, {Griffith}, {Guo}, {Guy}, {Hadjara}, {Harding},
  {Hasselquist}, {Hayes}, {Hearty}, {Hern{\'a}ndez}, {Hill}, {Hogg},
  {Holtzman}, {Horta}, {Hsieh}, {Hsu}, {Hsu}, {Huber}, {Huertas-Company},
  {Hutchinson}, {Hwang}, {Ibarra-Medel}, {Chitham}, {Ilha}, {Imig}, {Jaekle},
  {Jayasinghe}, {Ji}, {Johnson}, {Jones}, {J{\"o}nsson}, {Katkov}, {Khalatyan},
  {Kinemuchi}, {Kisku}, {Knapen}, {Kneib}, {Kollmeier}, {Kong}, {Kounkel},
  {Kreckel}, {Krishnarao}, {Lacerna}, {Lane}, {Langgin}, {Lavender}, {Law},
  {Lazarz}, {Leung}, {Leung}, {Lewis}, {Li}, {Li}, {Lian}, {Liang}, {Lin},
  {Lin}, {Lin}, {Lintott}, {Long}, {Longa-Pe{\~n}a}, {L{\'o}pez-Cob{\'a}},
  {Lu}, {Lundgren}, {Luo}, {Mackereth}, {de la Macorra}, {Mahadevan},
  {Majewski}, {Manchado}, {Mandeville}, {Maraston}, {Margalef-Bentabol},
  {Masseron}, {Masters}, {Mathur}, {McDermid}, {Mckay}, {Merloni},
  {Merrifield}, {Meszaros}, {Miglio}, {Di Mille}, {Minniti}, {Minsley},
  {Monachesi}, {Moon}, {Mosser}, {Mulchaey}, {Muna}, {Mu{\~n}oz}, {Myers},
  {Myers}, {Nadathur}, {Nair}, {Nandra}, {Neumann}, {Newman}, {Nidever},
  {Nikakhtar}, {Nitschelm}, {O'Connell}, {Garma-Oehmichen}, {Luan Souza de
  Oliveira}, {Olney}, {Oravetz}, {Ortigoza-Urdaneta}, {Osorio}, {Otter},
  {Pace}, {Padilla}, {Pan}, {Pan}, {Parikh}, {Parker}, {Peirani}, {Pe{\~n}a
  Ram{\'\i}rez}, {Penny}, {Percival}, {Perez-Fournon}, {Pinsonneault},
  {Poidevin}, {Poovelil}, {Price-Whelan}, {B{\'a}rbara de Andrade Queiroz},
  {Raddick}, {Ray}, {Rembold}, {Riddle}, {Riffel}, {Riffel}, {Rix}, {Robin},
  {Rodr{\'\i}guez-Puebla}, {Roman-Lopes}, {Rom{\'a}n-Z{\'u}{\~n}iga}, {Rose},
  {Ross}, {Rossi}, {Rubin}, {Salvato}, {S{\'a}nchez}, {S{\'a}nchez-Gallego},
  {Sanderson}, {Santana Rojas}, {Sarceno}, {Sarmiento}, {Sayres}, {Sazonova},
  {Schaefer}, {Schiavon}, {Schlegel}, {Schneider}, {Schultheis}, {Schwope},
  {Serenelli}, {Serna}, {Shao}, {Shapiro}, {Sharma}, {Shen}, {Shetrone}, {Shu},
  {Simon}, {Skrutskie}, {Smethurst}, {Smith}, {Sobeck}, {Spoo}, {Sprague},
  {Stark}, {Stassun}, {Steinmetz}, {Stello}, {Stone-Martinez},
  {Storchi-Bergmann}, {Stringfellow}, {Stutz}, {Su}, {Taghizadeh-Popp},
  {Talbot}, {Tayar}, {Telles}, {Teske}, {Thakar}, {Theissen}, {Tkachenko},
  {Thomas}, {Tojeiro}, {Hernandez Toledo}, {Troup}, {Trump}, {Trussler},
  {Turner}, {Tuttle}, {Unda-Sanzana}, {V{\'a}zquez-Mata}, {Valentini},
  {Valenzuela}, {Vargas-Gonz{\'a}lez}, {Vargas-Maga{\~n}a}, {Alfaro},
  {Villanova}, {Vincenzo}, {Wake}, {Warfield}, {Washington}, {Weaver},
  {Weijmans}, {Weinberg}, {Weiss}, {Westfall}, {Wild}, {Wilde}, {Wilson},
  {Wilson}, {Wilson}, {Wolf}, {Wood-Vasey}, {Yan}, {Zamora}, {Zasowski},
  {Zhang}, {Zhao}, {Zheng}, {Zheng}, \& {Zhu}}]{abdurro}
{Abdurro'uf}, {Accetta}, K., {Aerts}, C., {et~al.} 2022, \apjs, 259, 35

\bibitem[{{Adibekyan} {et~al.}(2012){Adibekyan}, {Sousa}, {Santos}, {Delgado
  Mena}, {Gonz{\'a}lez Hern{\'a}ndez}, {Israelian}, {Mayor}, \&
  {Khachatryan}}]{Adibekyan2012}
{Adibekyan}, V.~Z., {Sousa}, S.~G., {Santos}, N.~C., {et~al.} 2012, \aap, 545,
  A32

\bibitem[{{Agertz} {et~al.}(2021){Agertz}, {Renaud}, {Feltzing}, {Read},
  {Ryde}, {Andersson}, {Rey}, {Bensby}, \& {Feuillet}}]{agertz21}
{Agertz}, O., {Renaud}, F., {Feltzing}, S., {et~al.} 2021, \mnras, 503, 5826

\bibitem[{{Anders} {et~al.}(2018){Anders}, {Chiappini}, {Santiago},
  {Matijevi{\v{c}}}, {Queiroz}, {Steinmetz}, \& {Guiglion}}]{anders18}
{Anders}, F., {Chiappini}, C., {Santiago}, B.~X., {et~al.} 2018, \aap, 619,
  A125

\bibitem[{{Aparicio} \& {Gallart}(2004)}]{aparicio04}
{Aparicio}, A. \& {Gallart}, C. 2004, \aj, 128, 1465

\bibitem[{{Bellazzini} {et~al.}(2024){Bellazzini}, {Massari}, {Ceccarelli},
  {Mucciarelli}, {Bragaglia}, {Riello}, {De Angeli}, \&
  {Montegriffo}}]{bellazzini24}
{Bellazzini}, M., {Massari}, D., {Ceccarelli}, E., {et~al.} 2024, \aap, 683,
  A136

\bibitem[{{Belokurov} {et~al.}(2018){Belokurov}, {Erkal}, {Evans}, {Koposov},
  \& {Deason}}]{belokurov18}
{Belokurov}, V., {Erkal}, D., {Evans}, N.~W., {Koposov}, S.~E., \& {Deason},
  A.~J. 2018, \mnras, 478, 611

\bibitem[{{Belokurov} \& {Kravtsov}(2022)}]{BK22}
{Belokurov}, V. \& {Kravtsov}, A. 2022, \mnras, 514, 689

\bibitem[{{Belokurov} {et~al.}(2020){Belokurov}, {Sanders}, {Fattahi}, {Smith},
  {Deason}, {Evans}, \& {Grand}}]{belokurov20}
{Belokurov}, V., {Sanders}, J.~L., {Fattahi}, A., {et~al.} 2020, \mnras, 494,
  3880

\bibitem[{{Bensby} {et~al.}(2003){Bensby}, {Feltzing}, \&
  {Lundstr{\"o}m}}]{bensby03}
{Bensby}, T., {Feltzing}, S., \& {Lundstr{\"o}m}, I. 2003, \aap, 410, 527

\bibitem[{{Bensby} {et~al.}(2004){Bensby}, {Feltzing}, \&
  {Lundstr{\"o}m}}]{bensby04}
{Bensby}, T., {Feltzing}, S., \& {Lundstr{\"o}m}, I. 2004, \aap, 421, 969

\bibitem[{{Bernard}(2018)}]{Bernard2018IAUS}
{Bernard}, E.~J. 2018, in IAU Symposium, Vol. 330, Astrometry and Astrophysics
  in the Gaia Sky, ed. A.~{Recio-Blanco}, P.~{de Laverny}, A.~G.~A. {Brown}, \&
  T.~{Prusti}, 148--151

\bibitem[{{Bernard} {et~al.}(2007){Bernard}, {Aparicio}, {Gallart},
  {Padilla-Torres}, \& {Panniello}}]{bernard07}
{Bernard}, E.~J., {Aparicio}, A., {Gallart}, C., {Padilla-Torres}, C.~P., \&
  {Panniello}, M. 2007, \aj, 134, 1124

\bibitem[{{Bernard} {et~al.}(2018){Bernard}, {Schultheis}, {Di Matteo}, {Hill},
  {Haywood}, \& {Calamida}}]{Bernard2018bulge_TheStorm}
{Bernard}, E.~J., {Schultheis}, M., {Di Matteo}, P., {et~al.} 2018, \mnras,
  477, 3507

\bibitem[{{Bertelli} \& {Nasi}(2001)}]{bertelli2001AJ....121.1013B}
{Bertelli}, G. \& {Nasi}, E. 2001, \aj, 121, 1013

\bibitem[{{Bland-Hawthorn} \& {Gerhard}(2016)}]{BG16}
{Bland-Hawthorn}, J. \& {Gerhard}, O. 2016, \araa, 54, 529

\bibitem[{{Bonaca} {et~al.}(2020){Bonaca}, {Conroy}, {Cargile}, {Naidu},
  {Johnson}, {Zaritsky}, {Ting}, {Caldwell}, {Han}, \& {van Dokkum}}]{bonaca20}
{Bonaca}, A., {Conroy}, C., {Cargile}, P.~A., {et~al.} 2020, \apjl, 897, L18

\bibitem[{{Bovy} {et~al.}(2012){Bovy}, {Rix}, \& {Hogg}}]{bovy12}
{Bovy}, J., {Rix}, H.-W., \& {Hogg}, D.~W. 2012, \apj, 751, 131

\bibitem[{{Bovy} {et~al.}(2016){Bovy}, {Rix}, {Schlafly}, {Nidever},
  {Holtzman}, {Shetrone}, \& {Beers}}]{bovy16}
{Bovy}, J., {Rix}, H.-W., {Schlafly}, E.~F., {et~al.} 2016, \apj, 823, 30

\bibitem[{{Brook} {et~al.}(2004){Brook}, {Kawata}, {Gibson}, \&
  {Freeman}}]{brook04}
{Brook}, C.~B., {Kawata}, D., {Gibson}, B.~K., \& {Freeman}, K.~C. 2004, \apj,
  612, 894

\bibitem[{{Buder} {et~al.}(2021){Buder}, {Sharma}, {Kos}, {Amarsi},
  {Nordlander}, {Lind}, {Martell}, {Asplund}, {Bland-Hawthorn}, {Casey}, {de
  Silva}, {D'Orazi}, {Freeman}, {Hayden}, {Lewis}, {Lin}, {Schlesinger},
  {Simpson}, {Stello}, {Zucker}, {Zwitter}, {Beeson}, {Buck}, {Casagrande},
  {Clark}, {{\v{C}}otar}, {da Costa}, {de Grijs}, {Feuillet}, {Horner},
  {Kafle}, {Khanna}, {Kobayashi}, {Liu}, {Montet}, {Nandakumar}, {Nataf},
  {Ness}, {Spina}, {Tepper-Garc{\'\i}a}, {Ting}, {Traven},
  {Vogrin{\v{c}}i{\v{c}}}, {Wittenmyer}, {Wyse}, {{\v{Z}}erjal}, \& {Galah
  Collaboration}}]{galah21}
{Buder}, S., {Sharma}, S., {Kos}, J., {et~al.} 2021, \mnras, 506, 150

\bibitem[{{Cantat-Gaudin} {et~al.}(2023){Cantat-Gaudin}, {Fouesneau}, {Rix},
  {Brown}, {Castro-Ginard}, {Kostrzewa-Rutkowska}, {Drimmel}, {Hogg}, {Casey},
  {Khanna}, {Oh}, {Price-Whelan}, {Belokurov}, {Saydjari}, \&
  {Green}}]{2023A&A...669A..55C}
{Cantat-Gaudin}, T., {Fouesneau}, M., {Rix}, H.-W., {et~al.} 2023, \aap, 669,
  A55

\bibitem[{{Carbajo-Hijarrubia} {et~al.}(2024){Carbajo-Hijarrubia},
  {Casamiquela}, {Carrera}, {Balaguer-N{\'u}{\~n}ez}, {Jordi}, {Anders},
  {Gallart}, {Pancino}, {Drazdauskas}, {Stonkut{\.{e}}},
  {Tautvai{\v{s}}ien{\.{e}}}, {Carrasco}, {Masana}, {Cantat-Gaudin}, \&
  {Blanco-Cuaresma}}]{CarbajoHijarrubia_2024_OCCASO}
{Carbajo-Hijarrubia}, J., {Casamiquela}, L., {Carrera}, R., {et~al.} 2024,
  \aap, 687, A239

\bibitem[{{Chaplin} \& {Miglio}(2013)}]{chaplin13}
{Chaplin}, W.~J. \& {Miglio}, A. 2013, \araa, 51, 353

\bibitem[{{Chiappini} {et~al.}(2015){Chiappini}, {Anders}, {Rodrigues},
  {Miglio}, {Montalb{\'a}n}, {Mosser}, {Girardi}, {Valentini}, {Noels},
  {Morel}, {Minchev}, {Steinmetz}, {Santiago}, {Schultheis}, {Martig}, {da
  Costa}, {Maia}, {Allende Prieto}, {de Assis Peralta}, {Hekker},
  {Theme{\ss}l}, {Kallinger}, {Garc{\'\i}a}, {Mathur}, {Baudin}, {Beers},
  {Cunha}, {Harding}, {Holtzman}, {Majewski}, {M{\'e}sz{\'a}ros}, {Nidever},
  {Pan}, {Schiavon}, {Shetrone}, {Schneider}, \& {Stassun}}]{Chiappini2015}
{Chiappini}, C., {Anders}, F., {Rodrigues}, T.~S., {et~al.} 2015, \aap, 576,
  L12

\bibitem[{{Chiappini} {et~al.}(1997){Chiappini}, {Matteucci}, \&
  {Gratton}}]{chiappini97}
{Chiappini}, C., {Matteucci}, F., \& {Gratton}, R. 1997, \apj, 477, 765

\bibitem[{{Chiba} \& {Beers}(2000)}]{chiba2000}
{Chiba}, M. \& {Beers}, T.~C. 2000, \aj, 119, 2843

\bibitem[{{Cignoni} {et~al.}(2013){Cignoni}, {Cole}, {Tosi}, {Gallagher},
  {Sabbi}, {Anderson}, {Grebel}, \& {Nota}}]{Cignoni2013}
{Cignoni}, M., {Cole}, A.~A., {Tosi}, M., {et~al.} 2013, \apj, 775, 83

\bibitem[{{Cignoni} {et~al.}(2006){Cignoni}, {Degl'Innocenti}, {Prada Moroni},
  \& {Shore}}]{cignoni2006}
{Cignoni}, M., {Degl'Innocenti}, S., {Prada Moroni}, P.~G., \& {Shore}, S.~N.
  2006, \aap, 459, 783

\bibitem[{{Cignoni} \& {Tosi}(2010)}]{cignonitosi2010}
{Cignoni}, M. \& {Tosi}, M. 2010, Advances in Astronomy, 2010, 158568

\bibitem[{{Ciuc{\u{a}}} {et~al.}(2021){Ciuc{\u{a}}}, {Kawata}, {Miglio},
  {Davies}, \& {Grand}}]{ciuca21}
{Ciuc{\u{a}}}, I., {Kawata}, D., {Miglio}, A., {Davies}, G.~R., \& {Grand}, R.
  J.~J. 2021, \mnras, 503, 2814

\bibitem[{{Cole} {et~al.}(2014){Cole}, {Weisz}, {Dolphin}, {Skillman},
  {McConnachie}, {Brooks}, \& {Leaman}}]{Cole2014}
{Cole}, A.~A., {Weisz}, D.~R., {Dolphin}, A.~E., {et~al.} 2014, \apj, 795, 54

\bibitem[{{Comer{\'o}n} {et~al.}(2011){Comer{\'o}n}, {Elmegreen}, {Knapen},
  {Salo}, {Laurikainen}, {Laine}, {Athanassoula}, {Bosma}, {Sheth}, {Regan},
  {Hinz}, {Gil de Paz}, {Men{\'e}ndez-Delmestre}, {Mizusawa},
  {Mu{\~n}oz-Mateos}, {Seibert}, {Kim}, {Elmegreen}, {Gadotti}, {Ho},
  {Holwerda}, {Lappalainen}, {Schinnerer}, \& {Skibba}}]{comeron11}
{Comer{\'o}n}, S., {Elmegreen}, B.~G., {Knapen}, J.~H., {et~al.} 2011, \apj,
  741, 28

\bibitem[{{Comer{\'o}n} {et~al.}(2014){Comer{\'o}n}, {Elmegreen}, {Salo},
  {Laurikainen}, {Holwerda}, \& {Knapen}}]{comeron14}
{Comer{\'o}n}, S., {Elmegreen}, B.~G., {Salo}, H., {et~al.} 2014, \aap, 571,
  A58

\bibitem[{{Creevey} {et~al.}(2024){Creevey}, {Cassisi}, {Th{\'e}venin},
  {Salaris}, \& {Pietrinferni}}]{Creevey2024}
{Creevey}, O.~L., {Cassisi}, S., {Th{\'e}venin}, F., {Salaris}, M., \&
  {Pietrinferni}, A. 2024, \aap, 689, A243

\bibitem[{{Dalcanton} \& {Bernstein}(2002)}]{dalcanton2002}
{Dalcanton}, J.~J. \& {Bernstein}, R.~A. 2002, \aj, 124, 1328

\bibitem[{{Di Cintio} {et~al.}(2021){Di Cintio}, {Mostoghiu}, {Knebe}, \&
  {Navarro}}]{dicintio21}
{Di Cintio}, A., {Mostoghiu}, R., {Knebe}, A., \& {Navarro}, J.~F. 2021,
  \mnras, 506, 531

\bibitem[{{Di Matteo} {et~al.}(2019){Di Matteo}, {Haywood}, {Lehnert}, {Katz},
  {Khoperskov}, {Snaith}, {G{\'o}mez}, \& {Robichon}}]{DM19}
{Di Matteo}, P., {Haywood}, M., {Lehnert}, M.~D., {et~al.} 2019, \aap, 632, A4

\bibitem[{{Dillamore} {et~al.}(2024){Dillamore}, {Belokurov}, {Kravtsov}, \&
  {Font}}]{dillamore24}
{Dillamore}, A.~M., {Belokurov}, V., {Kravtsov}, A., \& {Font}, A.~S. 2024,
  \mnras, 527, 7070

\bibitem[{{Dolphin}(2002)}]{Dolphin2002MNRAS_method}
{Dolphin}, A.~E. 2002, \mnras, 332, 91

\bibitem[{{Elmegreen} \& {Elmegreen}(2006)}]{elmegreen06}
{Elmegreen}, B.~G. \& {Elmegreen}, D.~M. 2006, \apj, 650, 644

\bibitem[{{Esteban} \& {Garc{\'\i}a-Rojas}(2018)}]{esteban18}
{Esteban}, C. \& {Garc{\'\i}a-Rojas}, J. 2018, \mnras, 478, 2315

\bibitem[{{Fabricius} {et~al.}(2021){Fabricius}, {Luri}, {Arenou}, {Babusiaux},
  {Helmi}, {Muraveva}, {Reyl{\'e}}, {Spoto}, {Vallenari}, {Antoja}, {Balbinot},
  {Barache}, {Bauchet}, {Bragaglia}, {Busonero}, {Cantat-Gaudin}, {Carrasco},
  {Diakit{\'e}}, {Fabrizio}, {Figueras}, {Garcia-Gutierrez}, {Garofalo},
  {Jordi}, {Kervella}, {Khanna}, {Leclerc}, {Licata}, {Lambert}, {Marrese},
  {Masip}, {Ramos}, {Robichon}, {Robin}, {Romero-G{\'o}mez}, {Rubele}, \&
  {Weiler}}]{2021A&A...649A...5F}
{Fabricius}, C., {Luri}, X., {Arenou}, F., {et~al.} 2021, \aap, 649, A5

\bibitem[{{Fern{\'a}ndez-Alvar} {et~al.}(2024){Fern{\'a}ndez-Alvar},
  {Kordopatis}, {Hill}, {Battaglia}, {Gallart}, {Gonz{\'a}lez Rivera de la
  Vernhe}, {Thomas}, {Sestito}, {Ardern-Arentsen}, {Martin}, {Viswanathan}, \&
  {Starkenburg}}]{FA24}
{Fern{\'a}ndez-Alvar}, E., {Kordopatis}, G., {Hill}, V., {et~al.} 2024, \aap,
  685, A151

\bibitem[{{Fern{\'a}ndez-Alvar} {et~al.}(2021){Fern{\'a}ndez-Alvar},
  {Kordopatis}, {Hill}, {Starkenburg}, {Viswanathan}, {Martin}, {Thomas},
  {Navarro}, {Malhan}, {Sestito}, {Gonz{\'a}lez Hern{\'a}ndez}, \&
  {Carlberg}}]{FA21}
{Fern{\'a}ndez-Alvar}, E., {Kordopatis}, G., {Hill}, V., {et~al.} 2021, \mnras,
  508, 1509

\bibitem[{{Fern{\'a}ndez-Trincado} {et~al.}(2022){Fern{\'a}ndez-Trincado},
  {Beers}, {Barbuy}, {Minniti}, {Chiappini}, {Garro}, {Tang}, {Alves-Brito},
  {Villanova}, {Geisler}, {Lane}, \& {Diaz}}]{FT22}
{Fern{\'a}ndez-Trincado}, J.~G., {Beers}, T.~C., {Barbuy}, B., {et~al.} 2022,
  \aap, 663, A126

\bibitem[{{Fern{\'a}ndez-Trincado} {et~al.}(2021){Fern{\'a}ndez-Trincado},
  {Beers}, {Queiroz}, {Chiappini}, {Minniti}, {Barbuy}, {Majewski},
  {Ortigoza-Urdaneta}, {Moni Bidin}, {Robin}, {Moreno}, {Chaves-Velasquez},
  {Villanova}, {Lane}, {Pan}, \& {Bizyaev}}]{FT21}
{Fern{\'a}ndez-Trincado}, J.~G., {Beers}, T.~C., {Queiroz}, A. B.~A., {et~al.}
  2021, \apjl, 918, L37

\bibitem[{{Feuillet} {et~al.}(2018){Feuillet}, {Bovy}, {Holtzman}, {Weinberg},
  {Garc{\'\i}a-Hern{\'a}ndez}, {Hearty}, {Majewski}, {Roman-Lopes}, {Rybizki},
  \& {Zamora}}]{Feuillet2018}
{Feuillet}, D.~K., {Bovy}, J., {Holtzman}, J., {et~al.} 2018, \mnras, 477, 2326

\bibitem[{{Fitzpatrick} {et~al.}(2019){Fitzpatrick}, {Massa}, {Gordon},
  {Bohlin}, \& {Clayton}}]{Fitz19}
{Fitzpatrick}, E.~L., {Massa}, D., {Gordon}, K.~D., {Bohlin}, R., \& {Clayton},
  G.~C. 2019, \apj, 886, 108

\bibitem[{{Freeman} \& {Bland-Hawthorn}(2002)}]{freeman02}
{Freeman}, K. \& {Bland-Hawthorn}, J. 2002, \araa, 40, 487

\bibitem[{{Fuhrmann}(1998)}]{Fuhrmann1998}
{Fuhrmann}, K. 1998, \aap, 338, 161

\bibitem[{{Fuhrmann} \& {Chini}(2021)}]{Fuhrmann2021}
{Fuhrmann}, K. \& {Chini}, R. 2021, \mnras, 501, 4903

\bibitem[{{Gaia Collaboration} {et~al.}(2016){Gaia Collaboration}, {Prusti},
  {de Bruijne}, {Brown}, {Vallenari}, {Babusiaux}, {Bailer-Jones}, {Bastian},
  {Biermann}, {Evans}, {Eyer}, {Jansen}, {Jordi}, {Klioner}, {Lammers},
  {Lindegren}, {Luri}, {Mignard}, {Milligan}, {Panem}, {Poinsignon},
  {Pourbaix}, {Randich}, {Sarri}, {Sartoretti}, {Siddiqui}, {Soubiran},
  {Valette}, {van Leeuwen}, {Walton}, {Aerts}, {Arenou}, {Cropper}, {Drimmel},
  {H{\o}g}, {Katz}, {Lattanzi}, {O'Mullane}, {Grebel}, {Holland}, {Huc},
  {Passot}, {Bramante}, {Cacciari}, {Casta{\~n}eda}, {Chaoul}, {Cheek}, {De
  Angeli}, {Fabricius}, {Guerra}, {Hern{\'a}ndez}, {Jean-Antoine-Piccolo},
  {Masana}, {Messineo}, {Mowlavi}, {Nienartowicz}, {Ord{\'o}{\~n}ez-Blanco},
  {Panuzzo}, {Portell}, {Richards}, {Riello}, {Seabroke}, {Tanga},
  {Th{\'e}venin}, {Torra}, {Els}, {Gracia-Abril}, {Comoretto},
  {Garcia-Reinaldos}, {Lock}, {Mercier}, {Altmann}, {Andrae}, {Astraatmadja},
  {Bellas-Velidis}, {Benson}, {Berthier}, {Blomme}, {Busso}, {Carry},
  {Cellino}, {Clementini}, {Cowell}, {Creevey}, {Cuypers}, {Davidson}, {De
  Ridder}, {de Torres}, {Delchambre}, {Dell'Oro}, {Ducourant}, {Fr{\'e}mat},
  {Garc{\'\i}a-Torres}, {Gosset}, {Halbwachs}, {Hambly}, {Harrison}, {Hauser},
  {Hestroffer}, {Hodgkin}, {Huckle}, {Hutton}, {Jasniewicz}, {Jordan},
  {Kontizas}, {Korn}, {Lanzafame}, {Manteiga}, {Moitinho}, {Muinonen},
  {Osinde}, {Pancino}, {Pauwels}, {Petit}, {Recio-Blanco}, {Robin}, {Sarro},
  {Siopis}, {Smith}, {Smith}, {Sozzetti}, {Thuillot}, {van Reeven}, {Viala},
  {Abbas}, {Abreu Aramburu}, {Accart}, {Aguado}, {Allan}, {Allasia},
  {Altavilla}, {{\'A}lvarez}, {Alves}, {Anderson}, {Andrei}, {Anglada Varela},
  {Antiche}, {Antoja}, {Ant{\'o}n}, {Arcay}, {Atzei}, {Ayache}, {Bach},
  {Baker}, {Balaguer-N{\'u}{\~n}ez}, {Barache}, {Barata}, {Barbier}, {Barblan},
  {Baroni}, {Barrado y Navascu{\'e}s}, {Barros}, {Barstow}, {Becciani},
  {Bellazzini}, {Bellei}, {Bello Garc{\'\i}a}, {Belokurov}, {Bendjoya},
  {Berihuete}, {Bianchi}, {Bienaym{\'e}}, {Billebaud}, {Blagorodnova},
  {Blanco-Cuaresma}, {Boch}, {Bombrun}, {Borrachero}, {Bouquillon}, {Bourda},
  {Bouy}, {Bragaglia}, {Breddels}, {Brouillet}, {Br{\"u}semeister},
  {Bucciarelli}, {Budnik}, {Burgess}, {Burgon}, {Burlacu}, {Busonero}, {Buzzi},
  {Caffau}, {Cambras}, {Campbell}, {Cancelliere}, {Cantat-Gaudin}, {Carlucci},
  {Carrasco}, {Castellani}, {Charlot}, {Charnas}, {Charvet}, {Chassat},
  {Chiavassa}, {Clotet}, {Cocozza}, {Collins}, {Collins}, {Costigan}, {Crifo},
  {Cross}, {Crosta}, {Crowley}, {Dafonte}, {Damerdji}, {Dapergolas}, {David},
  {David}, {De Cat}, {de Felice}, {de Laverny}, {De Luise}, {De March}, {de
  Martino}, {de Souza}, {Debosscher}, {del Pozo}, {Delbo}, {Delgado},
  {Delgado}, {di Marco}, {Di Matteo}, {Diakite}, {Distefano}, {Dolding}, {Dos
  Anjos}, {Drazinos}, {Dur{\'a}n}, {Dzigan}, {Ecale}, {Edvardsson}, {Enke},
  {Erdmann}, {Escolar}, {Espina}, {Evans}, {Eynard Bontemps}, {Fabre},
  {Fabrizio}, {Faigler}, {Falc{\~a}o}, {Farr{\`a}s Casas}, {Faye}, {Federici},
  {Fedorets}, {Fern{\'a}ndez-Hern{\'a}ndez}, {Fernique}, {Fienga}, {Figueras},
  {Filippi}, {Findeisen}, {Fonti}, {Fouesneau}, {Fraile}, {Fraser}, {Fuchs},
  {Furnell}, {Gai}, {Galleti}, {Galluccio}, {Garabato}, {Garc{\'\i}a-Sedano},
  {Gar{\'e}}, {Garofalo}, {Garralda}, {Gavras}, {Gerssen}, {Geyer}, {Gilmore},
  {Girona}, {Giuffrida}, {Gomes}, {Gonz{\'a}lez-Marcos},
  {Gonz{\'a}lez-N{\'u}{\~n}ez}, {Gonz{\'a}lez-Vidal}, {Granvik}, {Guerrier},
  {Guillout}, {Guiraud}, {G{\'u}rpide}, {Guti{\'e}rrez-S{\'a}nchez}, {Guy},
  {Haigron}, {Hatzidimitriou}, {Haywood}, {Heiter}, {Helmi}, {Hobbs},
  {Hofmann}, {Holl}, {Holland}, {Hunt}, {Hypki}, {Icardi}, {Irwin}, {Jevardat
  de Fombelle}, {Jofr{\'e}}, {Jonker}, {Jorissen}, {Julbe}, {Karampelas},
  {Kochoska}, {Kohley}, {Kolenberg}, {Kontizas}, {Koposov}, {Kordopatis},
  {Koubsky}, {Kowalczyk}, {Krone-Martins}, {Kudryashova}, {Kull}, {Bachchan},
  {Lacoste-Seris}, {Lanza}, {Lavigne}, {Le Poncin-Lafitte}, {Lebreton},
  {Lebzelter}, {Leccia}, {Leclerc}, {Lecoeur-Taibi}, {Lemaitre}, {Lenhardt},
  {Leroux}, {Liao}, {Licata}, {Lindstr{\o}m}, {Lister}, {Livanou}, {Lobel},
  {L{\"o}ffler}, {L{\'o}pez}, {Lopez-Lozano}, {Lorenz}, {Loureiro},
  {MacDonald}, {Magalh{\~a}es Fernandes}, {Managau}, {Mann}, {Mantelet},
  {Marchal}, {Marchant}, {Marconi}, {Marie}, {Marinoni}, {Marrese},
  {Marschalk{\'o}}, {Marshall}, {Mart{\'\i}n-Fleitas}, {Martino}, {Mary},
  {Matijevi{\v{c}}}, {Mazeh}, {McMillan}, {Messina}, {Mestre}, {Michalik},
  {Millar}, {Miranda}, {Molina}, {Molinaro}, {Molinaro}, {Moln{\'a}r},
  {Moniez}, {Montegriffo}, {Monteiro}, {Mor}, {Mora}, {Morbidelli}, {Morel},
  {Morgenthaler}, {Morley}, {Morris}, {Mulone}, {Muraveva}, {Musella},
  {Narbonne}, {Nelemans}, {Nicastro}, {Noval}, {Ord{\'e}novic},
  {Ordieres-Mer{\'e}}, {Osborne}, {Pagani}, {Pagano}, {Pailler}, {Palacin},
  {Palaversa}, {Parsons}, {Paulsen}, {Pecoraro}, {Pedrosa}, {Pentik{\"a}inen},
  {Pereira}, {Pichon}, {Piersimoni}, {Pineau}, {Plachy}, {Plum}, {Poujoulet},
  {Pr{\v{s}}a}, {Pulone}, {Ragaini}, {Rago}, {Rambaux}, {Ramos-Lerate},
  {Ranalli}, {Rauw}, {Read}, {Regibo}, {Renk}, {Reyl{\'e}}, {Ribeiro},
  {Rimoldini}, {Ripepi}, {Riva}, {Rixon}, {Roelens}, {Romero-G{\'o}mez},
  {Rowell}, {Royer}, {Rudolph}, {Ruiz-Dern}, {Sadowski}, {Sagrist{\`a}
  Sell{\'e}s}, {Sahlmann}, {Salgado}, {Salguero}, {Sarasso}, {Savietto},
  {Schnorhk}, {Schultheis}, {Sciacca}, {Segol}, {Segovia}, {Segransan},
  {Serpell}, {Shih}, {Smareglia}, {Smart}, {Smith}, {Solano}, {Solitro},
  {Sordo}, {Soria Nieto}, {Souchay}, {Spagna}, {Spoto}, {Stampa}, {Steele},
  {Steidelm{\"u}ller}, {Stephenson}, {Stoev}, {Suess}, {S{\"u}veges}, {Surdej},
  {Szabados}, {Szegedi-Elek}, {Tapiador}, {Taris}, {Tauran}, {Taylor},
  {Teixeira}, {Terrett}, {Tingley}, {Trager}, {Turon}, {Ulla}, {Utrilla},
  {Valentini}, {van Elteren}, {Van Hemelryck}, {van Leeuwen}, {Varadi},
  {Vecchiato}, {Veljanoski}, {Via}, {Vicente}, {Vogt}, {Voss}, {Votruba},
  {Voutsinas}, {Walmsley}, {Weiler}, {Weingrill}, {Werner}, {Wevers},
  {Whitehead}, {Wyrzykowski}, {Yoldas}, {{\v{Z}}erjal}, {Zucker}, {Zurbach},
  {Zwitter}, {Alecu}, {Allen}, {Allende Prieto}, {Amorim},
  {Anglada-Escud{\'e}}, {Arsenijevic}, {Azaz}, {Balm}, {Beck}, {Bernstein},
  {Bigot}, {Bijaoui}, {Blasco}, {Bonfigli}, {Bono}, {Boudreault}, {Bressan},
  {Brown}, {Brunet}, {Bunclark}, {Buonanno}, {Butkevich}, {Carret}, {Carrion},
  {Chemin}, {Ch{\'e}reau}, {Corcione}, {Darmigny}, {de Boer}, {de Teodoro}, {de
  Zeeuw}, {Delle Luche}, {Domingues}, {Dubath}, {Fodor}, {Fr{\'e}zouls},
  {Fries}, {Fustes}, {Fyfe}, {Gallardo}, {Gallegos}, {Gardiol}, {Gebran},
  {Gomboc}, {G{\'o}mez}, {Grux}, {Gueguen}, {Heyrovsky}, {Hoar}, {Iannicola},
  {Isasi Parache}, {Janotto}, {Joliet}, {Jonckheere}, {Keil}, {Kim},
  {Klagyivik}, {Klar}, {Knude}, {Kochukhov}, {Kolka}, {Kos}, {Kutka}, {Lainey},
  {LeBouquin}, {Liu}, {Loreggia}, {Makarov}, {Marseille}, {Martayan},
  {Martinez-Rubi}, {Massart}, {Meynadier}, {Mignot}, {Munari}, {Nguyen},
  {Nordlander}, {Ocvirk}, {O'Flaherty}, {Olias Sanz}, {Ortiz}, {Osorio},
  {Oszkiewicz}, {Ouzounis}, {Palmer}, {Park}, {Pasquato}, {Peltzer}, {Peralta},
  {P{\'e}turaud}, {Pieniluoma}, {Pigozzi}, {Poels}, {Prat}, {Prod'homme},
  {Raison}, {Rebordao}, {Risquez}, {Rocca-Volmerange}, {Rosen}, {Ruiz-Fuertes},
  {Russo}, {Sembay}, {Serraller Vizcaino}, {Short}, {Siebert}, {Silva},
  {Sinachopoulos}, {Slezak}, {Soffel}, {Sosnowska}, {Strai{\v{z}}ys}, {ter
  Linden}, {Terrell}, {Theil}, {Tiede}, {Troisi}, {Tsalmantza}, {Tur},
  {Vaccari}, {Vachier}, {Valles}, {Van Hamme}, {Veltz}, {Virtanen}, {Wallut},
  {Wichmann}, {Wilkinson}, {Ziaeepour}, \& {Zschocke}}]{gaia16b}
{Gaia Collaboration}, {Prusti}, T., {de Bruijne}, J.~H.~J., {et~al.} 2016,
  \aap, 595, A1

\bibitem[{{Gaia Collaboration} {et~al.}(2021){Gaia Collaboration}, {Smart},
  {Sarro}, {Rybizki}, {Reyl{\'e}}, {Robin}, {Hambly}, {Abbas}, {Barstow}, {de
  Bruijne}, {Bucciarelli}, {Carrasco}, {Cooper}, {Hodgkin}, {Masana},
  {Michalik}, {Sahlmann}, {Sozzetti}, {Brown}, {Vallenari}, {Prusti},
  {Babusiaux}, {Biermann}, {Creevey}, {Evans}, {Eyer}, {Hutton}, {Jansen},
  {Jordi}, {Klioner}, {Lammers}, {Lindegren}, {Luri}, {Mignard}, {Panem},
  {Pourbaix}, {Randich}, {Sartoretti}, {Soubiran}, {Walton}, {Arenou},
  {Bailer-Jones}, {Bastian}, {Cropper}, {Drimmel}, {Katz}, {Lattanzi}, {van
  Leeuwen}, {Bakker}, {Casta{\~n}eda}, {De Angeli}, {Ducourant}, {Fabricius},
  {Fouesneau}, {Fr{\'e}mat}, {Guerra}, {Guerrier}, {Guiraud}, {Jean-Antoine
  Piccolo}, {Messineo}, {Mowlavi}, {Nicolas}, {Nienartowicz}, {Pailler},
  {Panuzzo}, {Riclet}, {Roux}, {Seabroke}, {Sordo}, {Tanga}, {Th{\'e}venin},
  {Gracia-Abril}, {Portell}, {Teyssier}, {Altmann}, {Andrae}, {Bellas-Velidis},
  {Benson}, {Berthier}, {Blomme}, {Brugaletta}, {Burgess}, {Busso}, {Carry},
  {Cellino}, {Cheek}, {Clementini}, {Damerdji}, {Davidson}, {Delchambre},
  {Dell'Oro}, {Fern{\'a}ndez-Hern{\'a}ndez}, {Galluccio}, {Garc{\'\i}a-Lario},
  {Garcia-Reinaldos}, {Gonz{\'a}lez-N{\'u}{\~n}ez}, {Gosset}, {Haigron},
  {Halbwachs}, {Harrison}, {Hatzidimitriou}, {Heiter}, {Hern{\'a}ndez},
  {Hestroffer}, {Holl}, {Jan{\ss}en}, {Jevardat de Fombelle}, {Jordan},
  {Krone-Martins}, {Lanzafame}, {L{\"o}ffler}, {Lorca}, {Manteiga}, {Marchal},
  {Marrese}, {Moitinho}, {Mora}, {Muinonen}, {Osborne}, {Pancino}, {Pauwels},
  {Recio-Blanco}, {Richards}, {Riello}, {Rimoldini}, {Roegiers}, {Siopis},
  {Smith}, {Ulla}, {Utrilla}, {van Leeuwen}, {van Reeven}, {Abreu Aramburu},
  {Accart}, {Aerts}, {Aguado}, {Ajaj}, {Altavilla}, {{\'A}lvarez}, {{\'A}lvarez
  Cid-Fuentes}, {Alves}, {Anderson}, {Anglada Varela}, {Antoja}, {Audard},
  {Baines}, {Baker}, {Balaguer-N{\'u}{\~n}ez}, {Balbinot}, {Balog}, {Barache},
  {Barbato}, {Barros}, {Bartolom{\'e}}, {Bassilana}, {Bauchet},
  {Baudesson-Stella}, {Becciani}, {Bellazzini}, {Bernet}, {Bertone}, {Bianchi},
  {Blanco-Cuaresma}, {Boch}, {Bombrun}, {Bossini}, {Bouquillon}, {Bragaglia},
  {Bramante}, {Breedt}, {Bressan}, {Brouillet}, {Burlacu}, {Busonero},
  {Butkevich}, {Buzzi}, {Caffau}, {Cancelliere}, {C{\'a}novas},
  {Cantat-Gaudin}, {Carballo}, {Carlucci}, {Carnerero}, {Casamiquela},
  {Castellani}, {Castro-Ginard}, {Castro Sampol}, {Chaoul}, {Charlot},
  {Chemin}, {Chiavassa}, {Cioni}, {Comoretto}, {Cornez}, {Cowell}, {Crifo},
  {Crosta}, {Crowley}, {Dafonte}, {Dapergolas}, {David}, {David}, {de Laverny},
  {De Luise}, {De March}, {De Ridder}, {de Souza}, {de Teodoro}, {de Torres},
  {del Peloso}, {del Pozo}, {Delgado}, {Delgado}, {Delisle}, {Di Matteo},
  {Diakite}, {Diener}, {Distefano}, {Dolding}, {Eappachen}, {Edvardsson},
  {Enke}, {Esquej}, {Fabre}, {Fabrizio}, {Faigler}, {Fedorets}, {Fernique},
  {Fienga}, {Figueras}, {Fouron}, {Fragkoudi}, {Fraile}, {Franke}, {Gai},
  {Garabato}, {Garcia-Gutierrez}, {Garc{\'\i}a-Torres}, {Garofalo}, {Gavras},
  {Gerlach}, {Geyer}, {Giacobbe}, {Gilmore}, {Girona}, {Giuffrida}, {Gomel},
  {Gomez}, {Gonzalez-Santamaria}, {Gonz{\'a}lez-Vidal}, {Granvik},
  {Guti{\'e}rrez-S{\'a}nchez}, {Guy}, {Hauser}, {Haywood}, {Helmi}, {Hidalgo},
  {Hilger}, {H{\l}adczuk}, {Hobbs}, {Holland}, {Huckle}, {Jasniewicz},
  {Jonker}, {Juaristi Campillo}, {Julbe}, {Karbevska}, {Kervella}, {Khanna},
  {Kochoska}, {Kontizas}, {Kordopatis}, {Korn}, {Kostrzewa-Rutkowska},
  {Kruszy{\'n}ska}, {Lambert}, {Lanza}, {Lasne}, {Le Campion}, {Le Fustec},
  {Lebreton}, {Lebzelter}, {Leccia}, {Leclerc}, {Lecoeur-Taibi}, {Liao},
  {Licata}, {Lindstr{\o}m}, {Lister}, {Livanou}, {Lobel}, {Madrero Pardo},
  {Managau}, {Mann}, {Marchant}, {Marconi}, {Marcos Santos}, {Marinoni},
  {Marocco}, {Marshall}, {Martin Polo}, {Mart{\'\i}n-Fleitas}, {Masip},
  {Massari}, {Mastrobuono-Battisti}, {Mazeh}, {McMillan}, {Messina}, {Millar},
  {Mints}, {Molina}, {Molinaro}, {Moln{\'a}r}, {Montegriffo}, {Mor},
  {Morbidelli}, {Morel}, {Morris}, {Mulone}, {Munoz}, {Muraveva}, {Murphy},
  {Musella}, {Noval}, {Ord{\'e}novic}, {Orr{\`u}}, {Osinde}, {Pagani},
  {Pagano}, {Palaversa}, {Palicio}, {Panahi}, {Pawlak}, {Pe{\~n}alosa
  Esteller}, {Penttil{\"a}}, {Piersimoni}, {Pineau}, {Plachy}, {Plum},
  {Poggio}, {Poretti}, {Poujoulet}, {Pr{\v{s}}a}, {Pulone}, {Racero},
  {Ragaini}, {Rainer}, {Raiteri}, {Rambaux}, {Ramos}, {Ramos-Lerate}, {Re
  Fiorentin}, {Regibo}, {Ripepi}, {Riva}, {Rixon}, {Robichon}, {Robin},
  {Roelens}, {Rohrbasser}, {Romero-G{\'o}mez}, {Rowell}, {Royer}, {Rybicki},
  {Sadowski}, {Sagrist{\`a} Sell{\'e}s}, {Salgado}, {Salguero}, {Samaras},
  {Sanchez Gimenez}, {Sanna}, {Santove{\~n}a}, {Sarasso}, {Schultheis},
  {Sciacca}, {Segol}, {Segovia}, {S{\'e}gransan}, {Semeux}, {Shahaf},
  {Siddiqui}, {Siebert}, {Siltala}, {Slezak}, {Solano}, {Solitro}, {Souami},
  {Souchay}, {Spagna}, {Spoto}, {Steele}, {Steidelm{\"u}ller}, {Stephenson},
  {S{\"u}veges}, {Szabados}, {Szegedi-Elek}, {Taris}, {Tauran}, {Taylor},
  {Teixeira}, {Thuillot}, {Tonello}, {Torra}, {Torra}, {Turon}, {Unger},
  {Vaillant}, {van Dillen}, {Vanel}, {Vecchiato}, {Viala}, {Vicente},
  {Voutsinas}, {Weiler}, {Wevers}, {Wyrzykowski}, {Yoldas}, {Yvard}, {Zhao},
  {Zorec}, {Zucker}, {Zurbach}, \& {Zwitter}}]{GCNS21}
{Gaia Collaboration}, {Smart}, R.~L., {Sarro}, L.~M., {et~al.} 2021, \aap, 649,
  A6

\bibitem[{{Gallart} {et~al.}(1996){Gallart}, {Aparicio}, {Bertelli}, \&
  {Chiosi}}]{gallart96b}
{Gallart}, C., {Aparicio}, A., {Bertelli}, G., \& {Chiosi}, C. 1996, \aj, 112,
  1950

\bibitem[{{Gallart} {et~al.}(2019){Gallart}, {Bernard}, {Brook}, {Ruiz-Lara},
  {Cassisi}, {Hill}, \& {Monelli}}]{gallart19}
{Gallart}, C., {Bernard}, E.~J., {Brook}, C.~B., {et~al.} 2019, Nature
  Astronomy, 3, 932

\bibitem[{{Gallart} {et~al.}(1999{\natexlab{a}}){Gallart}, {Freedman},
  {Aparicio}, {Bertelli}, \& {Chiosi}}]{gallart99}
{Gallart}, C., {Freedman}, W.~L., {Aparicio}, A., {Bertelli}, G., \& {Chiosi},
  C. 1999{\natexlab{a}}, \aj, 118, 2245

\bibitem[{{Gallart} {et~al.}(1999{\natexlab{b}}){Gallart}, {Freedman}, {Mateo},
  {Chiosi}, {Thompson}, {Aparicio}, {Bertelli}, {Hodge}, {Lee}, {Olszewski},
  {Saha}, {Stetson}, \& {Suntzeff}}]{1999ApJ...514..665G}
{Gallart}, C., {Freedman}, W.~L., {Mateo}, M., {et~al.} 1999{\natexlab{b}},
  \apj, 514, 665

\bibitem[{{Gallart} {et~al.}(2015){Gallart}, {Monelli}, {Mayer}, {Aparicio},
  {Battaglia}, {Bernard}, {Cassisi}, {Cole}, {Dolphin}, {Drozdovsky},
  {Hidalgo}, {Navarro}, {Salvadori}, {Skillman}, {Stetson}, \&
  {Weisz}}]{gallart15}
{Gallart}, C., {Monelli}, M., {Mayer}, L., {et~al.} 2015, \apjl, 811, L18

\bibitem[{{Gallart} {et~al.}(2024){Gallart}, {Surot}, {Cassisi},
  {Fern{\'a}ndez-Alvar}, {Mirabal}, {Rivero}, {Ruiz-Lara}, {Santos-Torres},
  {Aznar-Menargues}, {Battaglia}, {Queiroz}, {Monelli}, {Vasiliev},
  {Chiappini}, {Helmi}, {Hill}, {Massari}, \& {Thomas}}]{2024A&A...687A.168G}
{Gallart}, C., {Surot}, F., {Cassisi}, S., {et~al.} 2024, \aap, 687, A168

\bibitem[{{Gilmore} \& {Reid}(1983)}]{GR83}
{Gilmore}, G. \& {Reid}, N. 1983, \mnras, 202, 1025

\bibitem[{{Gonz{\'a}lez Rivera de La Vernhe} {et~al.}(2024){Gonz{\'a}lez Rivera
  de La Vernhe}, {Hill}, {Kordopatis}, {Gran}, {Fern{\'a}ndez-Alvar},
  {Ardern-Arentsen}, {Thomas}, {Sestito}, {Navarrete}, {Martin}, {Starkenburg},
  {Viswanathan}, {Battaglia}, {Venn}, \& {Vitali}}]{GZ24}
{Gonz{\'a}lez Rivera de La Vernhe}, I., {Hill}, V., {Kordopatis}, G., {et~al.}
  2024, arXiv e-prints, arXiv:2406.05728

\bibitem[{{Grand} {et~al.}(2018){Grand}, {Bustamante}, {G{\'o}mez}, {Kawata},
  {Marinacci}, {Pakmor}, {Rix}, {Simpson}, {Sparre}, \&
  {Springel}}]{Grand2018_chemdisks}
{Grand}, R. J.~J., {Bustamante}, S., {G{\'o}mez}, F.~A., {et~al.} 2018, \mnras,
  474, 3629

\bibitem[{{GRAVITY Collaboration} {et~al.}(2019){GRAVITY Collaboration},
  {Abuter}, {Amorim}, {Baub{\"o}ck}, {Berger}, {Bonnet}, {Brandner},
  {Cl{\'e}net}, {Coud{\'e} Du Foresto}, {de Zeeuw}, {Dexter}, {Duvert},
  {Eckart}, {Eisenhauer}, {F{\"o}rster Schreiber}, {Garcia}, {Gao}, {Gendron},
  {Genzel}, {Gerhard}, {Gillessen}, {Habibi}, {Haubois}, {Henning}, {Hippler},
  {Horrobin}, {Jim{\'e}nez-Rosales}, {Jocou}, {Kervella}, {Lacour},
  {Lapeyr{\`e}re}, {Le Bouquin}, {L{\'e}na}, {Ott}, {Paumard}, {Perraut},
  {Perrin}, {Pfuhl}, {Rabien}, {Rodriguez Coira}, {Rousset}, {Scheithauer},
  {Sternberg}, {Straub}, {Straubmeier}, {Sturm}, {Tacconi}, {Vincent}, {von
  Fellenberg}, {Waisberg}, {Widmann}, {Wieprecht}, {Wiezorrek}, {Woillez}, \&
  {Yazici}}]{gravity19}
{GRAVITY Collaboration}, {Abuter}, R., {Amorim}, A., {et~al.} 2019, \aap, 625,
  L10

\bibitem[{{Green} {et~al.}(2019){Green}, {Schlafly}, {Zucker}, {Speagle}, \&
  {Finkbeiner}}]{green19}
{Green}, G.~M., {Schlafly}, E., {Zucker}, C., {Speagle}, J.~S., \&
  {Finkbeiner}, D. 2019, \apj, 887, 93

\bibitem[{{Grisoni} {et~al.}(2024){Grisoni}, {Chiappini}, {Miglio}, {Brogaard},
  {Casali}, {Willett}, {Montalb{\'a}n}, {Stokholm}, {Thomsen}, {Tailo},
  {Matteuzzi}, {Valentini}, {Elsworth}, \& {Mosser}}]{Grisoni2024}
{Grisoni}, V., {Chiappini}, C., {Miglio}, A., {et~al.} 2024, \aap, 683, A111

\bibitem[{{Grisoni} {et~al.}(2017){Grisoni}, {Spitoni}, {Matteucci},
  {Recio-Blanco}, {de Laverny}, {Hayden}, {Mikolaitis}, \&
  {Worley}}]{grisoni17}
{Grisoni}, V., {Spitoni}, E., {Matteucci}, F., {et~al.} 2017, \mnras, 472, 3637

\bibitem[{{Hayden} {et~al.}(2015){Hayden}, {Bovy}, {Holtzman}, {Nidever},
  {Bird}, {Weinberg}, {Andrews}, {Majewski}, {Allende Prieto}, {Anders},
  {Beers}, {Bizyaev}, {Chiappini}, {Cunha}, {Frinchaboy},
  {Garc{\'\i}a-Her{\'n}andez}, {Garc{\'\i}a P{\'e}rez}, {Girardi}, {Harding},
  {Hearty}, {Johnson}, {M{\'e}sz{\'a}ros}, {Minchev}, {O'Connell}, {Pan},
  {Robin}, {Schiavon}, {Schneider}, {Schultheis}, {Shetrone}, {Skrutskie},
  {Steinmetz}, {Smith}, {Wilson}, {Zamora}, \&
  {Zasowski}}]{Hayden2015_chemicalStructure}
{Hayden}, M.~R., {Bovy}, J., {Holtzman}, J.~A., {et~al.} 2015, \apj, 808, 132

\bibitem[{{Haywood}(2008)}]{haywood08}
{Haywood}, M. 2008, \mnras, 388, 1175

\bibitem[{{Haywood} {et~al.}(2013){Haywood}, {Di Matteo}, {Lehnert}, {Katz}, \&
  {G{\'o}mez}}]{haywood13}
{Haywood}, M., {Di Matteo}, P., {Lehnert}, M.~D., {Katz}, D., \& {G{\'o}mez},
  A. 2013, \aap, 560, A109

\bibitem[{{Helmi} {et~al.}(2018){Helmi}, {Babusiaux}, {Koppelman}, {Massari},
  {Veljanoski}, \& {Brown}}]{helmi18}
{Helmi}, A., {Babusiaux}, C., {Koppelman}, H.~H., {et~al.} 2018, \nat, 563, 85

\bibitem[{{Hernandez} {et~al.}(1999){Hernandez}, {Valls-Gabaud}, \&
  {Gilmore}}]{Hernandez1999}
{Hernandez}, X., {Valls-Gabaud}, D., \& {Gilmore}, G. 1999, \mnras, 304, 705

\bibitem[{{Hernandez} {et~al.}(2000){Hernandez}, {Valls-Gabaud}, \&
  {Gilmore}}]{Hernandez2000_Hipparcos}
{Hernandez}, X., {Valls-Gabaud}, D., \& {Gilmore}, G. 2000, \mnras, 316, 605

\bibitem[{{Hidalgo} {et~al.}(2009){Hidalgo}, {Aparicio},
  {Mart{\'\i}nez-Delgado}, \& {Gallart}}]{hidalgo09}
{Hidalgo}, S.~L., {Aparicio}, A., {Mart{\'\i}nez-Delgado}, D., \& {Gallart}, C.
  2009, \apj, 705, 704

\bibitem[{{Hidalgo} {et~al.}(2013){Hidalgo}, {Monelli}, {Aparicio}, {Gallart},
  {Skillman}, {Cassisi}, {Bernard}, {Mayer}, {Stetson}, {Cole}, \&
  {Dolphin}}]{hidalgo13}
{Hidalgo}, S.~L., {Monelli}, M., {Aparicio}, A., {et~al.} 2013, \apj, 778, 103

\bibitem[{{Hidalgo} {et~al.}(2018){Hidalgo}, {Pietrinferni}, {Cassisi},
  {Salaris}, {Mucciarelli}, {Savino}, {Aparicio}, {Silva Aguirre}, \&
  {Verma}}]{hidalgo18}
{Hidalgo}, S.~L., {Pietrinferni}, A., {Cassisi}, S., {et~al.} 2018, \apj, 856,
  125

\bibitem[{{Jofr{\'e}} {et~al.}(2023){Jofr{\'e}}, {Jorissen},
  {Aguilera-G{\'o}mez}, {Van Eck}, {Tayar}, {Pinsonneault}, {Zinn}, {Goriely},
  \& {Van Winckel}}]{Jofre2023}
{Jofr{\'e}}, P., {Jorissen}, A., {Aguilera-G{\'o}mez}, C., {et~al.} 2023, \aap,
  671, A21

\bibitem[{{Jofr{\'e}} {et~al.}(2016){Jofr{\'e}}, {Jorissen}, {Van Eck},
  {Izzard}, {Masseron}, {Hawkins}, {Gilmore}, {Paladini}, {Escorza},
  {Blanco-Cuaresma}, \& {Manick}}]{Jofre2016}
{Jofr{\'e}}, P., {Jorissen}, A., {Van Eck}, S., {et~al.} 2016, \aap, 595, A60

\bibitem[{{Juri{\'c}} {et~al.}(2008){Juri{\'c}}, {Ivezi{\'c}}, {Brooks},
  {Lupton}, {Schlegel}, {Finkbeiner}, {Padmanabhan}, {Bond}, {Sesar},
  {Rockosi}, {Knapp}, {Gunn}, {Sumi}, {Schneider}, {Barentine}, {Brewington},
  {Brinkmann}, {Fukugita}, {Harvanek}, {Kleinman}, {Krzesinski}, {Long},
  {Neilsen}, {Nitta}, {Snedden}, \& {York}}]{Juric2008}
{Juri{\'c}}, M., {Ivezi{\'c}}, {\v{Z}}., {Brooks}, A., {et~al.} 2008, \apj,
  673, 864

\bibitem[{{Kawata} \& {Chiappini}(2016)}]{KawataChiappini2016}
{Kawata}, D. \& {Chiappini}, C. 2016, Astronomische Nachrichten, 337, 976

\bibitem[{{Kepley} {et~al.}(2007){Kepley}, {Morrison}, {Helmi}, {Kinman}, {Van
  Duyne}, {Martin}, {Harding}, {Norris}, \& {Freeman}}]{Kepley2007HS}
{Kepley}, A.~A., {Morrison}, H.~L., {Helmi}, A., {et~al.} 2007, \aj, 134, 1579

\bibitem[{{Khanna} {et~al.}(2024){Khanna}, {Yu}, {Drimmel}, {Poggio},
  {Cantat-Gaudin}, {Castro-Ginard}, {Kurbatov}, {Belokurov}, {Brown},
  {Fouesneau}, {Casey}, \& {Rix}}]{khannaa24}
{Khanna}, S., {Yu}, J., {Drimmel}, R., {et~al.} 2024, arXiv e-prints,
  arXiv:2410.22036

\bibitem[{{Kobayashi} {et~al.}(2011){Kobayashi}, {Karakas}, \&
  {Umeda}}]{kobayashi11}
{Kobayashi}, C., {Karakas}, A.~I., \& {Umeda}, H. 2011, \mnras, 414, 3231

\bibitem[{{Koppelman} {et~al.}(2019){Koppelman}, {Helmi}, {Massari},
  {Roelenga}, \& {Bastian}}]{Koppelman2019HS}
{Koppelman}, H.~H., {Helmi}, A., {Massari}, D., {Roelenga}, S., \& {Bastian},
  U. 2019, \aap, 625, A5

\bibitem[{{Kroupa} {et~al.}(1993){Kroupa}, {Tout}, \& {Gilmore}}]{kroupa93}
{Kroupa}, P., {Tout}, C.~A., \& {Gilmore}, G. 1993, \mnras, 262, 545

\bibitem[{{Lagarde} {et~al.}(2021){Lagarde}, {Reyl{\'e}}, {Chiappini}, {Mor},
  {Anders}, {Figueras}, {Miglio}, {Romero-G{\'o}mez}, {Antoja}, {Cabral},
  {Salomon}, {Robin}, {Bienaym{\'e}}, {Soubiran}, {Cornu}, \&
  {Montillaud}}]{Lagarde2021}
{Lagarde}, N., {Reyl{\'e}}, C., {Chiappini}, C., {et~al.} 2021, \aap, 654, A13

\bibitem[{{Lallement} {et~al.}(2022){Lallement}, {Vergely}, {Babusiaux}, \&
  {Cox}}]{lallement22}
{Lallement}, R., {Vergely}, J.~L., {Babusiaux}, C., \& {Cox}, N.~L.~J. 2022,
  \aap, 661, A147

\bibitem[{{Laporte} {et~al.}(2018){Laporte}, {Johnston}, {G{\'o}mez},
  {Garavito-Camargo}, \& {Besla}}]{laporte18}
{Laporte}, C. F.~P., {Johnston}, K.~V., {G{\'o}mez}, F.~A., {Garavito-Camargo},
  N., \& {Besla}, G. 2018, \mnras, 481, 286

\bibitem[{{Law} \& {Majewski}(2010)}]{law10}
{Law}, D.~R. \& {Majewski}, S.~R. 2010, \apj, 714, 229

\bibitem[{{Lebreton} \& {Goupil}(2014)}]{lebreton14}
{Lebreton}, Y. \& {Goupil}, M.~J. 2014, \aap, 569, A21

\bibitem[{{Lemasle} {et~al.}(2018){Lemasle}, {Hajdu}, {Kovtyukh}, {Inno},
  {Grebel}, {Catelan}, {Bono}, {Fran{\c{c}}ois}, {Kniazev}, {da Silva}, \&
  {Storm}}]{lemasle18}
{Lemasle}, B., {Hajdu}, G., {Kovtyukh}, V., {et~al.} 2018, \aap, 618, A160

\bibitem[{{Lian} {et~al.}(2023){Lian}, {Bergemann}, {Pillepich}, {Zasowski}, \&
  {Lane}}]{lian23}
{Lian}, J., {Bergemann}, M., {Pillepich}, A., {Zasowski}, G., \& {Lane}, R.~R.
  2023, Nature Astronomy, 7, 951

\bibitem[{{Lindegren} {et~al.}(2021){Lindegren}, {Bastian}, {Biermann},
  {Bombrun}, {de Torres}, {Gerlach}, {Geyer}, {Hern{\'a}ndez}, {Hilger},
  {Hobbs}, {Klioner}, {Lammers}, {McMillan}, {Ramos-Lerate},
  {Steidelm{\"u}ller}, {Stephenson}, \& {van Leeuwen}}]{lindegren21}
{Lindegren}, L., {Bastian}, U., {Biermann}, M., {et~al.} 2021, \aap, 649, A4

\bibitem[{{Mackereth} {et~al.}(2019){Mackereth}, {Bovy}, {Leung}, {Schiavon},
  {Trick}, {Chaplin}, {Cunha}, {Feuillet}, {Majewski}, {Martig}, {Miglio},
  {Nidever}, {Pinsonneault}, {Aguirre}, {Sobeck}, {Tayar}, \&
  {Zasowski}}]{mackereth19}
{Mackereth}, J.~T., {Bovy}, J., {Leung}, H.~W., {et~al.} 2019, \mnras, 489, 176

\bibitem[{{Majewski} {et~al.}(2017){Majewski}, {Schiavon}, {Frinchaboy},
  {Allende Prieto}, {Barkhouser}, {Bizyaev}, {Blank}, {Brunner}, {Burton},
  {Carrera}, {Chojnowski}, {Cunha}, {Epstein}, {Fitzgerald}, {Garc{\'\i}a
  P{\'e}rez}, {Hearty}, {Henderson}, {Holtzman}, {Johnson}, {Lam}, {Lawler},
  {Maseman}, {M{\'e}sz{\'a}ros}, {Nelson}, {Nguyen}, {Nidever}, {Pinsonneault},
  {Shetrone}, {Smee}, {Smith}, {Stolberg}, {Skrutskie}, {Walker}, {Wilson},
  {Zasowski}, {Anders}, {Basu}, {Beland}, {Blanton}, {Bovy}, {Brownstein},
  {Carlberg}, {Chaplin}, {Chiappini}, {Eisenstein}, {Elsworth}, {Feuillet},
  {Fleming}, {Galbraith-Frew}, {Garc{\'\i}a}, {Garc{\'\i}a-Hern{\'a}ndez},
  {Gillespie}, {Girardi}, {Gunn}, {Hasselquist}, {Hayden}, {Hekker}, {Ivans},
  {Kinemuchi}, {Klaene}, {Mahadevan}, {Mathur}, {Mosser}, {Muna}, {Munn},
  {Nichol}, {O'Connell}, {Parejko}, {Robin}, {Rocha-Pinto}, {Schultheis},
  {Serenelli}, {Shane}, {Silva Aguirre}, {Sobeck}, {Thompson}, {Troup},
  {Weinberg}, \& {Zamora}}]{majewski17}
{Majewski}, S.~R., {Schiavon}, R.~P., {Frinchaboy}, P.~M., {et~al.} 2017, \aj,
  154, 94

\bibitem[{{Martig} {et~al.}(2015){Martig}, {Rix}, {Silva Aguirre}, {Hekker},
  {Mosser}, {Elsworth}, {Bovy}, {Stello}, {Anders}, {Garc{\'\i}a}, {Tayar},
  {Rodrigues}, {Basu}, {Carrera}, {Ceillier}, {Chaplin}, {Chiappini},
  {Frinchaboy}, {Garc{\'\i}a-Hern{\'a}ndez}, {Hearty}, {Holtzman}, {Johnson},
  {Majewski}, {Mathur}, {M{\'e}sz{\'a}ros}, {Miglio}, {Nidever}, {Pan},
  {Pinsonneault}, {Schiavon}, {Schneider}, {Serenelli}, {Shetrone}, \&
  {Zamora}}]{Martig2015}
{Martig}, M., {Rix}, H.-W., {Silva Aguirre}, V., {et~al.} 2015, \mnras, 451,
  2230

\bibitem[{{Matteucci} \& {Greggio}(1986)}]{matteucci86}
{Matteucci}, F. \& {Greggio}, L. 1986, \aap, 154, 279

\bibitem[{{McQuinn} {et~al.}(2024){McQuinn}, {B. Newman}, {Savino}, {Dolphin},
  {Weisz}, {Williams}, {Boyer}, {Cohen}, {Correnti}, {Cole}, {Geha}, {Gennaro},
  {Kallivayalil}, {Sandstrom}, {Skillman}, {Anderson}, {Bolatto},
  {Boylan-Kolchin}, {Garling}, {Gilbert}, {Girardi}, {Kalirai}, {Mazzi},
  {Pastorelli}, {Richstein}, \& {Warfield}}]{McQuinn2024}
{McQuinn}, K. B.~W., {B. Newman}, M.~J., {Savino}, A., {et~al.} 2024, \apj,
  961, 16

\bibitem[{{M{\'e}ndez-Delgado} {et~al.}(2022){M{\'e}ndez-Delgado}, {Amayo},
  {Arellano-C{\'o}rdova}, {Esteban}, {Garc{\'\i}a-Rojas}, {Carigi}, \&
  {Delgado-Inglada}}]{MendezDelgado2022}
{M{\'e}ndez-Delgado}, J.~E., {Amayo}, A., {Arellano-C{\'o}rdova}, K.~Z.,
  {et~al.} 2022, \mnras, 510, 4436

\bibitem[{{Miglio} {et~al.}(2021){Miglio}, {Chiappini}, {Mackereth}, {Davies},
  {Brogaard}, {Casagrande}, {Chaplin}, {Girardi}, {Kawata}, {Khan}, {Izzard},
  {Montalb{\'a}n}, {Mosser}, {Vincenzo}, {Bossini}, {Noels}, {Rodrigues},
  {Valentini}, \& {Mandel}}]{miglio21}
{Miglio}, A., {Chiappini}, C., {Mackereth}, J.~T., {et~al.} 2021, \aap, 645,
  A85

\bibitem[{{Minchev} {et~al.}(2013){Minchev}, {Chiappini}, \&
  {Martig}}]{minchev13}
{Minchev}, I., {Chiappini}, C., \& {Martig}, M. 2013, \aap, 558, A9

\bibitem[{{Minchev} {et~al.}(2014){Minchev}, {Chiappini}, \&
  {Martig}}]{minchev14}
{Minchev}, I., {Chiappini}, C., \& {Martig}, M. 2014, \aap, 572, A92

\bibitem[{{Monelli} {et~al.}(2010{\natexlab{a}}){Monelli}, {Hidalgo},
  {Stetson}, {Aparicio}, {Gallart}, {Dolphin}, {Cole}, {Weisz}, {Skillman},
  {Bernard}, {Mayer}, {Navarro}, {Cassisi}, {Drozdovsky}, \&
  {Tolstoy}}]{monelli10}
{Monelli}, M., {Hidalgo}, S.~L., {Stetson}, P.~B., {et~al.} 2010{\natexlab{a}},
  \apj, 720, 1225

\bibitem[{{Monelli} {et~al.}(2010{\natexlab{b}}){Monelli}, {Hidalgo},
  {Stetson}, {Aparicio}, {Gallart}, {Dolphin}, {Cole}, {Weisz}, {Skillman},
  {Bernard}, {Mayer}, {Navarro}, {Cassisi}, {Drozdovsky}, \&
  {Tolstoy}}]{2010ApJ...720.1225M}
{Monelli}, M., {Hidalgo}, S.~L., {Stetson}, P.~B., {et~al.} 2010{\natexlab{b}},
  \apj, 720, 1225

\bibitem[{{Montalb{\'a}n} {et~al.}(2021){Montalb{\'a}n}, {Mackereth}, {Miglio},
  {Vincenzo}, {Chiappini}, {Buldgen}, {Mosser}, {Noels}, {Scuflaire}, {Vrard},
  {Willett}, {Davies}, {Hall}, {Nielsen}, {Khan}, {Rendle}, {van Rossem},
  {Ferguson}, \& {Chaplin}}]{montalban21}
{Montalb{\'a}n}, J., {Mackereth}, J.~T., {Miglio}, A., {et~al.} 2021, Nature
  Astronomy, 5, 640

\bibitem[{{Morrison} {et~al.}(1994){Morrison}, {Boroson}, \&
  {Harding}}]{morrison94}
{Morrison}, H.~L., {Boroson}, T.~A., \& {Harding}, P. 1994, \aj, 108, 1191

\bibitem[{{Nepal} {et~al.}(2024{\natexlab{a}}){Nepal}, {Chiappini}, {Guiglion},
  {Steinmetz}, {P{\'e}rez-Villegas}, {Queiroz}, {Miglio}, {Dohme}, \&
  {Khalatyan}}]{Nepal2024youngbar}
{Nepal}, S., {Chiappini}, C., {Guiglion}, G., {et~al.} 2024{\natexlab{a}},
  \aap, 681, L8

\bibitem[{{Nepal} {et~al.}(2024{\natexlab{b}}){Nepal}, {Chiappini}, {Queiroz},
  {Guiglion}, {Montalb{\'a}n}, {Steinmetz}, {Miglio}, \&
  {Khalatyan}}]{nepal24disk}
{Nepal}, S., {Chiappini}, C., {Queiroz}, A.~B., {et~al.} 2024{\natexlab{b}},
  \aap, 688, A167

\bibitem[{{No{\"e}l} {et~al.}(2007){No{\"e}l}, {Gallart}, {Costa}, \&
  {M{\'e}ndez}}]{noel07}
{No{\"e}l}, N. E.~D., {Gallart}, C., {Costa}, E., \& {M{\'e}ndez}, R.~A. 2007,
  \aj, 133, 2037

\bibitem[{{Orkney} {et~al.}(2022){Orkney}, {Laporte}, {Grand}, {G{\'o}mez},
  {van de Voort}, {Marinacci}, {Fragkoudi}, {Pakmor}, \& {Springel}}]{orkney22}
{Orkney}, M. D.~A., {Laporte}, C. F.~P., {Grand}, R. J.~J., {et~al.} 2022,
  \mnras, 517, L138

\bibitem[{{Pagel} \& {Tautvaisiene}(1995)}]{pagel95}
{Pagel}, B.~E.~J. \& {Tautvaisiene}, G. 1995, \mnras, 276, 505

\bibitem[{{Pietrinferni} {et~al.}(2021){Pietrinferni}, {Hidalgo}, {Cassisi},
  {Salaris}, {Savino}, {Mucciarelli}, {Verma}, {Silva Aguirre}, {Aparicio}, \&
  {Ferguson}}]{pietri21}
{Pietrinferni}, A., {Hidalgo}, S., {Cassisi}, S., {et~al.} 2021, \apj, 908, 102

\bibitem[{{Pinna} {et~al.}(2024){Pinna}, {Walo-Mart{\'\i}n}, {Grand}, {Martig},
  {Fragkoudi}, {G{\'o}mez}, {Marinacci}, \& {Pakmor}}]{pinna24a}
{Pinna}, F., {Walo-Mart{\'\i}n}, D., {Grand}, R. J.~J., {et~al.} 2024, \aap,
  683, A236

\bibitem[{{Pinsonneault} {et~al.}(2024){Pinsonneault}, {Zinn}, {Tayar},
  {Serenelli}, {Garcia}, {Mathur}, {Vrard}, {Elsworth}, {Mosser}, {Stello},
  {Bell}, {Bugnet}, {Corsaro}, {Gaulme}, {Hekker}, {Hon}, {Huber}, {Kallinger},
  {Cao}, {Johnson}, {Liagre}, {Patton}, {Santos}, {Basu}, {Beck}, {Beers},
  {Chaplin}, {Cunha}, {Frinchaboy}, {Girardi}, {Godoy-Rivera}, {Holtzman},
  {Jonsson}, {Meszaros}, {Reyes}, {Rix}, {Shetrone}, {Smith}, {Spoo},
  {Stassun}, \& {Wang}}]{Pinsonneault2024arXiv}
{Pinsonneault}, M.~H., {Zinn}, J.~C., {Tayar}, J., {et~al.} 2024, arXiv
  e-prints, arXiv:2410.00102

\bibitem[{{Prantzos} {et~al.}(2023){Prantzos}, {Abia}, {Chen}, {de Laverny},
  {Recio-Blanco}, {Athanassoula}, {Roberti}, {Vescovi}, {Limongi}, {Chieffi},
  \& {Cristallo}}]{prantzos23}
{Prantzos}, N., {Abia}, C., {Chen}, T., {et~al.} 2023, \mnras, 523, 2126

\bibitem[{{Queiroz} {et~al.}(2023){Queiroz}, {Anders}, {Chiappini},
  {Khalatyan}, {Santiago}, {Nepal}, {Steinmetz}, {Gallart}, {Valentini}, {Dal
  Ponte}, {Barbuy}, {P{\'e}rez-Villegas}, {Masseron}, {Fern{\'a}ndez-Trincado},
  {Khoperskov}, {Minchev}, {Fern{\'a}ndez-Alvar}, {Lane}, \&
  {Nitschelm}}]{queiroz23}
{Queiroz}, A.~B.~A., {Anders}, F., {Chiappini}, C., {et~al.} 2023, \aap, 673,
  A155

\bibitem[{{Queiroz} {et~al.}(2020){Queiroz}, {Anders}, {Chiappini},
  {Khalatyan}, {Santiago}, {Steinmetz}, {Valentini}, {Miglio}, {Bossini},
  {Barbuy}, {Minchev}, {Minniti}, {Garc{\'\i}a Hern{\'a}ndez}, {Schultheis},
  {Beaton}, {Beers}, {Bizyaev}, {Brownstein}, {Cunha},
  {Fern{\'a}ndez-Trincado}, {Frinchaboy}, {Lane}, {Majewski}, {Nataf},
  {Nitschelm}, {Pan}, {Roman-Lopes}, {Sobeck}, {Stringfellow}, \&
  {Zamora}}]{Queiroz2020}
{Queiroz}, A.~B.~A., {Anders}, F., {Chiappini}, C., {et~al.} 2020, \aap, 638,
  A76

\bibitem[{{Ram{\'\i}rez} {et~al.}(2013){Ram{\'\i}rez}, {Allende Prieto}, \&
  {Lambert}}]{ramirez13}
{Ram{\'\i}rez}, I., {Allende Prieto}, C., \& {Lambert}, D.~L. 2013, \apj, 764,
  78

\bibitem[{{Randich} {et~al.}(2022){Randich}, {Gilmore}, {Magrini}, {Sacco},
  {Jackson}, {Jeffries}, {Worley}, {Hourihane}, {Gonneau}, {Viscasillas
  Vazquez}, {Franciosini}, {Lewis}, {Alfaro}, {Allende Prieto}, {Bensby},
  {Blomme}, {Bragaglia}, {Flaccomio}, {Fran{\c{c}}ois}, {Irwin}, {Koposov},
  {Korn}, {Lanzafame}, {Pancino}, {Recio-Blanco}, {Smiljanic}, {Van Eck},
  {Zwitter}, {Asplund}, {Bonifacio}, {Feltzing}, {Binney}, {Drew}, {Ferguson},
  {Micela}, {Negueruela}, {Prusti}, {Rix}, {Vallenari}, {Bayo}, {Bergemann},
  {Biazzo}, {Carraro}, {Casey}, {Damiani}, {Frasca}, {Heiter}, {Hill},
  {Jofr{\'e}}, {de Laverny}, {Lind}, {Marconi}, {Martayan}, {Masseron},
  {Monaco}, {Morbidelli}, {Prisinzano}, {Sbordone}, {Sousa}, {Zaggia},
  {Adibekyan}, {Bonito}, {Caffau}, {Daflon}, {Feuillet}, {Gebran}, {Gonzalez
  Hernandez}, {Guiglion}, {Herrero}, {Lobel}, {Maiz Apellaniz}, {Merle},
  {Mikolaitis}, {Montes}, {Morel}, {Soubiran}, {Spina}, {Tabernero},
  {Tautvai{\v{s}}iene}, {Traven}, {Valentini}, {Van der Swaelmen}, {Villanova},
  {Wright}, {Abbas}, {Aguirre B{\o}rsen-Koch}, {Alves}, {Balaguer-Nunez},
  {Barklem}, {Barrado}, {Berlanas}, {Binks}, {Bressan}, {Capuzzo-Dolcetta},
  {Casagrande}, {Casamiquela}, {Collins}, {D'Orazi}, {Dantas}, {Debattista},
  {Delgado-Mena}, {Di Marcantonio}, {Drazdauskas}, {Evans}, {Famaey},
  {Franchini}, {Fr{\'e}mat}, {Friel}, {Fu}, {Geisler}, {Gerhard}, {Gonzalez
  Solares}, {Grebel}, {Gutierrez Albarran}, {Hatzidimitriou}, {Held},
  {Jim{\'e}nez-Esteban}, {J{\"o}nsson}, {Jordi}, {Khachaturyants},
  {Kordopatis}, {Kos}, {Lagarde}, {Mahy}, {Mapelli}, {Marfil}, {Martell},
  {Messina}, {Miglio}, {Minchev}, {Moitinho}, {Montalban}, {Monteiro},
  {Morossi}, {Mowlavi}, {Mucciarelli}, {Murphy}, {Nardetto}, {Ortolani},
  {Paletou}, {Palou{\v{s}}}, {Paunzen}, {Pickering}, {Quirrenbach}, {Re
  Fiorentin}, {Read}, {Romano}, {Ryde}, {Sanna}, {Santos}, {Seabroke},
  {Spagna}, {Steinmetz}, {Stonkut{\'e}}, {Sutorius}, {Th{\'e}venin}, {Tosi},
  {Tsantaki}, {Vink}, {Wright}, {Wyse}, {Zoccali}, {Zorec}, {Zucker}, \&
  {Walton}}]{gaiaeso22}
{Randich}, S., {Gilmore}, G., {Magrini}, L., {et~al.} 2022, \aap, 666, A121

\bibitem[{{Recio-Blanco} {et~al.}(2014){Recio-Blanco}, {de Laverny},
  {Kordopatis}, {Helmi}, {Hill}, {Gilmore}, {Wyse}, {Adibekyan}, {Randich},
  {Asplund}, {Feltzing}, {Jeffries}, {Micela}, {Vallenari}, {Alfaro}, {Allende
  Prieto}, {Bensby}, {Bragaglia}, {Flaccomio}, {Koposov}, {Korn}, {Lanzafame},
  {Pancino}, {Smiljanic}, {Jackson}, {Lewis}, {Magrini}, {Morbidelli},
  {Prisinzano}, {Sacco}, {Worley}, {Hourihane}, {Bergemann}, {Costado},
  {Heiter}, {Joffre}, {Lardo}, {Lind}, \& {Maiorca}}]{Recio_Blanco2014}
{Recio-Blanco}, A., {de Laverny}, P., {Kordopatis}, G., {et~al.} 2014, \aap,
  567, A5

\bibitem[{{Recio-Blanco} {et~al.}(2024{\natexlab{a}}){Recio-Blanco}, {de
  Laverny}, {Palicio}, {Cassisi}, {Pietrinferni}, \& {Lagarde}}]{RC24}
{Recio-Blanco}, A., {de Laverny}, P., {Palicio}, P.~A., {et~al.}
  2024{\natexlab{a}}, arXiv e-prints, arXiv:2402.01522

\bibitem[{{Recio-Blanco} {et~al.}(2024{\natexlab{b}}){Recio-Blanco}, {de
  Laverny}, {Palicio}, {Cassisi}, {Pietrinferni}, {Lagarde}, \&
  {Navarrete}}]{RB_golden}
{Recio-Blanco}, A., {de Laverny}, P., {Palicio}, P.~A., {et~al.}
  2024{\natexlab{b}}, arXiv e-prints, arXiv:2402.01522

\bibitem[{{Recio-Blanco} {et~al.}(2023){Recio-Blanco}, {de Laverny}, {Palicio},
  {Kordopatis}, {{\'A}lvarez}, {Schultheis}, {Contursi}, {Zhao}, {Torralba
  Elipe}, {Ordenovic}, {Manteiga}, {Dafonte}, {Oreshina-Slezak}, {Bijaoui},
  {Fr{\'e}mat}, {Seabroke}, {Pailler}, {Spitoni}, {Poggio}, {Creevey}, {Abreu
  Aramburu}, {Accart}, {Andrae}, {Bailer-Jones}, {Bellas-Velidis}, {Brouillet},
  {Brugaletta}, {Burlacu}, {Carballo}, {Casamiquela}, {Chiavassa}, {Cooper},
  {Dapergolas}, {Delchambre}, {Dharmawardena}, {Drimmel}, {Edvardsson},
  {Fouesneau}, {Garabato}, {Garc{\'\i}a-Lario}, {Garc{\'\i}a-Torres}, {Gavel},
  {Gomez}, {Gonz{\'a}lez-Santamar{\'\i}a}, {Hatzidimitriou}, {Heiter},
  {Jean-Antoine Piccolo}, {Kontizas}, {Korn}, {Lanzafame}, {Lebreton}, {Le
  Fustec}, {Licata}, {Lindstr{\o}m}, {Livanou}, {Lobel}, {Lorca}, {Magdaleno
  Romeo}, {Marocco}, {Marshall}, {Mary}, {Nicolas}, {Pallas-Quintela}, {Panem},
  {Pichon}, {Riclet}, {Robin}, {Rybizki}, {Santove{\~n}a}, {Silvelo}, {Smart},
  {Sarro}, {Sordo}, {Soubiran}, {S{\"u}veges}, {Ulla}, {Vallenari}, {Zorec},
  {Utrilla}, \& {Bakker}}]{RB23_RVS}
{Recio-Blanco}, A., {de Laverny}, P., {Palicio}, P.~A., {et~al.} 2023, \aap,
  674, A29

\bibitem[{{Reid} {et~al.}(2019){Reid}, {Menten}, {Brunthaler}, {Zheng}, {Dame},
  {Xu}, {Li}, {Sakai}, {Wu}, {Immer}, {Zhang}, {Sanna}, {Moscadelli}, {Rygl},
  {Bartkiewicz}, {Hu}, {Quiroga-Nu{\~n}ez}, \& {van Langevelde}}]{reid19}
{Reid}, M.~J., {Menten}, K.~M., {Brunthaler}, A., {et~al.} 2019, \apj, 885, 131

\bibitem[{{Renaud} {et~al.}(2021{\natexlab{a}}){Renaud}, {Agertz}, {Andersson},
  {Read}, {Ryde}, {Bensby}, {Rey}, \& {Feuillet}}]{renaud21b}
{Renaud}, F., {Agertz}, O., {Andersson}, E.~P., {et~al.} 2021{\natexlab{a}},
  \mnras, 503, 5868

\bibitem[{{Renaud} {et~al.}(2021{\natexlab{b}}){Renaud}, {Agertz}, {Read},
  {Ryde}, {Andersson}, {Bensby}, {Rey}, \& {Feuillet}}]{renaud21a}
{Renaud}, F., {Agertz}, O., {Read}, J.~I., {et~al.} 2021{\natexlab{b}}, \mnras,
  503, 5846

\bibitem[{{Riello} {et~al.}(2021){Riello}, {De Angeli}, {Evans}, {Montegriffo},
  {Carrasco}, {Busso}, {Palaversa}, {Burgess}, {Diener}, {Davidson}, {Rowell},
  {Fabricius}, {Jordi}, {Bellazzini}, {Pancino}, {Harrison}, {Cacciari}, {van
  Leeuwen}, {Hambly}, {Hodgkin}, {Osborne}, {Altavilla}, {Barstow}, {Brown},
  {Castellani}, {Cowell}, {De Luise}, {Gilmore}, {Giuffrida}, {Hidalgo},
  {Holland}, {Marinoni}, {Pagani}, {Piersimoni}, {Pulone}, {Ragaini}, {Rainer},
  {Richards}, {Sanna}, {Walton}, {Weiler}, \& {Yoldas}}]{2021A&A...649A...3R}
{Riello}, M., {De Angeli}, F., {Evans}, D.~W., {et~al.} 2021, \aap, 649, A3

\bibitem[{{Ruiz-Lara} {et~al.}(2020){Ruiz-Lara}, {Gallart}, {Bernard}, \&
  {Cassisi}}]{ruizlara20}
{Ruiz-Lara}, T., {Gallart}, C., {Bernard}, E.~J., \& {Cassisi}, S. 2020, Nature
  Astronomy, 4, 965

\bibitem[{{Ruiz-Lara} {et~al.}(2021){Ruiz-Lara}, {Gallart}, {Monelli}, {Fritz},
  {Battaglia}, {Cassisi}, {Aznar}, {Russo Cabrera},
  {Rodr{\'\i}guez-Mart{\'\i}n}, \&
  {Salazar-Gonz{\'a}lez}}]{2021MNRAS.501.3962R}
{Ruiz-Lara}, T., {Gallart}, C., {Monelli}, M., {et~al.} 2021, \mnras, 501, 3962

\bibitem[{{Ruiz-Lara} {et~al.}(2022){Ruiz-Lara}, {Helmi}, {Gallart}, {Surot},
  \& {Cassisi}}]{Ruiz-Lara2022HS}
{Ruiz-Lara}, T., {Helmi}, A., {Gallart}, C., {Surot}, F., \& {Cassisi}, S.
  2022, \aap, 668, L10

\bibitem[{{Sahlholdt} {et~al.}(2022){Sahlholdt}, {Feltzing}, \&
  {Feuillet}}]{sahlholdt22}
{Sahlholdt}, C.~L., {Feltzing}, S., \& {Feuillet}, D.~K. 2022, \mnras, 510,
  4669

\bibitem[{{Salaris} {et~al.}(1993){Salaris}, {Chieffi}, \&
  {Straniero}}]{salaris93}
{Salaris}, M., {Chieffi}, A., \& {Straniero}, O. 1993, \apj, 414, 580

\bibitem[{{Sch{\"o}nrich} {et~al.}(2010){Sch{\"o}nrich}, {Binney}, \&
  {Dehnen}}]{Schonrich2010}
{Sch{\"o}nrich}, R., {Binney}, J., \& {Dehnen}, W. 2010, \mnras, 403, 1829

\bibitem[{{Serenelli} {et~al.}(2017){Serenelli}, {Johnson}, {Huber},
  {Pinsonneault}, {Ball}, {Tayar}, {Silva Aguirre}, {Basu}, {Troup}, {Hekker},
  {Kallinger}, {Stello}, {Davies}, {Lund}, {Mathur}, {Mosser}, {Stassun},
  {Chaplin}, {Elsworth}, {Garc{\'\i}a}, {Handberg}, {Holtzman}, {Hearty},
  {Garc{\'\i}a-Hern{\'a}ndez}, {Gaulme}, \& {Zamora}}]{aldo17}
{Serenelli}, A., {Johnson}, J., {Huber}, D., {et~al.} 2017, \apjs, 233, 23

\bibitem[{{Sestito} {et~al.}(2021){Sestito}, {Buck}, {Starkenburg}, {Martin},
  {Navarro}, {Venn}, {Obreja}, {Jablonka}, \& {Macci{\`o}}}]{sestito21}
{Sestito}, F., {Buck}, T., {Starkenburg}, E., {et~al.} 2021, \mnras, 500, 3750

\bibitem[{{Sestito} {et~al.}(2019){Sestito}, {Longeard}, {Martin},
  {Starkenburg}, {Fouesneau}, {Gonz{\'a}lez Hern{\'a}ndez}, {Arentsen},
  {Ibata}, {Aguado}, {Carlberg}, {Jablonka}, {Navarro}, {Tolstoy}, \&
  {Venn}}]{sestito19}
{Sestito}, F., {Longeard}, N., {Martin}, N.~F., {et~al.} 2019, \mnras, 484,
  2166

\bibitem[{{Sestito} {et~al.}(2020){Sestito}, {Martin}, {Starkenburg},
  {Arentsen}, {Ibata}, {Longeard}, {Kielty}, {Youakim}, {Venn}, {Aguado},
  {Carlberg}, {Gonz{\'a}lez Hern{\'a}ndez}, {Hill}, {Jablonka}, {Kordopatis},
  {Malhan}, {Navarro}, {S{\'a}nchez-Janssen}, {Thomas}, {Tolstoy}, {Wilson},
  {Palicio}, {Bialek}, {Garcia-Dias}, {Lucchesi}, {North}, {Osorio}, {Patrick},
  \& {Peralta de Arriba}}]{sestito20}
{Sestito}, F., {Martin}, N.~F., {Starkenburg}, E., {et~al.} 2020, \mnras, 497,
  L7

\bibitem[{{Seth} {et~al.}(2005){Seth}, {Dalcanton}, \& {de Jong}}]{seth05}
{Seth}, A.~C., {Dalcanton}, J.~J., \& {de Jong}, R.~S. 2005, \aj, 130, 1574

\bibitem[{{Sharma} {et~al.}(2021){Sharma}, {Hayden}, \&
  {Bland-Hawthorn}}]{sharma21}
{Sharma}, S., {Hayden}, M.~R., \& {Bland-Hawthorn}, J. 2021, \mnras, 507, 5882

\bibitem[{{Silva Aguirre} {et~al.}(2018){Silva Aguirre}, {Bojsen-Hansen},
  {Slumstrup}, {Casagrande}, {Kawata}, {Ciuc{\v{a}}}, {Handberg}, {Lund},
  {Mosumgaard}, {Huber}, {Johnson}, {Pinsonneault}, {Serenelli}, {Stello},
  {Tayar}, {Bird}, {Cassisi}, {Hon}, {Martig}, {Nissen}, {Rix},
  {Sch{\"o}nrich}, {Sahlholdt}, {Trick}, \& {Yu}}]{Silva-Aguirre2018}
{Silva Aguirre}, V., {Bojsen-Hansen}, M., {Slumstrup}, D., {et~al.} 2018,
  \mnras, 475, 5487

\bibitem[{{Silva Aguirre} {et~al.}(2015){Silva Aguirre}, {Davies}, {Basu},
  {Christensen-Dalsgaard}, {Creevey}, {Metcalfe}, {Bedding}, {Casagrande},
  {Handberg}, {Lund}, {Nissen}, {Chaplin}, {Huber}, {Serenelli}, {Stello}, {Van
  Eylen}, {Campante}, {Elsworth}, {Gilliland}, {Hekker}, {Karoff}, {Kawaler},
  {Kjeldsen}, \& {Lundkvist}}]{silva15}
{Silva Aguirre}, V., {Davies}, G.~R., {Basu}, S., {et~al.} 2015, \mnras, 452,
  2127

\bibitem[{{Skillman} {et~al.}(2017){Skillman}, {Monelli}, {Weisz}, {Hidalgo},
  {Aparicio}, {Bernard}, {Boylan-Kolchin}, {Cassisi}, {Cole}, {Dolphin},
  {Ferguson}, {Gallart}, {Irwin}, {Martin}, {Mart{\'\i}nez-V{\'a}zquez},
  {Mayer}, {McConnachie}, {McQuinn}, {Navarro}, \& {Stetson}}]{Skilman2017}
{Skillman}, E.~D., {Monelli}, M., {Weisz}, D.~R., {et~al.} 2017, \apj, 837, 102

\bibitem[{{Snaith} {et~al.}(2014){Snaith}, {Haywood}, {Di Matteo}, {Lehnert},
  {Combes}, {Katz}, \& {G{\'o}mez}}]{snaith14}
{Snaith}, O.~N., {Haywood}, M., {Di Matteo}, P., {et~al.} 2014, \apjl, 781, L31

\bibitem[{{Soubiran} {et~al.}(2003){Soubiran}, {Bienaym{\'e}}, \&
  {Siebert}}]{Soubiran2003}
{Soubiran}, C., {Bienaym{\'e}}, O., \& {Siebert}, A. 2003, \aap, 398, 141

\bibitem[{{Spitoni} {et~al.}(2021){Spitoni}, {Verma}, {Silva Aguirre},
  {Vincenzo}, {Matteucci}, {Vai{\v{c}}ekauskait{\.{e}}}, {Palla}, {Grisoni}, \&
  {Calura}}]{spitoni21}
{Spitoni}, E., {Verma}, K., {Silva Aguirre}, V., {et~al.} 2021, \aap, 647, A73

\bibitem[{{Steinmetz} {et~al.}(2020){Steinmetz}, {Matijevi{\v{c}}}, {Enke},
  {Zwitter}, {Guiglion}, {McMillan}, {Kordopatis}, {Valentini}, {Chiappini},
  {Casagrande}, {Wojno}, {Anguiano}, {Bienaym{\'e}}, {Bijaoui}, {Binney},
  {Burton}, {Cass}, {de Laverny}, {Fiegert}, {Freeman}, {Fulbright}, {Gibson},
  {Gilmore}, {Grebel}, {Helmi}, {Kunder}, {Munari}, {Navarro}, {Parker},
  {Ruchti}, {Recio-Blanco}, {Reid}, {Seabroke}, {Siviero}, {Siebert}, {Stupar},
  {Watson}, {Williams}, {Wyse}, {Anders}, {Antoja}, {Birko}, {Bland-Hawthorn},
  {Bossini}, {Garc{\'\i}a}, {Carrillo}, {Chaplin}, {Elsworth}, {Famaey},
  {Gerhard}, {Jofre}, {Just}, {Mathur}, {Miglio}, {Minchev}, {Monari},
  {Mosser}, {Ritter}, {Rodrigues}, {Scholz}, {Sharma}, {Sysoliatina}, \& {RAVE
  Collaboration}}]{rave20}
{Steinmetz}, M., {Matijevi{\v{c}}}, G., {Enke}, H., {et~al.} 2020, \aj, 160, 82

\bibitem[{{Thomas} {et~al.}(2024){Thomas}, {Battaglia}, {Gran},
  {Fernandez-Alvar}, {Tsantaki}, {Pancino}, {Hill}, {Kordopatis}, {Gallart},
  {Turchi}, \& {Masseron}}]{thomas24}
{Thomas}, G.~F., {Battaglia}, G., {Gran}, F., {et~al.} 2024, arXiv e-prints,
  arXiv:2404.02578

\bibitem[{{Tissera} {et~al.}(2016){Tissera}, {Machado}, {Sanchez-Blazquez},
  {Pedrosa}, {S{\'a}nchez}, {Snaith}, \& {Vilchez}}]{tissera16}
{Tissera}, P.~B., {Machado}, R. E.~G., {Sanchez-Blazquez}, P., {et~al.} 2016,
  \aap, 592, A93

\bibitem[{{Tolstoy} \& {Saha}(1996)}]{Tolstoy1996_method}
{Tolstoy}, E. \& {Saha}, A. 1996, \apj, 462, 672

\bibitem[{{Tsikoudi}(1979)}]{T79}
{Tsikoudi}, V. 1979, \apj, 234, 842

\bibitem[{{Vergely} {et~al.}(2022){Vergely}, {Lallement}, \& {Cox}}]{vergely22}
{Vergely}, J.~L., {Lallement}, R., \& {Cox}, N.~L.~J. 2022, \aap, 664, A174

\bibitem[{{Vincenzo} {et~al.}(2019){Vincenzo}, {Spitoni}, {Calura},
  {Matteucci}, {Silva Aguirre}, {Miglio}, \& {Cescutti}}]{vincenzo19}
{Vincenzo}, F., {Spitoni}, E., {Calura}, F., {et~al.} 2019, \mnras, 487, L47

\bibitem[{{Weisz} {et~al.}(2014){Weisz}, {Dolphin}, {Skillman}, {Holtzman},
  {Gilbert}, {Dalcanton}, \& {Williams}}]{weisz14}
{Weisz}, D.~R., {Dolphin}, A.~E., {Skillman}, E.~D., {et~al.} 2014, \apj, 789,
  148

\bibitem[{{Weisz} {et~al.}(2012){Weisz}, {Zucker}, {Dolphin}, {Martin}, {de
  Jong}, {Holtzman}, {Dalcanton}, {Gilbert}, {Williams}, {Bell}, {Belokurov},
  \& {Evans}}]{Weisz2012}
{Weisz}, D.~R., {Zucker}, D.~B., {Dolphin}, A.~E., {et~al.} 2012, \apj, 748, 88

\bibitem[{{Xiang} \& {Rix}(2022)}]{xiang22}
{Xiang}, M. \& {Rix}, H.-W. 2022, \nat, 603, 599

\bibitem[{{Yanny} {et~al.}(2009){Yanny}, {Rockosi}, {Newberg}, {Knapp},
  {Adelman-McCarthy}, {Alcorn}, {Allam}, {Allende Prieto}, {An}, {Anderson},
  {Anderson}, {Bailer-Jones}, {Bastian}, {Beers}, {Bell}, {Belokurov},
  {Bizyaev}, {Blythe}, {Bochanski}, {Boroski}, {Brinchmann}, {Brinkmann},
  {Brewington}, {Carey}, {Cudworth}, {Evans}, {Evans}, {Gates}, {G{\"a}nsicke},
  {Gillespie}, {Gilmore}, {Nebot Gomez-Moran}, {Grebel}, {Greenwell}, {Gunn},
  {Jordan}, {Jordan}, {Harding}, {Harris}, {Hendry}, {Holder}, {Ivans},
  {Ivezi{\v{c}}}, {Jester}, {Johnson}, {Kent}, {Kleinman}, {Kniazev},
  {Krzesinski}, {Kron}, {Kuropatkin}, {Lebedeva}, {Lee}, {French Leger},
  {L{\'e}pine}, {Levine}, {Lin}, {Long}, {Loomis}, {Lupton}, {Malanushenko},
  {Malanushenko}, {Margon}, {Martinez-Delgado}, {McGehee}, {Monet}, {Morrison},
  {Munn}, {Neilsen}, {Nitta}, {Norris}, {Oravetz}, {Owen}, {Padmanabhan},
  {Pan}, {Peterson}, {Pier}, {Platson}, {Re Fiorentin}, {Richards}, {Rix},
  {Schlegel}, {Schneider}, {Schreiber}, {Schwope}, {Sibley}, {Simmons},
  {Snedden}, {Allyn Smith}, {Stark}, {Stauffer}, {Steinmetz}, {Stoughton},
  {SubbaRao}, {Szalay}, {Szkody}, {Thakar}, {Sivarani}, {Tucker}, {Uomoto},
  {Vanden Berk}, {Vidrih}, {Wadadekar}, {Watters}, {Wilhelm}, {Wyse}, {Yarger},
  \& {Zucker}}]{yanny09}
{Yanny}, B., {Rockosi}, C., {Newberg}, H.~J., {et~al.} 2009, \aj, 137, 4377

\bibitem[{{Yoachim} \& {Dalcanton}(2006{\natexlab{a}})}]{yoachim06}
{Yoachim}, P. \& {Dalcanton}, J.~J. 2006{\natexlab{a}}, \aj, 131, 226

\bibitem[{{Yoachim} \& {Dalcanton}(2006{\natexlab{b}})}]{YoachimDalcaton2006}
{Yoachim}, P. \& {Dalcanton}, J.~J. 2006{\natexlab{b}}, \aj, 131, 226

\bibitem[{{Yoachim} \& {Dalcanton}(2008)}]{YD08}
{Yoachim}, P. \& {Dalcanton}, J.~J. 2008, \apj, 683, 707

\bibitem[{{Yuan} {et~al.}(2023){Yuan}, {Li}, {Martin}, {Monari}, {Famaey},
  {Siebert}, {Ardern-Arentsen}, {Sestito}, {Thomas}, {Hill}, {Ibata},
  {Kordopatis}, {Starkenburg}, \& {Viswanathan}}]{yuan23}
{Yuan}, Z., {Li}, C., {Martin}, N.~F., {et~al.} 2023, arXiv e-prints,
  arXiv:2311.08464

\bibitem[{{Zhang} {et~al.}(2024){Zhang}, {Ardern-Arentsen}, \&
  {Belokurov}}]{zhang24}
{Zhang}, H., {Ardern-Arentsen}, A., \& {Belokurov}, V. 2024, \mnras, 533, 889

\end{thebibliography}

\begin{appendix}
\label{sec:Appendix}

\section{DisPar-{\it Gaia}, mimicking {\it Gaia} observational effects in the synthetic CMD}

\begin{figure}
\centering 
\includegraphics[width = 0.5\textwidth]{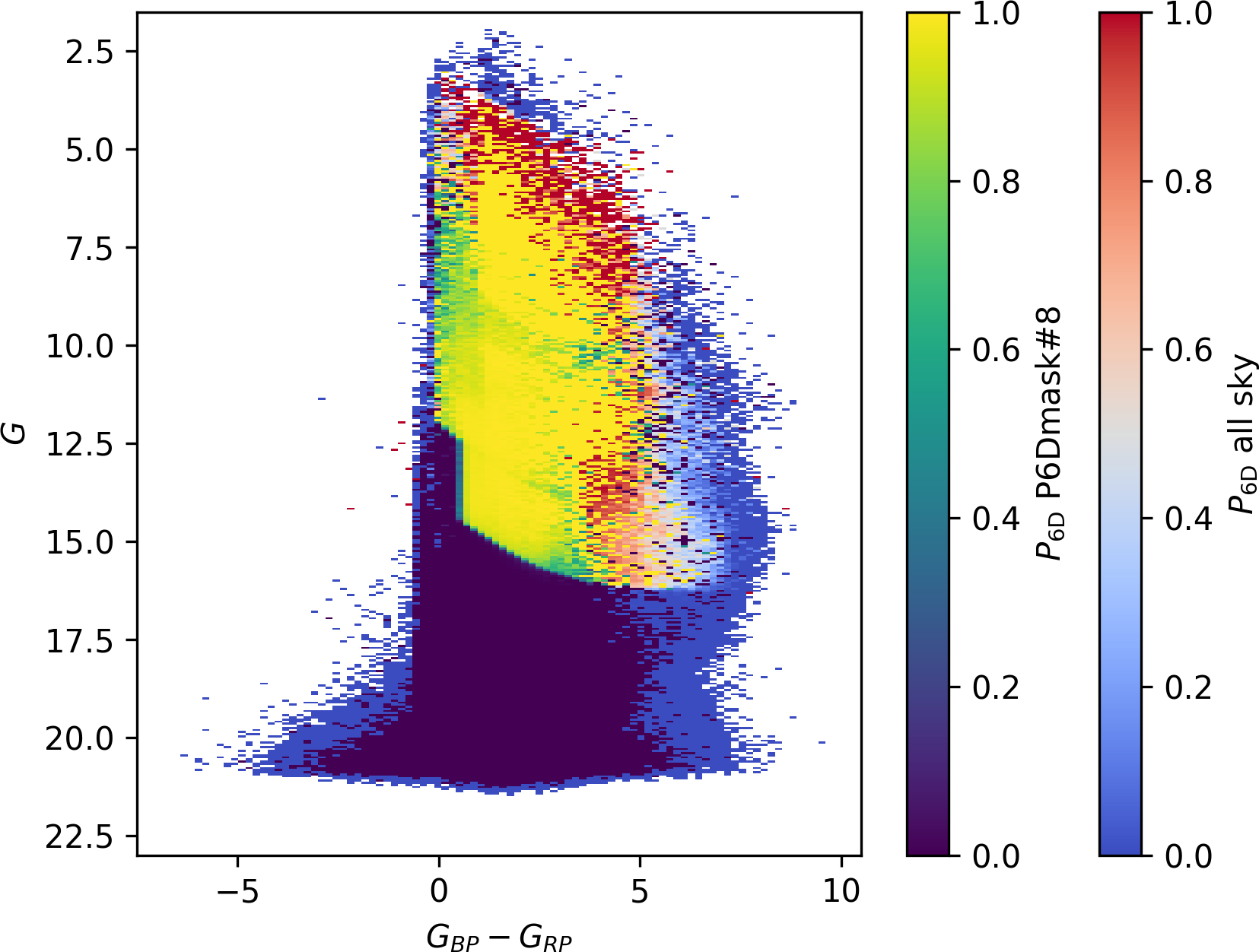} \\   
\caption{Example of a 6D completeness mask. This figure represents the ratio of the number of 6D stars and 5D stars (N$_{6D}$/N$_{5D}$ as a function of apparent colour and magnitude for healpixel number 8 ($P_{6D}$ P6Dmask$\#$8). Yellow colours denote regions where all 5D stars have a radial velocity measurement, whereas dark blue is linked to an abscence of radial velocity measurements. We show how, in areas in the CMD not covered by stars from this region, we use the P$_{6D}$ all sky. This is shown in this plot as background, where blue colours denote zero probability and red means 100$\%$ probability.} 
\label{fig:DisPar6D} 
\end{figure}

\begin{figure*}
\centering 
\includegraphics[width = 0.95\textwidth]{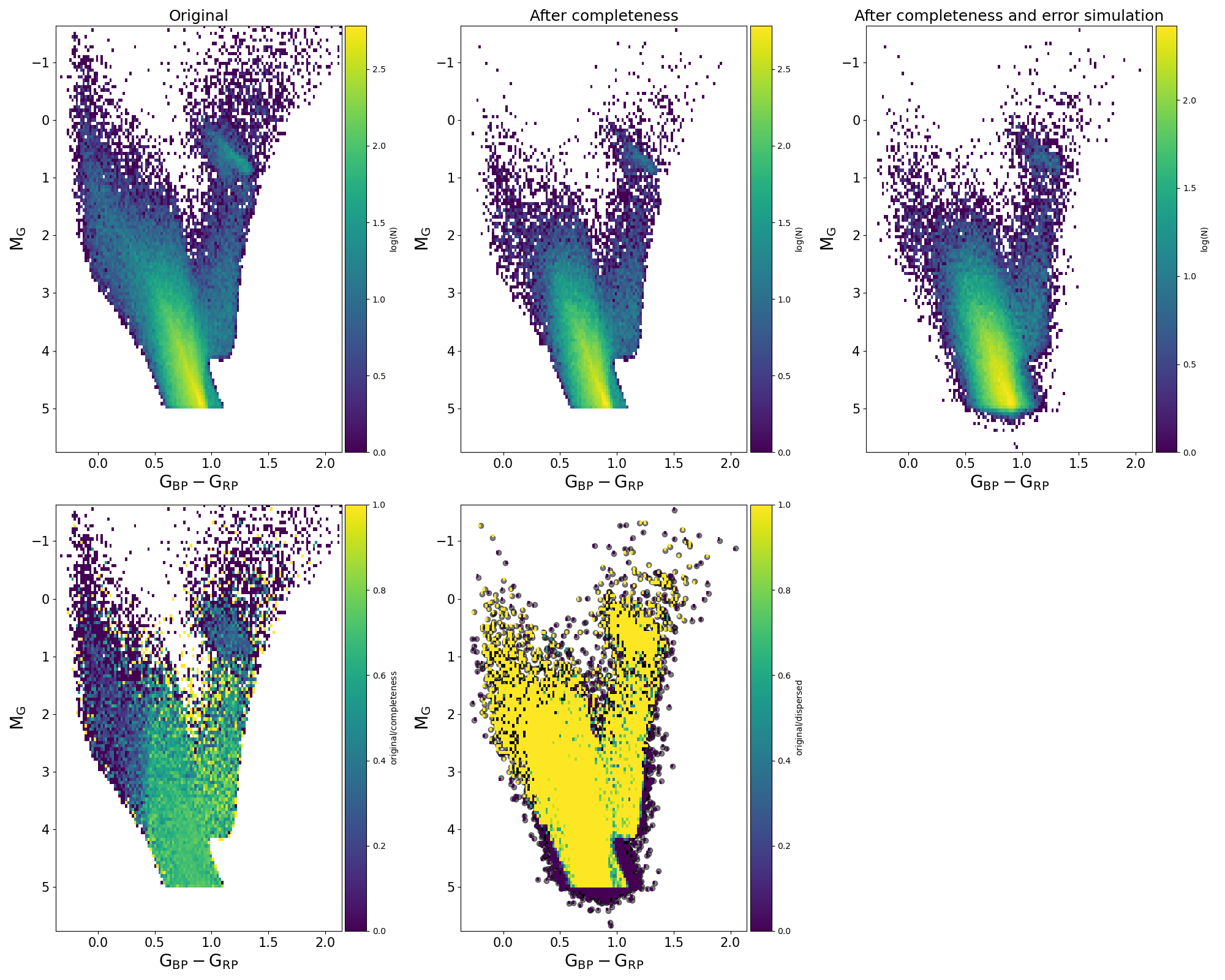} \\   
\caption{Error and completeness simulation using {\tt GAIA-DisPar}. Panel a: Original synthetic CMD, complete and not affected by observational errors. Panel b: Synthetic CMD after completeness has been simulated. Note that many stars in the upper main sequence are lost due to the blue sharp cut in the apparent color of the P6Dmask. Panel c: Final synthetic CMD, considering both completeness and error simulation. This will be the synthetic CMD to be used in the computation of the SFH using {\tt DirSFH}. Panel d: ratio of the synthetic CMD after completeness and the original one. Note the clear effect on the distribution of stars in the CMD due to the RVS sample selection. Panel e: ratio of the original synthetic CMD and the dispersed one to highlight the widening of some areas, especially at fainter magnitudes.} 
\label{fig:DisPar1} 
\end{figure*}

\begin{figure}
\centering 
\includegraphics[width = 0.43\textwidth]{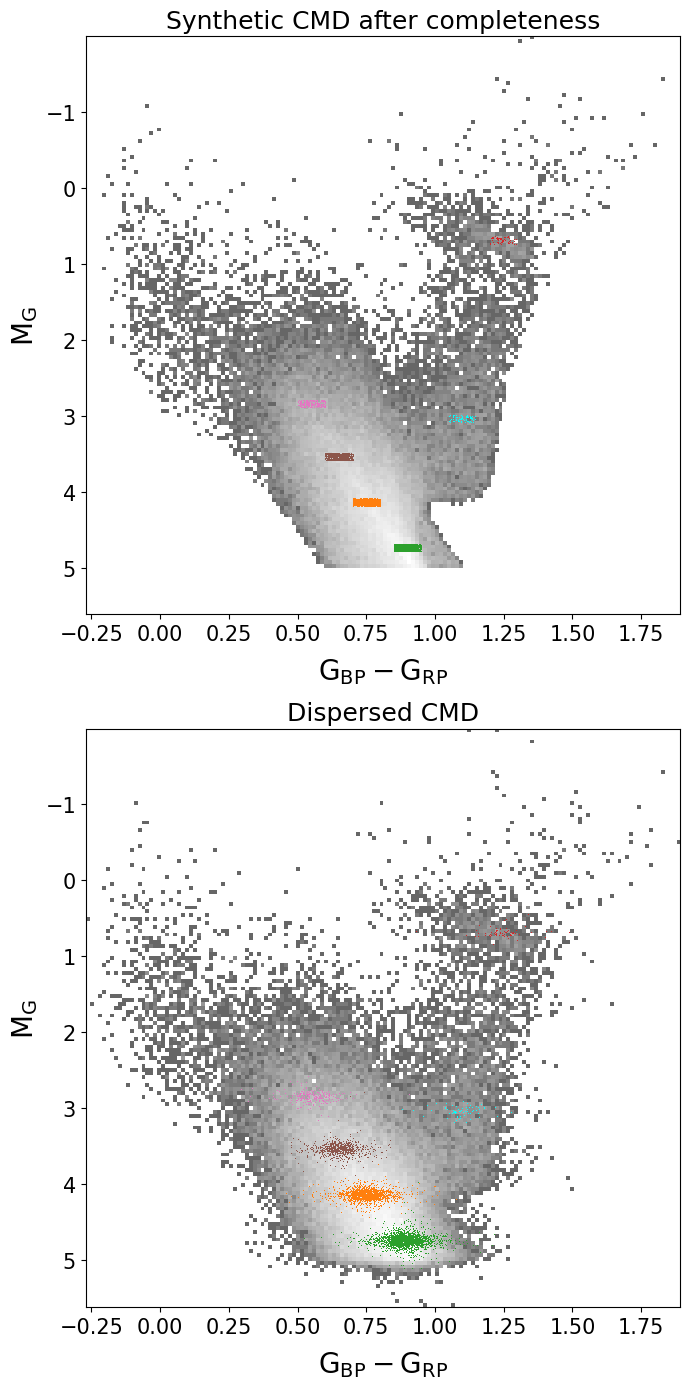} \\   
\caption{Error simulation using {\tt DisPar-Gaia}. Top panel: synthetic CMD before error simulation. To illustrate observational uncertainties some particular regions are highlighted in different colours. Bottom panel: same synthetic CMD after simulating photometric errors using {\tt GAIA-DisPar}. The coloured points represent the position of the synthetic stars highlighted in the top panel after the error simulation. As expected, fainter stars are more affected by observational uncertainties than brighter stars.} 
\label{fig:DisPar2} 
\end{figure}

As described in Sect.~\ref{sec:method_sfh}, our approach to compute SFHs is based on the comparison of {\it Gaia} observed CMDs in the absolute plane with those of synthetic model populations constructed from combinations of SSPs obtained from a mother CMD  containing a wide range of ages and metallicities \citep[see][for more details]{2024A&A...687A.168G}. However, whereas a synthetic CMD is free of errors and each star's properties are in principle fully determined\footnote{Although still affected by bolometric correction uncertainties when translating their luminosities into the desired photometric passband.} (age, metallicity, mass, evolutionary stage, absolute colour and magnitudes, etc.), {\it Gaia} observational datasets are affected by uncertainties and incompleteness due to various and complex sources. In the traditional application of CMD fitting techniques, i.e. for dwarf galaxies in the Local Group, this problem is tackled via Artificial Star Tests \citep[ASTs,][]{1999ApJ...514..665G, 2010ApJ...720.1225M}. Photometric errors and incompleteness ('observational') effects can be very successfully characterized by simply injecting artificial stars of known colours and magnitudes into the photometric images, and measuring their colours and magnitudes using the same method as for the determination of the photometry in the original images. The quantities measured for the artificial stars will be affected by the same observational effects as the real stars (in a statistical sense), and the stars lost in the process allow to characterize the incompleteness of the data.  Unfortunately, in the case of {\it Gaia} data this approach is not feasible. On the one hand, we do not have images into which we can inject synthetic stars for the ASTs computation. On the other hand, the sources of errors and incompleteness are more varied and complex, including not only errors in the photometry, but also due to the distance and reddening determination, crowding effects, {\it Gaia} scanning law, RVS observing limitations, etc. Within this context, and with the necessity of simulating {\it Gaia} observational uncertainties in synthetic CMDs, we developed {\tt DisPar-Gaia}, a new approach that mimicks the philosophy of our previous {\tt DisPar} \citep[e.g.][based on ASTs]{2021MNRAS.501.3962R}, but tailored for {\it Gaia} data.

The main idea behind {\tt DisPar-Gaia} is to simulate in the synthetic mother CMDs the observational conditions of particular samples drawn from {\it Gaia} data as if the stars in these synthetic CMDs were observed by {\it Gaia} itself. In this simulation we need to include i) the 3D distribution of the targeted stars in our Galaxy -given that uncertainties and completeness due to reddening, scanning law, RVS selection function, apparent magnitudes, etc., depend on distance or direction in the sky- and ii) the parameters (e.g. \texttt{phot\_bp\_rp\_excess\_factor}, photometric errors, reddening and reddening errors, \texttt{parallax\_error}, etc.) that are used in the quality cuts to obtain the final samples to be analysed (which we will call {\it QSHAG}). We thus define an observed sample that fulfills all physical requirements (geometry, radial velocity, integrals of motion, etc) but avoids the quality cuts, i.e. containing as much information on uncertainties as possible. We call this sample the {\it full} sample, and it will be used to simulate in the mother CMD the observational conditions of the sample under study. Sects.~\ref{sec:data} and~\ref{sec:method_selection} provide the information for the selection of the {\it full} and {\it QSHAG} samples for the particular science case in this paper.

In a first step, {\tt DisPar-Gaia} assigns to every synthetic star a value of the following properties: set1 = [\texttt{l}, \texttt{b}, \texttt{parallax}, \texttt{phot\_bp\_rp\_excess\_factor}]. This is done in such a way that the distribution of these parameters for the observed dataset ({\it full} sample) and for the synthetic mother CMD are identical without using necessarily the exact same values (avoiding discretisation). This allows us to locate each synthetic star in our Galaxy, estimate a value of reddening and its error, and finally move their absolute colours and magnitudes to the apparent plane (observational plane). For the reddening computation, {\tt DisPar-Gaia} uses the same extinction maps and recipes used while dereddening the observed sample. In this case, \cite{lallement22} and \cite{vergely22} dust maps with the \cite{Fitz19} recipes).

In a second step, we characterise a new set of parameters that will be crucial to mimic observational effects(we decided to restrict it to a radius of 1.3 kpc around the Sun) in the synthetic CMDs, and that are a function of the apparent magnitudes of the stars:  set2 = [\texttt{parallax\_over\_error}, \texttt{radial\_velocity\_error}, $\Delta m_{G}$, $\Delta m_{bp}$, $\Delta m_{rp}$], where $\Delta m_{G}$, $\Delta m_{bp}$, and $\Delta m_{rp}$ denote the simmetrised photometric errors in the G, G$_{\rm bp}$, and G$_{\rm rp}$ apparent magnitudes, respectively. In this second step, and to make sure that the expected/real behaviour prevails (i.e. faint stars are affected by larger uncertainties than brighter stars), we do not assign a random value following the global distribution but stick to specific procedures. For \texttt{parallax\_over\_error} and \texttt{radial\_velocity\_error} we characterise the observed distribution (from the sample we are analysing) of values of both parameters as a function of apparent colour and magnitude from the {\it full} observed dataset, and assign to each synthetic star a value according to its position in the apparent CMD. Those simulated values will be randomly generated from a Gaussian distribution centered at the average value of the {\it full} sample at that apparent colour and magnitude, using a $\sigma_{Gauss}$ that is equal to the dispersion of observed values in said position. In the case of the uncertainties in the photometric magnitudes we use a linear fit of the run of observed error in the apparent magnitude with respect to such magnitude \citep[i.e. $\Delta m_{x}$ vs. $m_x$ with x being G, G$_{\rm bp}$, and G$_{\rm rp}$; see][]{2021A&A...649A...3R, 2021A&A...649A...5F}.

In a third step, we refine all these parameters (set1 and set2), by finding counterparts of the synthetic stars in the observed CMD of the full sample. If a given synthetic star has 10 or more observed stars near its position in the CMD (Euclidean distance below 0.05 mag), then we link this synthetic star to one of the 10 closest observed stars, and assign new values for these parameters based on those of the linked observed stars (modified according to typical uncertainties\footnote{Given the huge number of synthetic stars compared to the number of observed stars, not assigning the exact observed values avoids repetition (as statistically each observed star will be chosen repeatedly) and allows to have a continuity of values in these parameters.}). If no counterpart is found, like for example in the hypothetical case of a synthetic star located in the bright main sequence if the observed sample is drawn from the (old) halo, then the previously simulated values are retained. This approach allows us to simultaneously have a synthetic mother CMD that is faithful to the properties of the particular observed sample ({\it full}) and display the typical behaviour of {\it Gaia} data uncertainties beyond the particular CMD coverage of the sample under study. 

To mimic a hypothetical observation of the stars in the synthetic CMD with the {\it Gaia} satellite, we still have to take into account the {\it Gaia} selection function. Due to technical limitations of the {\it Gaia} observations, we do not have parameters (5D or 6D) for every star in our Galaxy. \citet[][]{2023A&A...669A..55C} empirically quantified the completeness of the catalogue of {\it Gaia}'s third data release (DR3). The authors made their model available through the {\it gaiaunlimited} Python package. Given a location in our Galaxy (l, b and distance), as well as the apparent colour (G-G$_{\rm rp}$) and magnitude (G) of a star, this package provides the probability of such star being part of the 5D (P$_{5D}$) {\it Gaia} catalogue ({\tt DR3SelectionFunctionTCG}). In the previous steps, we simulated all these observed quantities for every synthetic star, and thus, we can compute the probability of each synthetic star to make it to the 5D catalogue (allowing the simulation of completeness effects in our synthetic CMD). For samples defined using the {\it Gaia} 5D sample the probability of synthetic star to belong to the {\it Gaia} catalogue is given by P = P$_{5D}$).

For samples defined based on RVS information (as in this paper), we also need to simulate the probability of a star to belong to the 6D {\it Gaia} catalogue ($P_{6D}$). For this, rather than using the implementation by \citet[][]{2023A&A...669A..55C}, we have decided to use an alternative, empirical approach. First, we assess the volume within which we can compute SFHs based on the RVS sample. Based on the apparent G-band magnitude limit of the RVS sample, and our need to get as faint as the oMSTO, we decided to restrict it to a radius of 1.3 kpc\footnote{Although for the mask definition in 6D we use all stars in the 3.5 kpc bubble to increase the statistics, for the sample definition we must restrict ourselves to a maximum distance of about 1.3 kpc to ensure that the faintest apparent magnitude with radial velocity measurements is below the oMSTO.} around the Sun. Then, given that we are mainly interested in describing how completeness and errors affect the overall distribution of stars in the CMD, we have computed $P_{6D}$ as the ratio of the number of stars with 6D and 5D information (N$_{6D}$/N$_{5D}$) in small cells across the apparent CMD (see Fig.~\ref{fig:DisPar6D}, we will call this P6Dmask\footnote{For the mask creation we restrict ourselves to the range of colours and magnitudes given by [-7.5, 10.5] and [1.5, 23], respectively, dividing those ranges in 100 and 400 equally-spaced bins.}). Given that the 6D completeness slightly depends on position in the sky, but also considering that for our method we need a large number of stars to properly estimate $P_{6D}$ with enough colour and magnitude resolution, we characterise such ratio using 12 different P6Dmasks by dividing the sky in 12 different healpixels (nside = 1) and an all-sky one. Each mask will contain the corresponding $P_{6D}$ of a given region of the sky across the apparent CMD. This way, for each synthetic star, we can associate its position in the sky with one of the masks, and assign a value of $P_{6D}$ based on its colour and magnitude. In the event that a synthetic star occupies a region in the CMD in which we do not have information, we use the overall mask mentioned before ($P_{6D}$ all sky in Fig.~\ref{fig:DisPar6D}). Once we have assigned to each synthetic star its $P_{5D}$ and/or $P_{6D}$, we can compute its total probability of being observed by Gaia as P = $P_{5D}$ (in the 5D case) or  P = $P_{5D}$ $\times$ $P_{6D}$ (in the case of a 6D sample). Finally, we give to each synthetic star a random value ranging from 0 to 1 (P$_\star$), if P~<~P$_\star$, then such synthetic star is kept, otherwise, we remove it mimicking the {\it Gaia} (5D and/or 6D) selection functions.

Finally, completeness is also affected by the quality cuts that transform our \textit{full} observed sample into the \textit{QSHAG} observed sample actually used for the fit. We need to simulate these cuts in our synthetic CMD as well. For the cuts that are generally applied -and in particular that have been applied in this work- all parameters involved have been simulated in the synthetic CMD. Thus, we can simply apply the same cuts as in the observed samples (see Sect.~\ref{sec:data} in the case of this paper), to automatically take into account the incompleteness derived from such quality cuts in our synthetic CMD.

Finally, for the synthetic stars that remain we alter their absolute magnitudes (M$_x$ with x being G, G$_{\rm bp}$, and G$_{\rm rp}$) in order to simulate the observational errors. For this, we analyse the following equation to compute absolute magnitudes in a given filter {\it x} (M$_x$):

\begin{equation}
   M_X = m_x-Ax+5.-5 \times log_{10}(1000.0/parallax) 
\end{equation}

where m$_x$ is the apparent magnitude in M$_G$, G$_{\rm bp}$, or G$_{\rm rp}$, A$_x$ the extinction, and {\tt parallax} its parallax\footnote{Note here that, given the volume we are considering in this work, we can accept that the inverse of the {\tt parallax} is a good estimate of the distance.}. By applying the corresponding derivatives (error propagation), the total error is given by:

\begin{equation}
   \Delta M_{x}^2 = \Delta M_{x, dist}^2 + \Delta m_x ^2 + \Delta A_x ^2 
\label{eq:err_M}
\end{equation}

where $\Delta$ m$_x$ correspond to the photometric uncertainties in the apparent magnitudes, $\Delta$ A$_x$ is the error in the extinction coefficient (including intrinsic error of the dust map, as well as the one derived from the uncertainty in the position/distance), and $\Delta M_{X, dist}$ is the error due to the uncertainty in the distance determination, which is computed as:

\begin{equation}
   \Delta M_{X, dist} = 2.17/parallax\_over\_error
\end{equation}

Finally, we modify the absolute magnitude in each {\it Gaia} passband based on this determination of the uncertainty in the absolute magnitudes. In this way, the modified magnitude (M'$_x$) in each filter will be given by:

\begin{equation}
   M'_X = M_x + \sigma_{M_x}
\end{equation}

where $\sigma_{M_x}$ is a correction generated randomly from a Gaussian distribution centered at M$_x$ and with a $\sigma_{gauss}$ of $\Delta$ M$_x$ (see Equation~\ref{eq:err_M}). The final outcome of applying {\tt DisPar-Gaia} to a synthetic CMD mimicking {\it Gaia} observational effects for a sample defined in 6D (RVS) is shown in Figs.~\ref{fig:DisPar1} and~\ref{fig:DisPar2}. Such synthetic CMD, including errors and completeness effects, can be directly compared to our observed {\it QSHAG} sample in order to compute SFHs using DirSFH. Along this and other works, we call this new synthetic mother CMD the ''{\it dispersed}'' mother CMD.


\section{The mask.}
\label{sec:Appendix_mask}
    \begin{figure}
        \centering
        \includegraphics[width=1\linewidth, trim= 0 50 0 100]{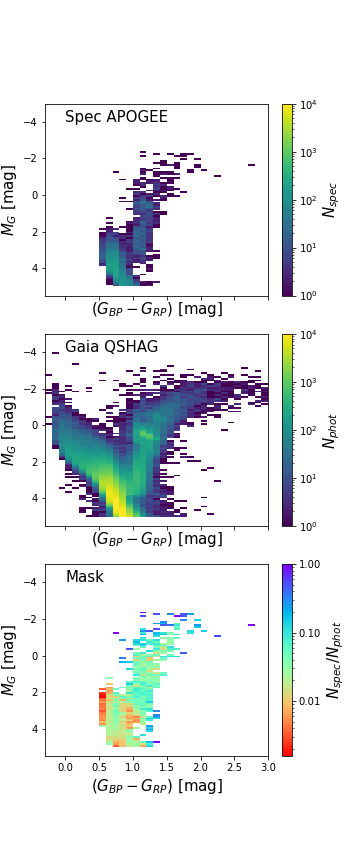}
        \caption{Left panel: Density map in the color-magnitud diagram, in bins of 0.1 mag in $M_{G}$ and G$_BP$-G$_RP$ of APOGEE spectroscopic observations within the volume analysed. Middle panel: Same as left panel by for Gaia photometric observations in the volume analysed. Right panel: Mask, i.e., the fraction of spectroscopic measurements with respect to the photometric observations in the same bins as left and middle panels.}
        \label{fig:mask}
    \end{figure}

    \begin{figure*}
    \centering
    \includegraphics[width=1\linewidth]{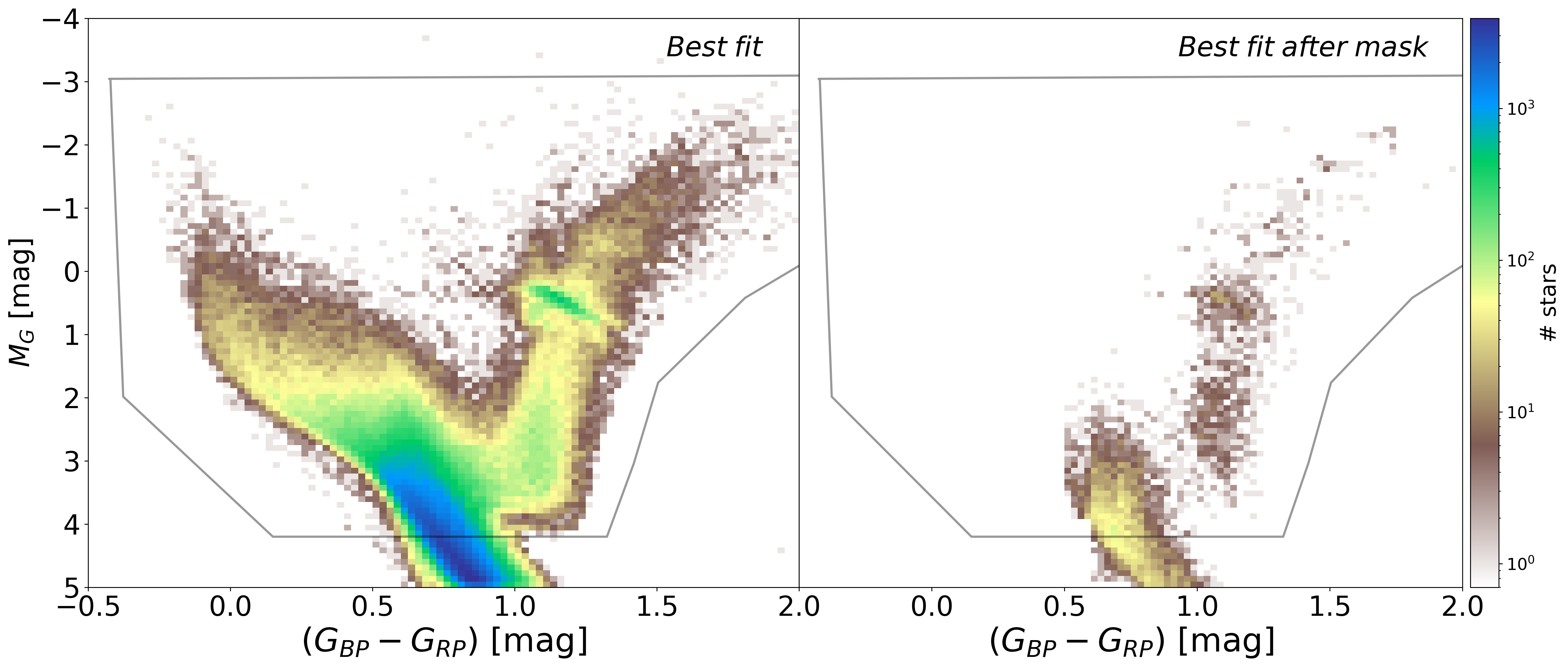}
    \caption{Left panel: Density map of the best fit solution for the \textit{ks\_thin} disc selection. Right panel: Density map of the best fit solution after applying the mask computed following the APOGEE spectroscopic observations in our cylindrical local volume.}
    \label{solution_after_mask}
    \end{figure*}

The comparison between spectroscopic measurements and our resulting metallicity distributions must account for the fact that spectroscopic observations are inherently less numerous and more incomplete than photometric observations. This discrepancy arises because the selection functions of spectroscopic surveys are limited to specific targets in the sky, whereas Gaia's photometric observations encompass all stars up to a magnitude $G \sim 21$ .

To address this issue, we compute a \textit{mask} that ensures consistency between the datasets. This mask is created by computing the fraction of stars observed spectroscopically with respect to the \textit{Gaia} photometric observations in bins of color and magnitude, i.e., $N_{spec}/N_{phot}$. The bins are square regions with sides of 0.1 magnitudes. Then, within our solution CMD we randomly select the corresponding fraction of stars in the same bins of color and magnitude.

Figure \ref{fig:mask} shows an example of the \textit{mask} computed for APOGEE observations in our volume which verifies the \textit{ks\_thin} disc selection. The upper and middle panels show the density of stars with spectroscopic and photometric observations, respectively, across the CMD. The bottom panel shows the division of the two, that is, the fraction of stars observed spectroscopically with respect to the photometric observations in each bin of color and magnitude. 

Figure \ref{solution_after_mask} displays on the left panel the best fit solution of the \textit{ks\_thin} disc. The right panel shows the distribution in the color-magnitude diagram of the solution after applying the mask computed following the APOGEE spectroscopic observations in our volume. These stars from the \textit{masked} solution are the ones whose metallicities would be compared with APOGEE spectroscopic data in Section \ref{sec:MDF}. The same method is computed for comparisons with the other spectroscopic surveys analyzed in Sections \ref{sec:MDF} and \ref{Section_afe}.

\section{Testing probabilities.}
\label{sec:other_probs}

\begin{figure}
\includegraphics[width=1\linewidth]{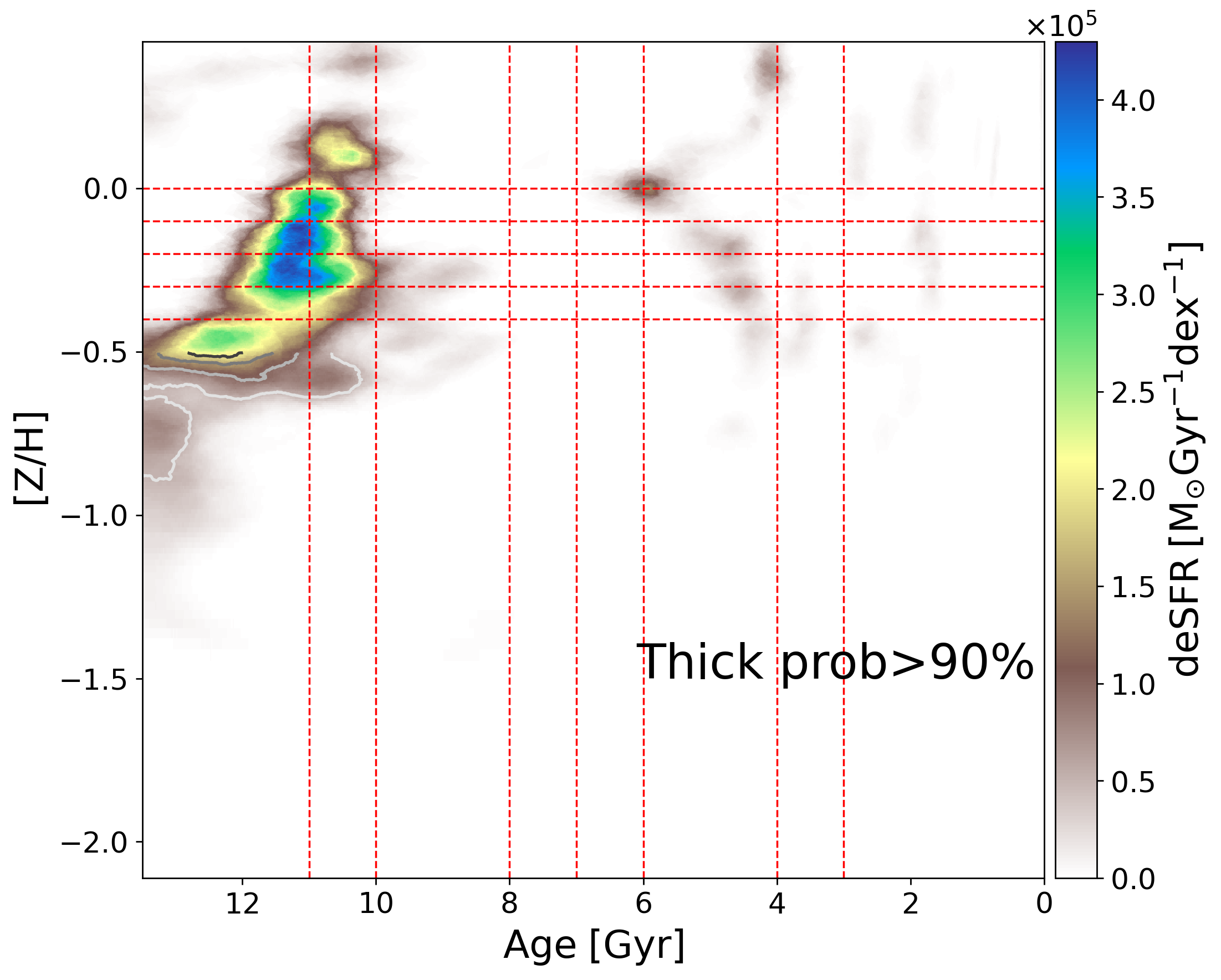}
\includegraphics[width=1\linewidth]{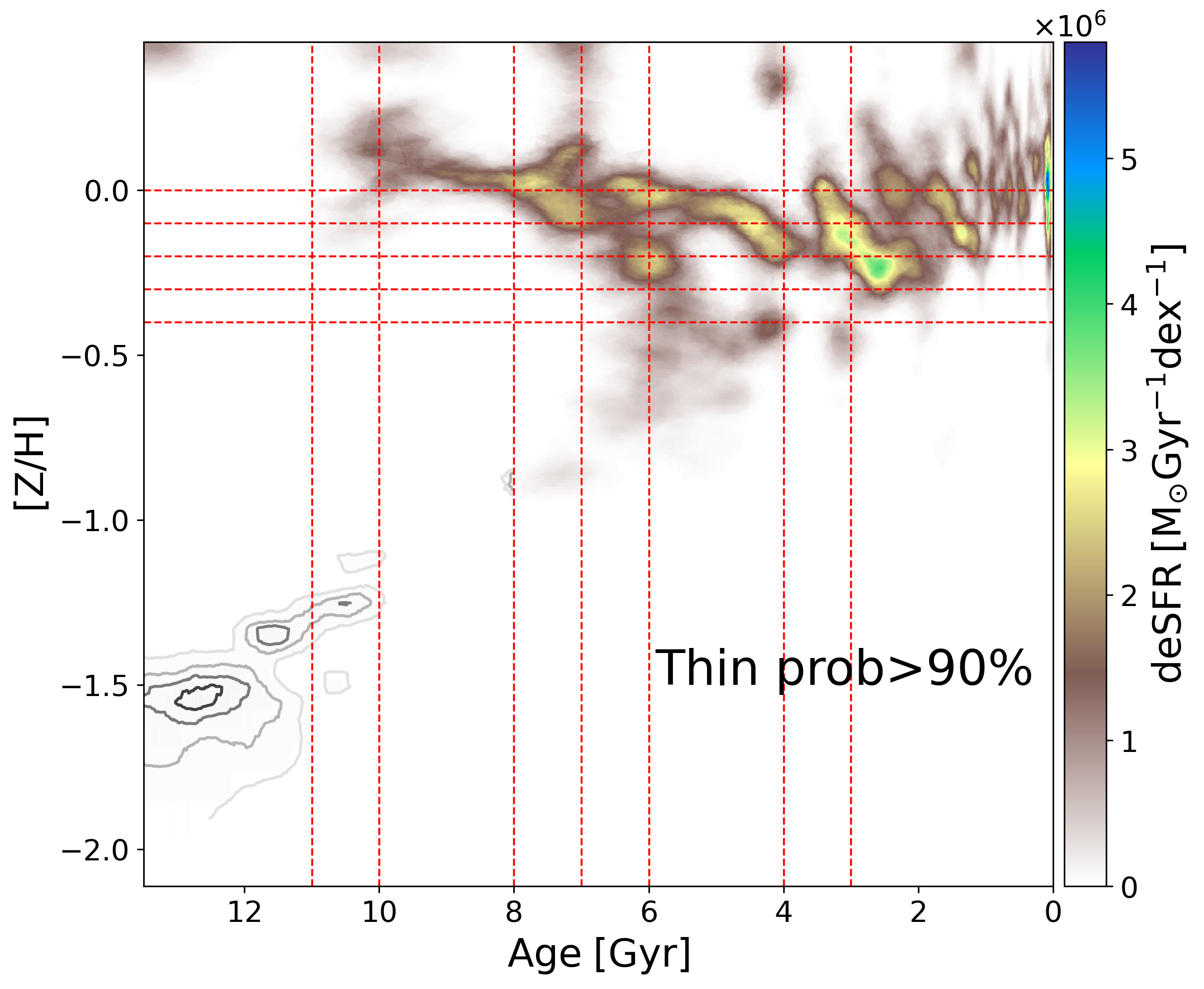}
\caption{Dynamically evolved star formation rates as a function of age and metallicity, as in Figure \ref{both_thinthick}, but in this case of stars with probability higher than 90\% of belonging to the \textit{ks\_thick} (top) and \textit{ks\_thin} (bottom) discs.}
\label{probM90}
\end{figure}

\begin{figure}
\includegraphics[width=1\linewidth]{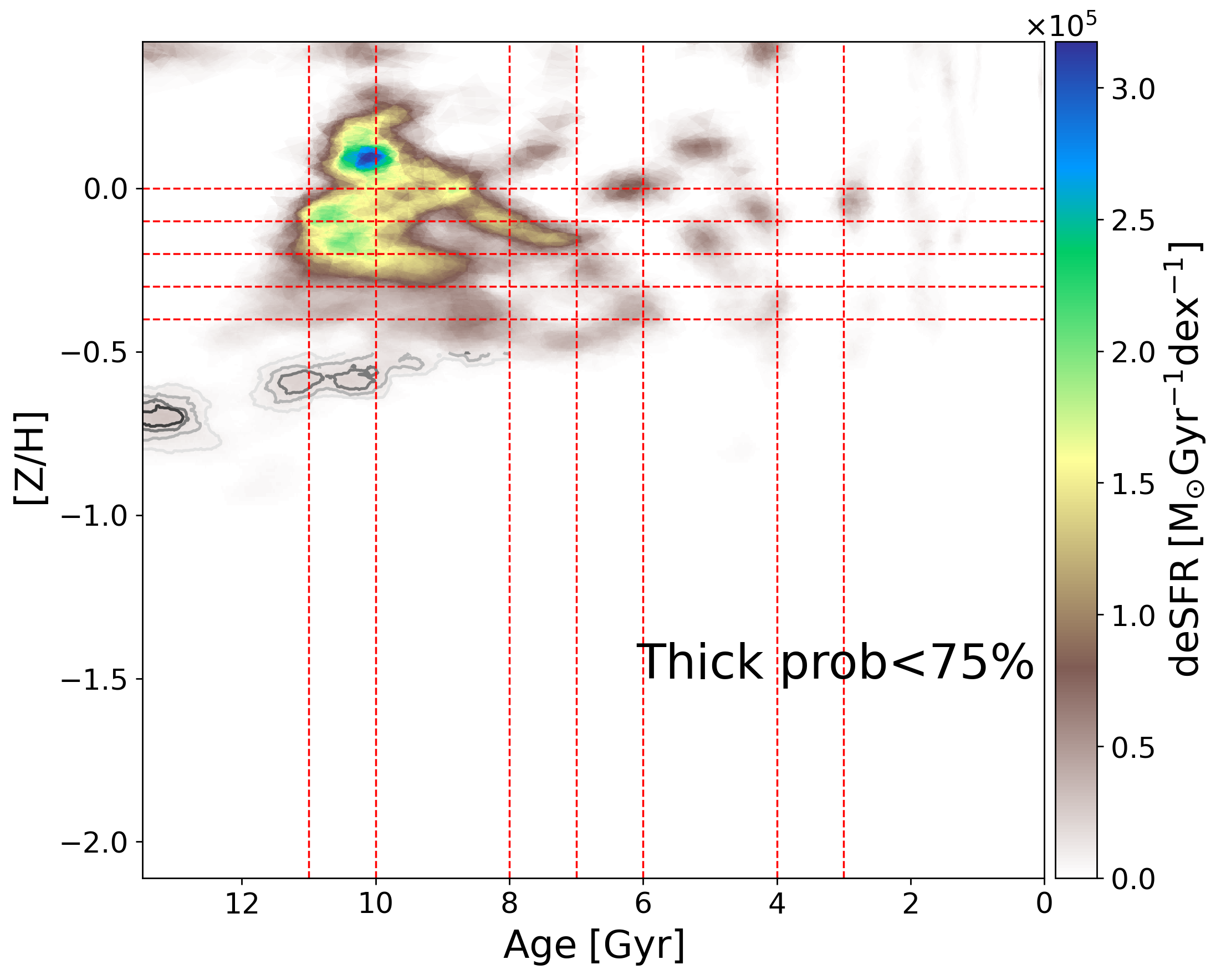}
\includegraphics[width=1\linewidth]{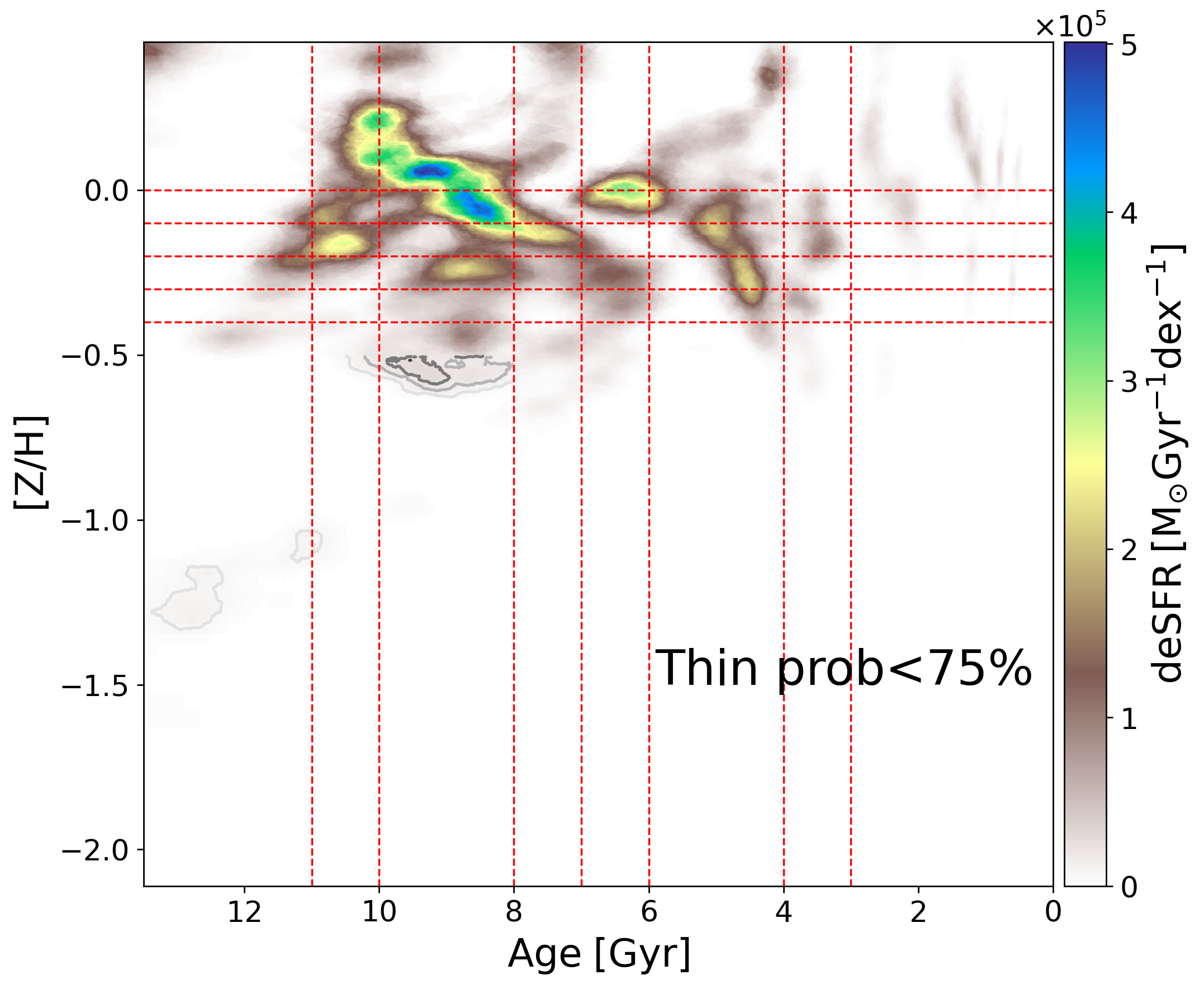}
\caption{Dynamically evolved star formation rates as a function of age and metallicities, as in Figure \ref{both_thinthick}, but in this case of stars with probability between 50\% and 75\% of belonging to the \textit{ks\_thick} (top) and \textit{ks\_thin} (bottom) discs.}
\label{probm75}
\end{figure}

\begin{figure*}
    \centering   \includegraphics[width=1\linewidth]{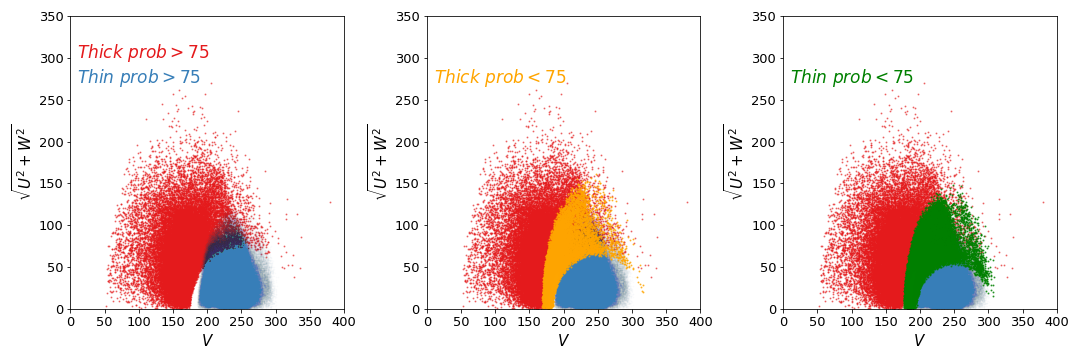}
    \caption{Left panel: Toomre diagram of the selected \textit{ks\_thin} disc (blue) and \textit{ks\_thick} disc (red) stellar samples, considering quality selection criteria, as in Figure \ref{kine}. Middle panel: Same as the left panel but with stars with probabilities between 50$\%$ and 75$\%$ of belonging to the kinematic thick disc overplotted in orange. Right panel: Same as left panel but with stars with probabilities between 50$\%$ and 75$\%$ of belonging to the kinematic thin disc overplotted in green.}
    \label{kine_transition}
\end{figure*}

\begin{figure}
    \centering   \includegraphics[width=1\linewidth]{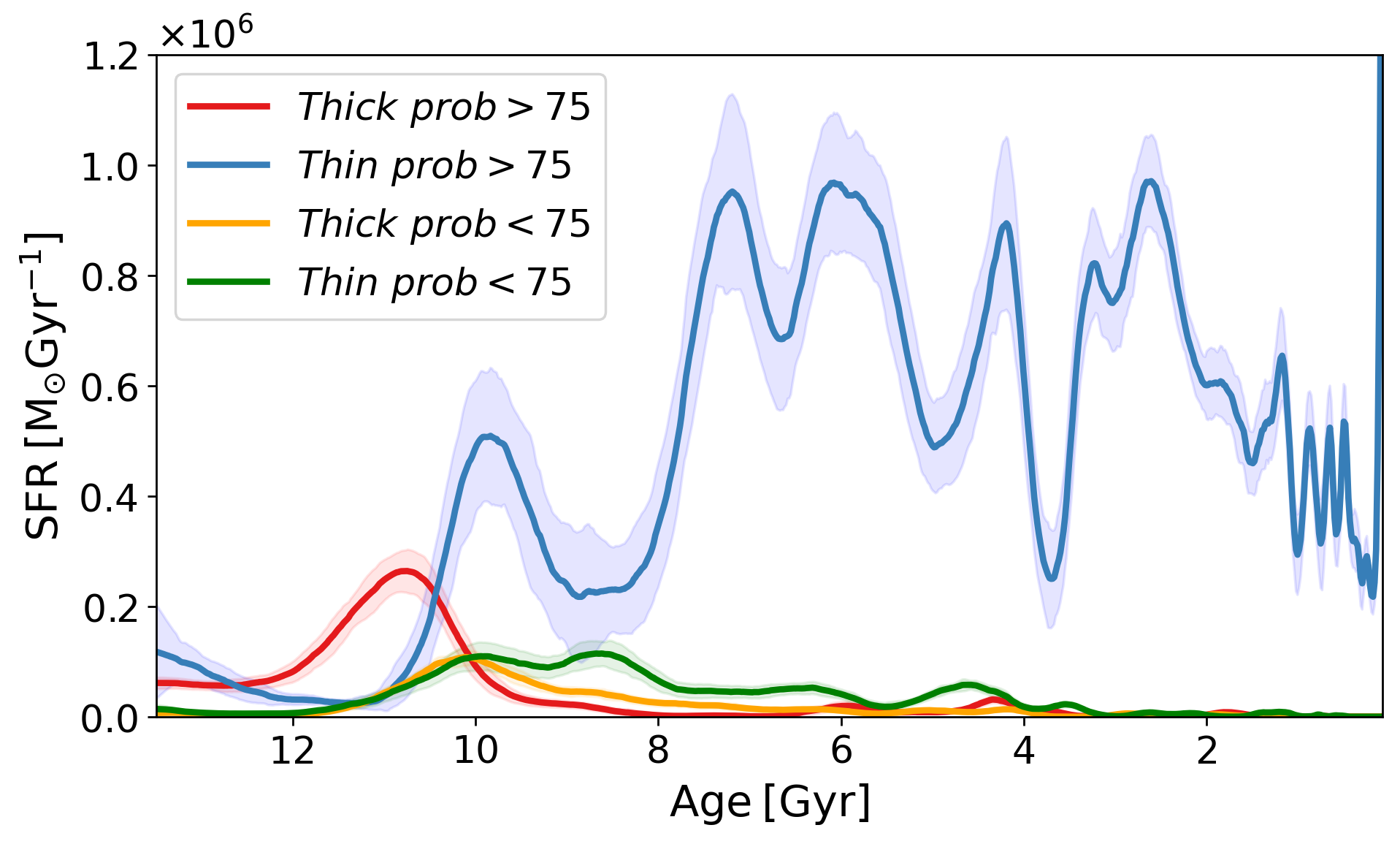}
    \caption{Dynamically evolved star formation rates as a function of stellar age derived for the \textit{ks\_thick} (blue), \textit{ks\_thin} (red) disks, and for  with probabilities between 50$\%$ and 75$\%$ of belonging to the kinematic thick disc (orange) and kinematic thin disc (green).}
    \label{SFH_probm75}
\end{figure}

In this appendix, we test how selection based on probability impacts the resulting deSFHs and explore what this reveals about the kinematic transition from hotter to cooler orbits over time.

Figure \ref{probM90} presents the deSFH obtained from stars with a probability higher than 90$\%$ of belonging to the \textit{ks\_thick} (top) and \textit{ks\_thin} (bottom) discs. The deSFHs closely resemble those derived from stars with probabilities higher than 75$\%$, particularly for the \textit{ks\_thick} disc. However, the younger episode of star formation with super-solar metallicities is weaker in this case. For the \textit{ks\_thin} disc, the deSFH of stars with probabilities exceeding 90$\%$ is more concentrated at ages around $\sim$3 Gyr and younger than 1 Gyr. Nonetheless, all the features identified in the deSFH of stars with probabilities greater than 75$\%$, as discussed in the main text, are also present here.

Figure \ref{probm75} shows the deSFH of stars with probabilities lower than 75$\%$ of belonging to the \textit{ks\_thick} (top) and \textit{ks\_thin} (bottom) discs (and higher than 50\% since they have been classified as belonging to these components). For the \textit{ks\_thick} disc, we exclude the stars with probabilities higher than 25$\%$ of belonging to the \textit{ks\_halo}. These stars exhibit kinematics between those of the final selection for \textit{ks\_thin} and \textit{ks\_thick} disc stars, as illustrated in Figure \ref{kine_transition}. The deSFH indicates that \textit{ks\_thick} disc stars with cooler orbits align the younger and more metal-rich end of the hotter \textit{ks\_thick} disc deSFH, preserving the distinction between the two episodes of star formation below and above solar metallicity. The \textit{ks\_thin} disc with probabilities lower than 75$\%$ concentrates at older ages (between 11 and 6) than the cooler \textit{ks\_thin} disc, with only a few younger and less prominent star formation events occurring at later ages, and very few stars younger than 3 Gyr. This corroborates that we are seen a transition in age and metallicity with the kinematic cooling of the Milky Way disc. This analysis underscores the kinematic settlement that the Milky Way experienced over a relatively short period of time. 

Finally, it is worth noting that stars with probabilities lower than 75$\%$ constitute a minority of the total population in the \textit{ks\_thick} and \textit{ks\_thin} discs. They do not contribute significantly to the total deSFH, as Figure\ref{SFH_probm75} demonstrates.

\end{appendix}

\end{document}